\documentclass[%
reprint,
amsmath,amssymb,
aps,
pra,
floatfix,
]{revtex4-1}

\usepackage{graphicx}
\usepackage{dcolumn}
\usepackage{bm}
\usepackage{color}
\usepackage{soul}

\usepackage{float} 

\begin{document}

\title{Strong-field control and enhancement of chiral response in bi-elliptical high-order harmonic generation: an analytical model}

\author{David Ayuso$^1$} \email{david.ayuso@mbi-berlin.de}
\author{Piero Decleva$^2$}
\author{Serguei Patchkovskii$^{1}$}
\author{Olga Smirnova$^{1,3}$} \email{olga.smirnova@mbi-berlin.de}

\affiliation{$^1$Max-Born Institute for Nonlinear Optics and Short Pulse Spectroscopy, Max-Born-Stra{\ss}e 2A, D-12489 Berlin, Germany}
\affiliation{$^2$Dipartimento di Scienze Chimiche e Farmaceutiche, Universit\`a degli Studi di Trieste, via L. Giorgieri 1, 34127 Trieste, Italy}
\affiliation{$^3$Technische Universit\"at Berlin, Ernst-Ruska-Geb\"aude, Hardenbergstra{\ss}e 36A, 10623 Berlin, Germany}

\raggedbottom

\begin{abstract}

The generation of high-order harmonics in a medium of chiral molecules driven by intense bi-elliptical laser fields can lead to strong chiroptical response in a broad range of harmonic numbers and ellipticities [D. Ayuso et al, \emph{J. Phys. B} \textbf{51}, 06LT01 (2018)].
Here we present a comprehensive analytical model that can describe the most relevant features arising in the high-order harmonic spectra of chiral molecules driven by strong bi-elliptical fields. 
Our model recovers the physical picture underlying chiral high-order harmonic generation based on ultrafast chiral hole motion and identifies the rotationally invariant molecular pseudoscalars responsible for chiral dynamics.
Using the chiral molecule propylene oxide as an example, we show that one can control and enhance the chiral response in bi-elliptical high-order harmonic generation by tailoring the driving field, in particular by tuning its frequency, intensity and ellipticity, exploiting a suppression mechanism of achiral background based on the linear Stark effect.

\end{abstract}

\maketitle

High-order harmonic generation (HHG) is an extremely nonlinear process that converts intense radiation, usually in the infrared (IR) or mid-IR domain, into high-energy photons, with frequencies that are high-integer multiples of that of the driving field \cite{Ferray1998JPB,Krausz2009RevModPhys}.
The most established interpretation of HHG is based on the three step model \cite{Corkum1993PRL}.
The first step is tunnel ionization from an outer shell of an atom or a molecule, facilitated by the distortion of the electrostatic potential induced by the laser field.
It is followed by the second step: laser-driven electron propagation in the continuum.
The third step takes place when the electron is brought back to recombine with the core, releasing the energy accumulated during its round-trip in the form of a high-energy photon.
The three steps occur within one optical cycle of the driving field, thus leading to the formation of attosecond pulses.
The use of attosecond pulses generated via HHG in pump-probe experiments has enabled one to monitor ultrafast electron dynamics in atoms \cite{Uiberacker2007,Schultze2010Science,Goulielmakis2010Nature,Ott2014Nature,Gruson2016Science,Cirelli2018NatComm}, molecules \cite{Sansone2010,Ranitovic2014PNAS,Calegari2014Science,Trabattoni2015PRX,Calegari2015JSTQE} and condensed phases \cite{Cavalieri2007}.

The process of HHG is itself a pump-probe spectroscopic technique \cite{Lein2005PRL,Baker2006Science,Baker2008PRL,Smirnova2009Nature,Smirnova2009PNAS,Shafir2012,Pedatzur2015,Bruner2016FD,Cireasa2015NatPhys}.
The first step (tunnel ionization) acts as a pump, triggering an ultrafast response in the atomic or molecular target.
The initiated dynamics is probed in the third step, as harmonic emission is sensitive to the state of the core at the moment of recombination.
Since there is a well-defined relation between the duration of the electron excursion in the continuum and the energy of the emitted harmonics, these provide a series of snapshots of the dynamics in the ion.
The time resolution of the HHG camera is given by the delay in the emission of consecutive harmonics, which is usually on the order of a few tens of attoseconds.
This resolution can be tuned by adjusting the frequency and the intensity of the driving field.

Imaging sub-femtosecond chiral dynamics in chiral molecules is a recent achievement of HHG spectroscopy \cite{Cireasa2015NatPhys}.
A molecule is said to be chiral if it cannot be superimposed to its mirror image \cite{book_Wade_OrganicChemistry}.
The concept of chirality is of great importance, as, for instance, most biological molecules are chiral \cite{book_ComprehensiveChirality}.
Opposite enantiomers, i.e. mirror images of the same chiral molecule, present identical physical and chemical properties, unless they interact with another chiral entity.
The application of chiral light to the generation of high-order harmonics in a medium of chiral molecules has been recently demonstrated to be a powerful technique for chiral recognition and chiral discrimination \cite{Cireasa2015NatPhys,Smirnova2015JPB,Ayuso2018JPB}, open new directions in HHG spectroscopy.
The values of chiral dichroism in chiral HHG (cHHG) can compete with those from other well established chiroptical methods, such as photoabsorption circular dichroism \cite{book_ComprehensiveChiropticalSpectroscopy}, circular fluorescence \cite{Tinoco1976JACS,Castiglioni2014} or Raman optical activity spectroscopy \cite{Parchansky2014RSC}.
However, it has not yet reached the outstanding sensitivity of photoelectron circular dichroism \cite{Ritchie1976PRA,Powis2000JCP,Bowering2001PRL,Garcia2003JCP,Lux2012Angewandte,Garcia2013NatComm,Stefan2013JCP,Janssen2014PCCP,Lux2015ChemPhysChem}.
As cHHG is a time-resolved technique, it has the potential for probing ultrafast molecular processes, e.g. chemical reactions, at their natural time scales.
Other promising time-resolved chiroptical approaches developed in the last few years include vibrational circular dichroism spectroscopy \cite{Rhee2009Nature}, Coulomb explosion imaging \cite{Pitzer2013Science,Herwig2013Science}, microwave detection \cite{Patterson2013Nature}, chiral-sensitive 2D spectroscopy \cite{Fidler2014NatComm},
ultrafast resonant X-ray spectroscopy \cite{Rouxel2017SD,Zhang2017CS}, time-resolved photoelectron circular dichroism \cite{Beaulieu2016FD} and photoexcitation circular dichroism \cite{Beaulieu2016arXiv}.

The first implementation of cHHG used intense driving fields with weakly-elliptical polarization for driving and probing ultrafast electron dynamics in the chiral molecules propylene oxide and fenchone \cite{Cireasa2015NatPhys}.
Such fields can efficiently induce tunnel ionization from several valence shells in organic molecules with comparable probabilities, as the energy differences between them are usually on the order of one electron volt.
During the electron excursion in the continuum, the laser field can induce transitions between these ionic states, and thus the electron can recombine with a hole that is different from the one that was created upon tunneling, opening the so called cross HHG channels.
The harmonic emission associated with the cross channels can be enantiosensitive if both the ionic states and the driving field are chiral.
However, their intensity is usually weaker than that of the direct HHG channels, i.e. those resulting from ionization from and recombination to the same ionic state.
Direct channels are not enantiosensitive because they do not involve chiral electronic transitions in the core.
Therefore, in order to observe the chiral response of the cross channels in the harmonic spectrum, the achiral background associated with the direct channels needs to be suppressed.
Chiral dichroism was observed in \cite{Cireasa2015NatPhys} in the dynamical energy region of destructive interference between direct channels.

As the chirality of light increases with ellipticity, one would expect to maximize the chiroptical response of the system using circular polarization.
However, circularly polarized drivers do not allow electron-ion recombination, as the field component that is perpendicular to the direction of tunneling drives the electron away from the core.
As a result, the harmonic intensity rapidly drops with ellipticity.
Fortunately, light generation technology can allow one to tailor the driving field in a way that enhances the chiral response of the system while allowing the electron to recollide with the parent ion.
A promising example of a such field tailoring is the generation of two-color bi-circularly polarized radiation, which results from combining a circularly polarized driver with a counter-rotating second harmonic \cite{Eichmann1995PRA,Long1995PRA,Milosevic2000PRA,Milosevic2000PRA_2}.
Intense bi-circular fields can efficiently generate attosecond pulses with circular and elliptical polarization in the XUV domain \cite{Zuo1995JNOPM,Ivanov2014NatPhot,Fleischer2014NatPhot,Pisanty2014PRA,Kfir2015NatPhot,Milosevic2015OptLett,Medisauskas2015PRL,Mauger2016JPB,Bandrauk2016JPB,Pisanty2017PRA,Dorney2017PRL,Jimenez2018PRA,Bandrauk2018PRA}
and spin-polarized electron currents that recollide with the core \cite{MilosevicPRA2016,Ayuso2017NJP}, as well as probe molecular dynamical symmetry breaking \cite{Baykusheva2016PRL,Jimenez2017OptExpress}.

We have recently shown that intense bi-elliptical driving fields can induce strong chiral dichroism in the harmonic spectra of chiral molecules, in a broad range of harmonic numbers and ellipticities \cite{Smirnova2015JPB,Ayuso2018JPB}, exploiting a suppression mechanism of achiral background that does not rely on destructive interference between direct channels.
It is based on a fundamentally different principle: the variation of the energy levels of the system due to the presence of the intense field, i.e. the Stark effect, as already pointed out in \cite{Smirnova2015JPB}.
As a result of the interaction of the ionic states with the intense field, HHG channels accumulate an additional phase.
This additional phase depends on the relative orientation of the molecule with respect to the laser field, and induces a suppression of achiral background upon coherent orientational averaging.

Here we present an analytical model to evaluate the high-order harmonic spectra of chiral molecules in intense bi-elliptical laser fields, and illustrate how to exploit the suppression mechanism of achiral background based on linear Stark shift to control and enhance chiral response in HHG, using the chiral molecule propylene oxide as an example.
The purpose of this model is to recover the physical picture underlying chiral response in HHG based on chiral hole dynamics.
Our analytical model (1) quantifies the Stark suppression of direct channels,
(2) explains why the same mechanism does not lead to cancellation of the enantiosensitive cross channels,
(3) explicitly derives the rotationally invariant molecular pseudoscalars responsible for cHHG, and
(4) explicitly shows how the interference between electric dipole transitions and magnetic dipole transitions (identified as the main mechanism of cHHG in \cite{Cireasa2015NatPhys} for weakly elliptical fields) is controlled by the parameters of the bi-elliptical laser field and the molecular properties.

Atomic units are employed throughout the manuscript unless otherwise stated.

\section{Physical picture and model calculations}

Let us consider a bi-elliptically polarized laser field constituted by two counter-rotating elliptically polarized fields with the same ellipticity $\varepsilon$, whose electric field can be written as
\begin{equation}\label{eq_electricField}
\bold{F}(t) = F_{0} \Big[ f_{x}(t) \bold{\hat{x}} + \varepsilon f_{y}(t) \bold{\hat{y}} \Big]
\end{equation}
where the sub-cycle temporal structure is given by
\begin{align}
f_{x}(t) &= \cos{(\omega t)} + \cos{(2\omega t)} \label{eq_fx} \\
f_{y}(t) &= \sin{(\omega t)} - \sin{(2\omega t)} \label{eq_fy}
\end{align}
The corresponding vector potential, satisfying the condition $\bold{F}(t)=-d\bold{A}(t)/dt$, is given by
\begin{equation}\label{eq_vectorPotential}
\bold{A}(t) = \frac{F_{0}}{\omega} \Big[ a_{x}(t) \bold{\hat{x}} + \varepsilon a_{y}(t) \bold{\hat{y}} \Big]
\end{equation}
with
\begin{align}
a_{x}(t) &= - \sin{(\omega t)} - \sin{(2\omega t)} \label{eq_ax} \\
a_{y}(t) &=   \cos{(\omega t)} - \cos{(2\omega t)} \label{eq_ay}
\end{align}
In order to describe chiral effects in HHG, it is essential to account for the interaction of the system with the magnetic field of light, which can be written as
\begin{equation}\label{eq_magneticField}
\bold{B}(t)=\frac{1}{c}\hat{\bold{z}}\times\bold{F}(t)=\frac{F_0}{c} \Big[ -\varepsilon f_{y}(t) \bold{\hat{x}} + f_{x}(t) \bold{\hat{y}} \Big]
\end{equation}
where $c$ is the speed of light and $\bold{\hat{z}}$ is the direction of light propagation.

\begin{figure}
\centering
\includegraphics[width=\linewidth, keepaspectratio=true]{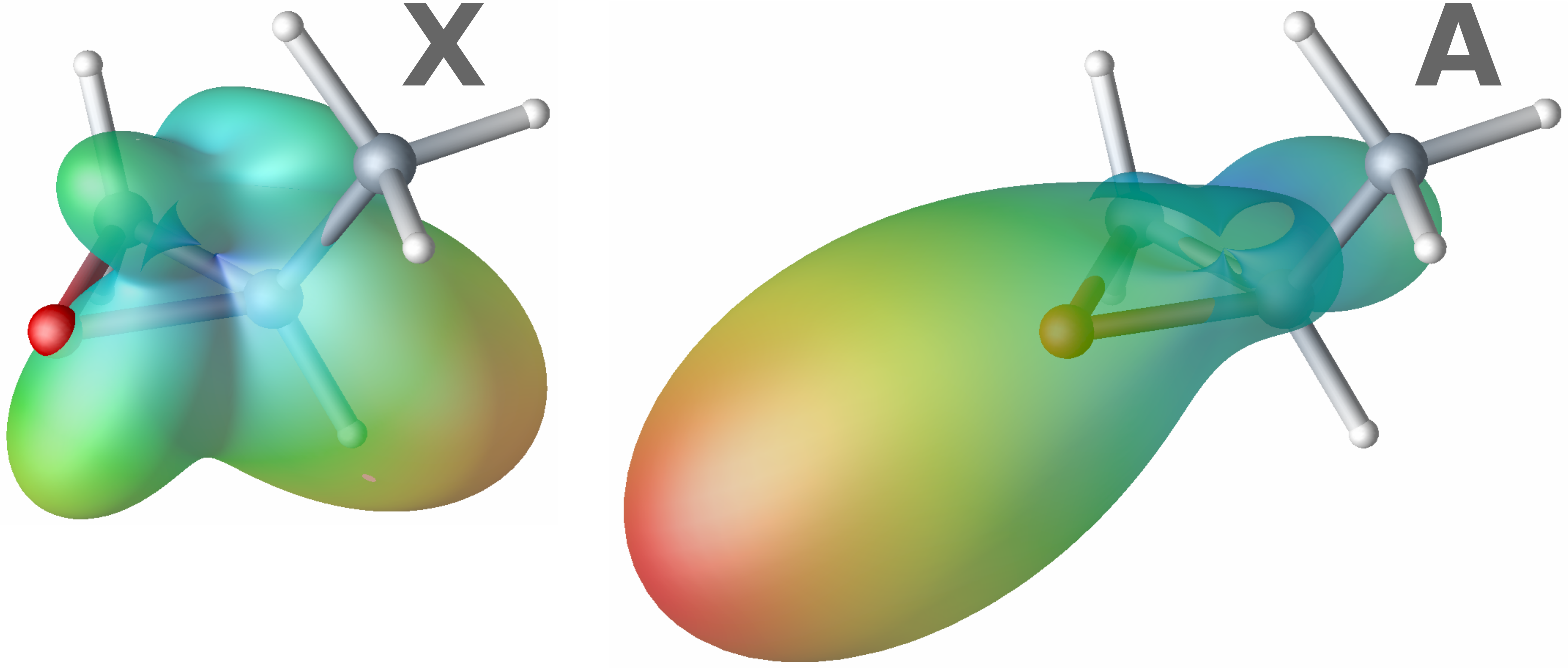}
\caption{Angular dependence of the ionization probability correlated to the ionic states X (left) and A (right) of propylene oxide, evaluated using the TDRIS method.
}
\label{fig_TDRIS}
\end{figure}

Intense driving fields in the near-IR or mid-IR domains can induce tunnel ionization from several molecular orbitals in organic molecules, thus generating superpositions of several ionic states.
In this work we have considered the electronic ground state (X) and the first excited state (A) of the propylene oxide cation.
We have calculated the strong-field ionization probabilities associated with these ionic states using the time-dependent resolution in ionic states (TDRIS) method \cite{Spanner2009PRA,Spanner2012PRA} (see \cite{Cireasa2015NatPhys}).
Fig. \ref{fig_TDRIS} contains the angular dependence of the tunneling probabilities correlated to the X and A states of the ion.
Our ab initio simulations show that both ionization channels exhibit a preferred direction.
This preference is especially pronounced in the case of the A state.

To keep things simple in the analytical analysis, we can assume that tunnel ionization occurs along the directions that maximize the ionization probability, that will be represented by $\bold{r}_{0}^{m}$ ($m=$ X, A).
In this case, ionized molecules contributing to the HHG signal will be oriented so that $\bold{r}_{0}^{m}$ points along the major component of the electric field.
In order to account for the experimental situation of randomly oriented molecules, one has to average over all possible molecular orientations.
Our model reduces full orientational averaging to one degree of freedom: that of molecular rotations around $\bold{r}_{0}^{m}$.
These assumptions allow for a simple analytical treatment that can qualitatively reproduce the most relevant features in the harmonic spectra.
A more quantitative analysis requires accurate description of recombination and sub-cycle dynamics of strong-field ionization \cite{Ayuso2018JPB}.

\subsection{Evaluation of high-order harmonic spectra}

The intensity of the harmonic signal is given by \cite{bookChapter_SmirnovaIvanov_AttosecondAndXUVPhysics}:
\begin{equation}\label{eq_I}
I(\varepsilon,N) \propto (N\omega)^4 \text{ } \Big|\sum_{nm}\bold{{D}}_{mn}(\varepsilon,N)\Big|^2
\end{equation}
where $N$ is the harmonic number and $\bold{D}_{mn}(\varepsilon,N)$ is the harmonic dipole in the frequency domain associated with a given HHG channel $mn$.
In this notation, $m$ represents the ionic state generated upon tunnel ionization and $n$ denotes the state with which the electron recombines.
If the molecules are not oriented, $\bold{{D}}_{mn}(\varepsilon,N)$ results from the coherent addition of all possible molecular orientations in the macroscopic sample.
However, if ionization has a strong preferred direction, as described above, it can be approximated by
\begin{equation}\label{eq_Dmn}
\bold{D}_{mn}(\varepsilon,N) \simeq \frac{1}{2\pi} \int_{0}^{2\pi} \bold{D}^{\alpha}_{mn}(\varepsilon,N) \text{ } d\alpha
\end{equation}
where $\bold{D}^{\alpha}_{mn}(\varepsilon,N)$ is the harmonic dipole associated with a specific molecular orientation, given by the angle $\alpha$, which accounts for molecular rotations around the direction of maximum ionization.
The application of the saddle-point method \cite{bookChapter_SmirnovaIvanov_AttosecondAndXUVPhysics} allows one to factorize the contribution from a given ionization burst into the three terms, associated with ionization, propagation and recombination, i.e. 
\begin{equation}
\bold{D}^{\alpha}_{mn}(\varepsilon,N) = a_{\text{ion},\bold{r}_{0}^{m}}^{mn} \cdot a_{\text{prop},\alpha}^{mn}(\varepsilon,N) \cdot \bold{a}_{\text{rec},\alpha}^{mn}(\varepsilon,N)
\end{equation}
where we have neglected the weak dependence of ionization amplitudes on ellipticity and harmonic number, as the laser field is essentially quasistatic with respect to ionization: within the span of ionization times, the variations in the field magnitude are very small.
Within the approximations above described, recombination occurs essentially along $-\bold{r}_{0}^{m}$.
The smalls deviations from this direction can be included via its weak dependence on ellipticity:
\begin{equation}\label{eq_a_rec}
\bold{a}_{\text{rec},\alpha}^{mn}(\varepsilon,N) \simeq \bold{a}_{\text{rec},-\bold{r}_{0}^{m}}^{mn\text{ }(0)}(N) \text{ } e^{i\varepsilon \Psi_{mn}^{\alpha(1)}(N)}
\end{equation}
where $\Psi_{mn}^{\alpha(1)}$ describes the $\varepsilon$-dependence of the phase of the recombination matrix elements.
The chiral response is contained in the propagation amplitudes \cite{Cireasa2015NatPhys}, which are given by
\begin{equation}\label{eq_a_prop}
a_{\text{prop},\alpha}^{mn} = \bigg(\frac{2\pi}{i(t_r-t_i)}\bigg)^{3/2} e^{-iS(t_r',t_i',\bold{p})} \text{ } a_{n\leftarrow m}^{\alpha}(t_r',t_i')
\end{equation}
with
\begin{equation}\label{eq_VolkovPhase}
S(t,t',\bold{p}) = \frac{1}{2} \int_{t'}^{t} \big[\bold{p}+\bold{A}(\tau)\big]^2 d\tau
\end{equation}
$t_i'$ and $t_r'$ are the real components of the (complex) ionization and recombination times (see \cite{bookChapter_SmirnovaIvanov_AttosecondAndXUVPhysics}),
and $a_{n\leftarrow m}^{\alpha}$ is the transition amplitude accounting for the laser-driven dynamics induced in the ion between ionization and recombination, which can be evaluated by propagating the initial wave function from $t_i'$ to $t_r'$ and projecting it onto the final state, i.e.
\begin{equation}\label{eq_Dmn_tilde}
a_{n\leftarrow m}^{\alpha}(t_r',t_i') = \langle n | U^\alpha(t_r',t_i')|m\rangle
\end{equation}
where $U^\alpha(t_r',t_i')$ is the evolution operator acting on the electronic coordinates of the ion (see \cite{bookChapter_SmirnovaIvanov_AttosecondAndXUVPhysics}).
The dependence of ionization and recombination times on $N$ and $\varepsilon$ has been omitted for the sake of clarity.
The harmonic dipole associated with a given HHG channel (eq. \ref{eq_Dmn}) can thus be written as
\begin{align}\label{eq_Dmn_approx}
\bold{D}_{mn}(\varepsilon,N) &= \bigg(\frac{2\pi}{i(t_r-t_i)}\bigg)^{3/2} e^{-iS_m(t_r',t_i',\bold{p})} \nonumber \\ 
& a_{\text{ion},\bold{r}_{0}^{m}}^{mn}\text{ } \bold{a}_{\text{rec},-\bold{r}_{0}^{m}}^{mn\text{ }(0)}(N) \text{ } \tilde{D}_{mn}(\varepsilon,N)
\end{align}
with
\begin{equation}\label{eq_Dmn_tilde}
\tilde{D}_{mn}(\varepsilon,N) = \frac{1}{2\pi} \int_{0}^{2\pi} a_{n\leftarrow m}^{\alpha}(t_r',t_i') \text{ } e^{i\varepsilon \Psi_{mn}^{\alpha(1)}(\varepsilon,N)} \text{ } d\alpha
\end{equation}
where, as already stated, $\Psi_{mn}^{\alpha(1)}$ describes the weak ellipticity dependence of the phase of the recombination matrix elements.
Since this contribution in general cannot be neglected, but does not have a unique dependence on $\alpha$, we postpone discussion of this term until the end of this section and set $\Psi_{mn}^{\alpha(1)}=0$ until then.
The key quantity to evaluate the relative contributions of the different HHG channels, their modulation with ellipticity and thus chiral dichroism is $\tilde{D}_{mn}$, the angle-averaged amplitude accounting for the chiral laser-driven dynamics in the ion.

\begin{figure}
\centering
\includegraphics[width=\linewidth, keepaspectratio=true]{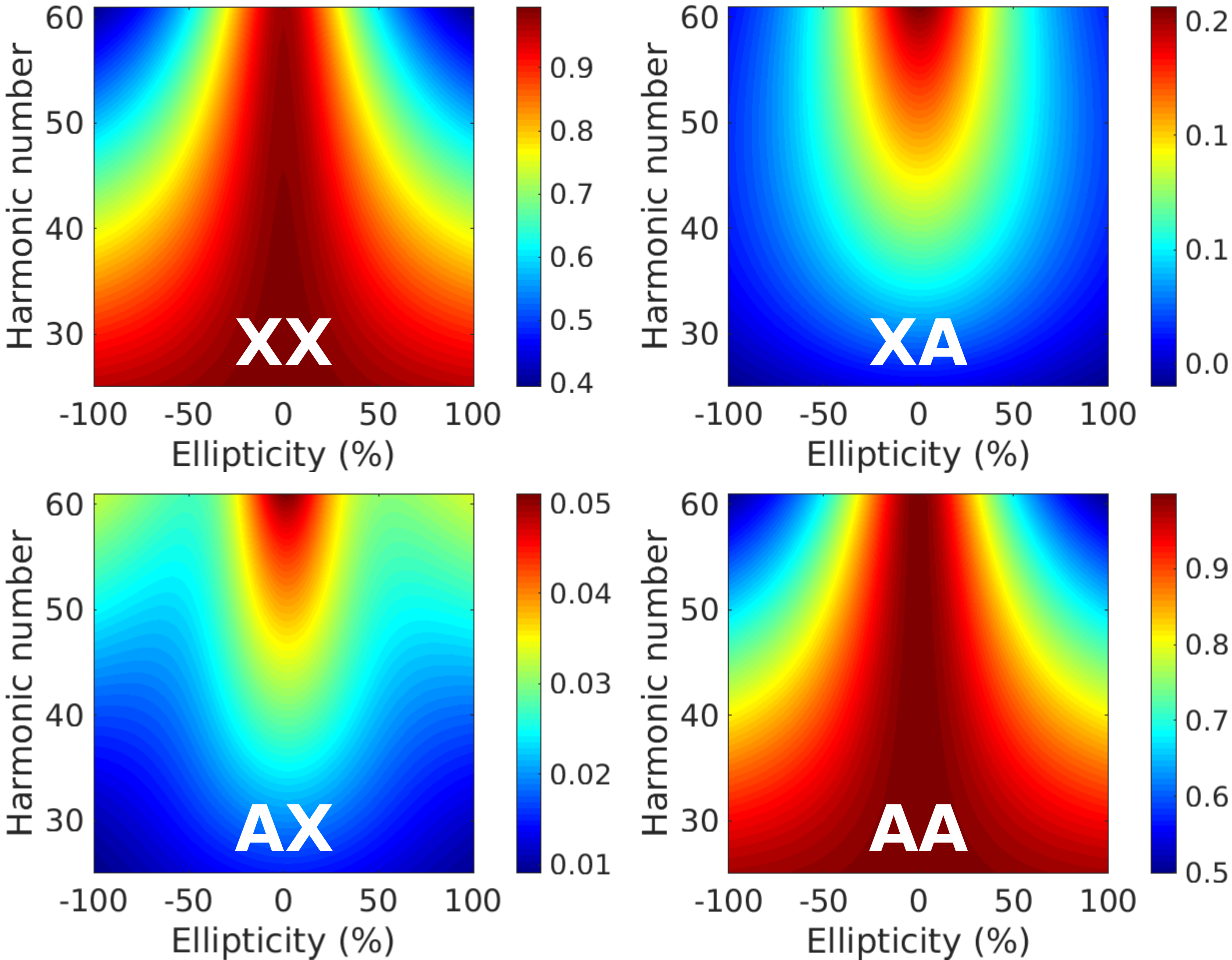}
\caption{Absolute values $|\tilde{D}_{mn}|$ of the harmonic dipole associated with the direct HHG channels XX and AA and with the cross HHG channels XA and AX as a function of ellipticity and harmonic number, obtained by numerical solution of the TDSE in the basis set of ionic states, 
for laser parameters: $F_0=0.04$ a.u. and $\omega=0.0224$ a.u.}
\label{fig_Dmn_TDSE}
\end{figure}

\begin{figure}
\centering
\includegraphics[width=\linewidth, keepaspectratio=true]{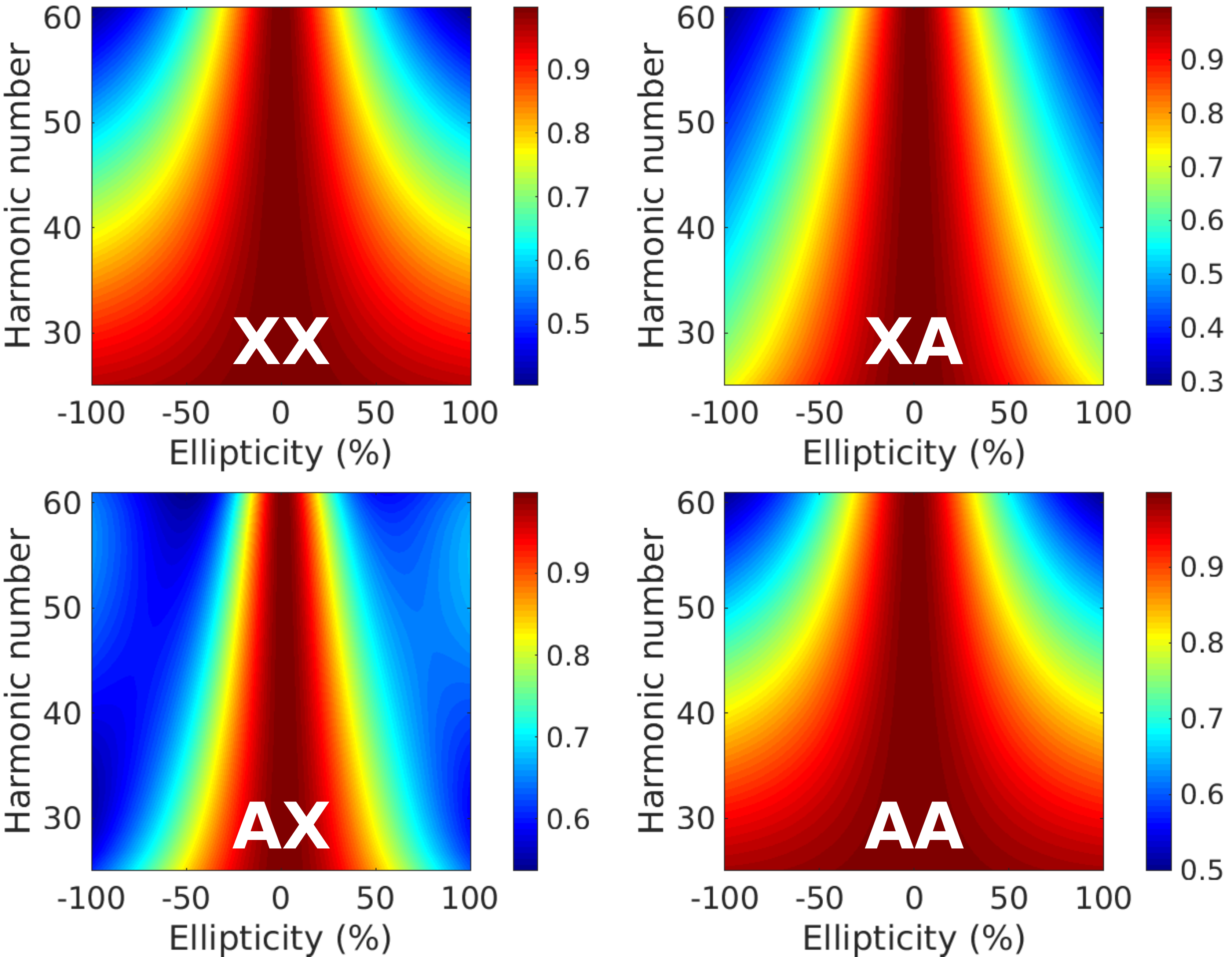}
\caption{Normalized values of $|\tilde{D}_{mn}|$ (absolute values are presented in fig. \ref{fig_Dmn_TDSE}).
For each harmonic number $N_0$, the amplitudes has been normalized to the maximum value of $|\tilde{D}_{mn}(\varepsilon,N_0)|$, with $\varepsilon\in[-1,1]$.}
\label{fig_Dmn_norm_TDSE}
\end{figure}

We have evaluated the harmonic spectrum of propylene oxide in bi-elliptical driving fields using this procedure, for the following laser parameters: field amplitude $F_0=0.04$ a.u., fundamental frequency $\omega=0.0224$ a.u. and ellipticity $\varepsilon$ varying from $-1$ to $1$. 
The absolute values of $\tilde{D}_{mn}$ resulting from solving the time-dependent Schr\"odinger equation (TDSE) in the basis of states X and A are presented in fig. \ref{fig_Dmn_TDSE}, for the direct HHG channels XX and AA and for the cross channels XA and AX, as a function of harmonic number and ellipticity.
In this notation, the first letter indicates the state generated upon tunnel ionization and the second letter denotes the state with which the electron recombines upon its round-trip.
For analysis, we show in fig. \ref{fig_Dmn_norm_TDSE} the values that result from normalizing $|\tilde{D}_{mn}(\varepsilon,N)|$, for each harmonic number, to its maximum value.

For the direct channels XX and AA, the normalized values of $|\tilde{D}_{mn}|$ are essentially identical to the non-normalized ones, as we can see in figs. \ref{fig_Dmn_TDSE} and \ref{fig_Dmn_norm_TDSE}.
The reason is that the probability of transition from the ionic state generated upon strong-field ionization to other ionic states is weak, and thus $|a_{mm}(\varepsilon,N)|^2\simeq1$ in the whole range of harmonic numbers and ellipticities.
However, their orientation-averaged values, i.e. the values of $\tilde{D}_{mn}(\varepsilon,N)$, drop with ellipticity as a result of a suppression mechanism based on the Stark effect.
This mechanism is explained in detail below.

The interaction of the ionic states with the strong field shifts their energy levels.
As a result, direct HHG channels accumulate an additional phase, given by
\begin{align}\label{StarkShift}
\phi_{\text{Stark}}^{mm} = F_0 d_{m,x} \int_{t_i'}^{t_r'} f_x(t)dt + \varepsilon F_0 d_{m,y} \int_{t_i'}^{t_r'} f_y(t)dt
\end{align}
where $d_{m,x}$ and $d_{m,y}$ are the projections of $\bold{d}_{m}$, the permanent dipole of the $m$ state, onto the laboratory frame directions $x$ and $y$.
The values of $d_{m,y}$ change for different molecular orientations.
Therefore, the additional phase accumulated in HHG channels due to the linear Stark shift changes as well.
As harmonic emission results from the coherent addition of radiation emitted from all the molecules in the medium, if this additional phase substantially changes, it has the potential to induce interferences and thus strongly suppress the achiral background associated with direct HHG channels.
We shall see that this does indeed happen for mid-IR drivers.

As expected, the intensity associated with the cross channels XA and AX increases with the harmonic number.
Higher-order harmonics are associated with longer excursion times, and thus the laser field has more time to induce an electronic transition in the core.
Interestingly, the modulation of the two cross channels with ellipticity is very different: whereas intensity associated with the XA channel decays with ellipticity as rapidly as for the direct channels, the AX channel exhibits a more complex behaviour.
We also note that, for low ellipticities, the absolute amplitude of the XA channel is approximately four times that of the AX channel.
These differences are a consequence of the different relative orientation of the transition dipole with respect to the direction that maximizes electron tunneling tunneling in each HHG channel, as we discuss in section \ref{section_crossChannels}.

\begin{figure}
\centering
\includegraphics[width=\linewidth, keepaspectratio=true]{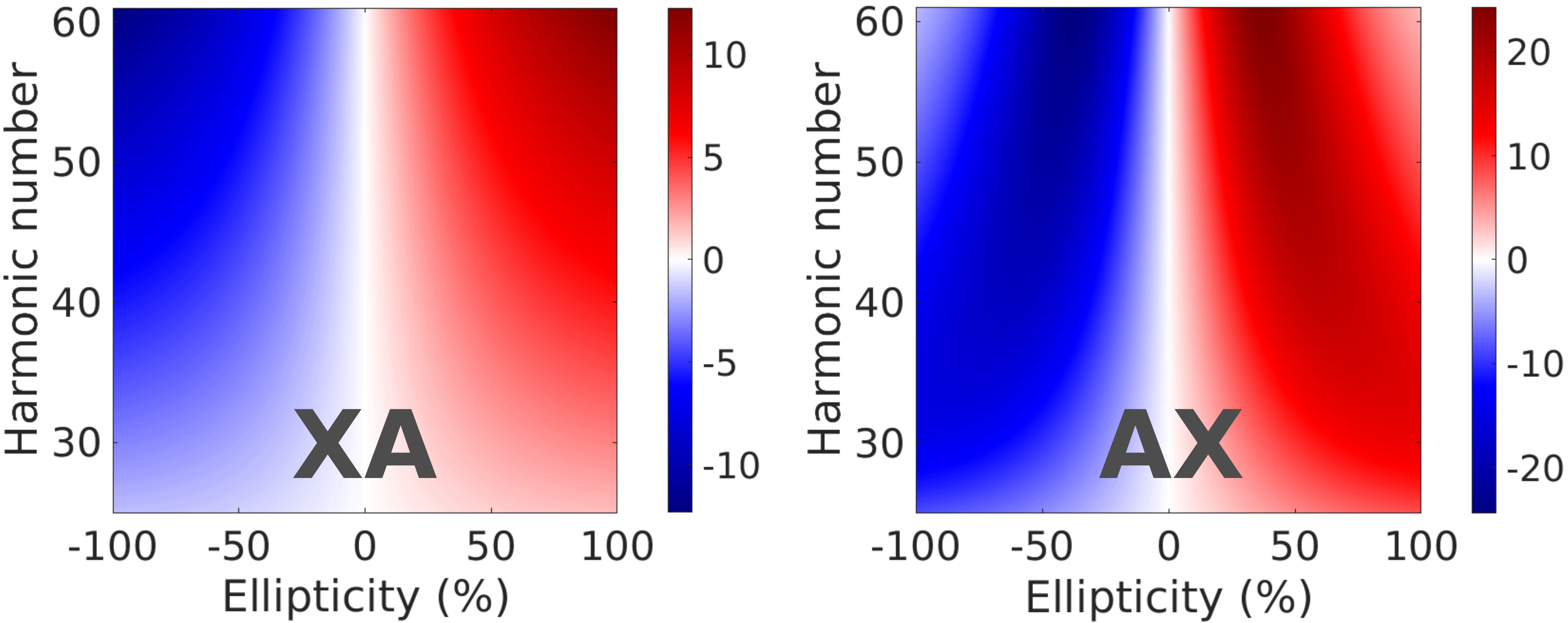}
\caption{Chiral dichroism ($\%$) in the harmonic dipole associated with the cross channels XA and AX.
Direct channels do not present chiral dichroism.}
\label{fig_CD_TDSE}
\end{figure}

Chiral dichroism in HHG is defined as
\begin{equation}\label{eq_CD}
\text{CD}(\varepsilon,N) = 2\text{ }\frac{I(\varepsilon,N)-I(-\varepsilon,N)}{I(\varepsilon,N)+I(-\varepsilon,N)}
\end{equation}
where $I$ is the harmonic intensity, given by eq. \ref{eq_I}.
Of course, the reversal of light polarization ($\varepsilon\leftrightarrow-\varepsilon$) is equivalent to the exchange of enantiomers ($R\leftrightarrow L$).
The values of chiral dichroism in the harmonic intensity associated with the cross HHG channels XA and AX are presented in fig. \ref{fig_CD_TDSE}, as a function of ellipticity and harmonic number.
Both channels present large values of dichroism, which exhibit an overall enhancement for higher-order harmonics and for high ellipticities.
These are precisely the regions of the spectra where the linear Stark shift mechanism described above induces stronger suppression of the achiral background associated with the direct HHG channels, thus enabling the possibility of observing strong chiral response.
We note, however, that the intensity of the direct channels is still about one order of magnitude larger than that of the cross channels.
Can we exploit the mechanism based on linear Stark shift to induce stronger suppression of the achiral background while keeping, or even enhancing, the chiral response of the cross channels?

In the following, we derive analytical expressions for $\tilde{D}_{mn}$, for the direct and for the cross HHG channels, in order to better understand the results presented in figs. \ref{fig_Dmn_TDSE}, \ref{fig_Dmn_norm_TDSE} and \ref{fig_CD_TDSE}, and to understand how to control and enhance chiral response in HHG.

\subsection{Direct HHG channels}

In order to derive a simple analytical expression for the HHG intensity associated with the direct channels, we can assume that the probability of transition from the state created upon ionization to other ionic states is weak, and therefore $|a_{mm}^{\alpha}(t_r',t_i')|^2\simeq1$.
This assumption is validated by the numerical results presented in figs. \ref{fig_Dmn_TDSE} and \ref{fig_Dmn_norm_TDSE}, and leads to
\begin{equation}\label{eq_a_mm}
a_{n\leftarrow m}^{\alpha}(t_r',t_i') \simeq e^{-i\int_{t_i'}^{t_r'}H_{mm}^{\alpha}(t)dt}
\end{equation}
where the diagonal term of the time-dependent Hamiltonian matrix is given by
\begin{equation}\label{eq_H_mm}
H_{mm}^{\alpha}(t) = E_{m} + \bold{F}(t)\cdot \bold{d}_{m}^{\alpha}
\end{equation}
and $\bold{d}_{m}^{\alpha}$ is the permanent dipole of the $m$ state in the laboratory frame, which depends on the molecular orientation through the angle $\alpha$.
Inserting eqs. \ref{eq_a_mm} and \ref{eq_H_mm} into eq. \ref{eq_Dmn_tilde} and using the definition of the electric field (eqs. \ref{eq_electricField}, \ref{eq_fx} and \ref{eq_fy}), we obtain
\begin{align}\label{eq_Dmm_model_1}
&\tilde{D}_{mm}(\varepsilon,N) \simeq e^{-iE_{m}(t_r'-t_i')} \text{ } e^{-i F_{0} d_{m}^{\parallel_m} \int_{t_i'}^{t_r'}f_x(t)dt} \nonumber \\
& \frac{1}{2\pi} \int_{0}^{2\pi} e^{-i \varepsilon F_{0} d_{m}^{\perp_m}\cos{(\alpha_{m,0}^{d_{m}}+\alpha)} \int_{t_i'}^{t_r'}f_y(t)d_{m,y}dt} d\alpha
\end{align}
where $d_{m}^{\parallel_m} = \langle\hat{\bold{x}}|\bold{d}_{m}^{\alpha}\rangle$ is the component of $\bold{d}_{m}$ in the direction that maximizes strong-field ionization from the $m$ state ($\bold{r}_0^m$ in the molecular frame), which coincides with the direction of the major component of the laser field ($\hat{\bold{x}}$ in the laboratory frame), and thus it is not affected by rotations around this axis (variations of $\alpha$).
Its perpendicular component is given by $\bold{d}_{m}^{\perp_m}(\alpha) = \langle\hat{\bold{y}}|\bold{d}_{m}^{\alpha}\rangle\hat{\bold{y}} + \langle\hat{\bold{z}}|\bold{d}_{m}^{\alpha}\rangle\hat{\bold{z}}$.
Note that only the direction of $\bold{d}_{m}^{\perp_m}(\alpha)$ depends on $\alpha$, as its modulus, given by $d_{m}^{\perp_m} = (|\bold{d}_{m}|^2-|d_{m}^{\parallel_m}|^2)^{1/2}$, is orientation-independent.
Thus, the component of $\bold{d}_{m}$ in the direction of the minor component of the laser field, $\langle\bold{d}_{m}^{\alpha}|\hat{\bold{y}}\rangle$, can be written as $d_{m}^{\perp_m}\cos{(\alpha_{m,0}^{d_{m}}+\alpha)}$,
where $\alpha_{m,0}^{d_{m}}$ is the offset angle between $\bold{d}_{m}^{\perp_m}(\alpha)$ and a reference direction in the $xy$ plane, e.g. $\hat{\bold{y}}$, for $\alpha=0$.
The electric and magnetic dipole matrix elements associated with the X and A electronic states of the core are shown in table \ref{table_DME}, expressed in the coordinates of the molecular frame.
They have been evaluated using the MCSCF method as described in \cite{Cireasa2015NatPhys}.
The directions that maximize the probability of strong-field ionization from these states ($\bold{r}_0^X$ and $\bold{r}_0^A$) are shown in table \ref{table_r0}.
table \ref{table_DME_projections} contains the projections of the electric and magnetic dipoles shown in table \ref{table_DME} onto $\bold{r}_0^X$ and $\bold{r}_0^A$ and onto the planes that are orthogonal to them.

\begin{table}
\begin{center}
\begin{tabular}{c|c|c|c|} \cline{2-4}
\multicolumn{1}{ c|}{  }                 & $\hat{\bold{x}}$ & $\hat{\bold{y}}$ & $\hat{\bold{z}}$ \\ \cline{1-4}
\multicolumn{1}{|c|}{$\bold{d}_{X}$ }    & $-1.603$         & $ 0.123$         & $-0.050$         \\ \cline{1-4}
\multicolumn{1}{|c|}{$\bold{d}_{A}$ }    & $-1.375$         & $-0.467$         & $-0.187$         \\ \cline{1-4}
\multicolumn{1}{|c|}{$\bold{d}_{XA}$}    & $-0.036$         & $-0.016$         & $-0.106$         \\ \cline{1-4}
\multicolumn{1}{|c|}{$\bold{m}_{XA}$}    & $-0.389i$        & $-0.224i$        & $ 0.212i$        \\ \cline{1-4}
\end{tabular}
\end{center}
\caption{Electric and magnetic matrix elements between the electronic ionic states X and A: permanent electric dipoles (first and second rows), electric transition dipole (third row) and magnetic transition dipole (fourth row).}
\label{table_DME}
\end{table}

\begin{table}
\begin{center}
\begin{tabular}{c|c|c|c|} \cline{2-4}
\multicolumn{1}{ c|}{  }                 & $\hat{\bold{x}}$ & $\hat{\bold{y}}$ & $\hat{\bold{z}}$ \\ \cline{1-4}
\multicolumn{1}{|c|}{$\bold{r}_{0}^{X}$} & $0.188$          & $ 0.515$         & $ 0.836$         \\ \cline{1-4}
\multicolumn{1}{|c|}{$\bold{r}_{0}^{A}$} & $0.317$          & $-0.926$         & $-0.203$         \\ \cline{1-4}
\end{tabular}
\end{center}
\caption{Directions that maximize strong-field ionization from the X and A states of the ionic core.}
\label{table_r0}
\end{table}

\begin{table}
\small
\begin{center}
\begin{tabular}{|c|c|c|c|c|c|c|c|c|c|c|} \cline{1-2}\cline{4-5}\cline{7-8}\cline{10-11}
$d_{X}^{ \parallel_X}$ & $-0.280$  & & $d_{X}^{ \perp_X}$ & $1.584$  & & $d_{X}^{ \parallel_A}$ & $-0.612$  & & $d_{X}^{ \perp_A}$ & $1.487$  \\ \cline{1-2}\cline{4-5}\cline{7-8}\cline{10-11}
$d_{A}^{ \parallel_X}$ & $-0.655$  & & $d_{A}^{ \perp_X}$ & $1.309$  & & $d_{A}^{ \parallel_A}$ & $ 0.034$  & & $d_{A}^{ \perp_A}$ & $1.464$  \\ \cline{1-2}\cline{4-5}\cline{7-8}\cline{10-11}
$d_{XA}^{\parallel_X}$ & $-0.104$  & & $d_{XA}^{\perp_X}$ & $0.045$  & & $d_{XA}^{\parallel_A}$ & $ 0.025$  & & $d_{XA}^{\perp_A}$ & $0.110$  \\ \cline{1-2}\cline{4-5}\cline{7-8}\cline{10-11}
$m_{XA}^{\parallel_X}$ & $ 0.011i$ & & $m_{XA}^{\perp_X}$ & $0.496i$ & & $m_{XA}^{\parallel_A}$ & $-0.041i$ & & $m_{XA}^{\perp_A}$ & $0.494i$ \\ \cline{1-2}\cline{4-5}\cline{7-8}\cline{10-11}
\end{tabular}
\end{center}
\caption{Parallel and absolute values of perpendicular components of the permanent and transition dipoles $\bold{d}_{X}$, $\bold{d}_{A}$, $\bold{d}_{XA}$ and $\bold{m}_{XA}$ (shown in table \ref{table_DME}) with respect to the directions that maximize strong-field ionization from the X and A states ($\bold{r}_{0}^{X}$ and $\bold{r}_{0}^{A}$, shown in table \ref{table_r0}).}
\label{table_DME_projections}
\end{table}

Eq. \ref{eq_Dmm_model_1} can be rewritten in a more compact form:
\begin{equation}\label{eq_Dmm_model_2}
\tilde{D}_{mm}(\varepsilon,N) \simeq e^{-i\phi_{m}^{E}} e^{i \phi_{m}^x} \frac{1}{2\pi} \int_{0}^{2\pi} e^{i \varepsilon\phi_{m}^{y\alpha}} d\alpha
\end{equation}
where we have introduced the following phase terms
\begin{align}
\phi_{m}^{E} &= E_{m}(t_r'-t_i')\\
\phi_{m}^x &= \frac{F_{0} d_{m}^{\parallel_m}}{\omega} \big[a_x(t_r')-a_x(t_i')\big] \\
\phi_{m}^{y\alpha} &= \frac{F_{0} d_{m}^{\perp_m}}{\omega} \cos{(\alpha_{m,0}^{d_{m}}+\alpha)} \big[a_y(t_r')-a_y(t_i')\big]
\end{align}
and used the definition of the vector potential (eqs. \ref{eq_vectorPotential}, \ref{eq_ax} and \ref{eq_ay}).
The term $\phi_{m}^{E}$ is the phase accumulated due to the field-free time evolution in the state $m$, and $\phi_{m}^x$ and $\phi_{m}^{y\alpha}$ are the additional phases accumulated in the direct channels due to linear Stark shift.
The dependence of $\phi_{m}^{E}$, $\phi_{m}^x$ and $\phi_{m}^{y\alpha}$ on ionization and recombination times (and therefore on ellipticity and harmonic number) has been dropped for the sake of simplicity.
Note that the terms outside the integral in eq. \ref{eq_Dmm_model_2} do not alter the intensity associated with a given HHG channel, they only add a global phase.
Eq. \ref{eq_Dmm_model_2} can be simplified by applying Euler's formula, $e^{i\theta}=\cos{\theta}+i\sin{\theta}$, to $e^{i \varepsilon\phi_{m}^{y\alpha}}$ and removing the sine contribution, as $\tilde{D}_{mm}$ is an even function with respect to $\varepsilon$ since direct channels are not chiral.
Thus, we have
\begin{equation}\label{eq_Dmm_model}
\tilde{D}_{mm}(\varepsilon,N) \simeq e^{-i\phi_{m}^{E}} e^{i \phi_{m}^x} \frac{1}{2\pi} \int_{0}^{2\pi} \cos{(\varepsilon\phi_{m}^{y\alpha}}) \text{ } d\alpha
\end{equation}
The integration in $\alpha$ can be solved analytically by performing a Taylor expansion of the cosine function.
By keeping the terms up to order 4, we obtain the following expression:
\begin{align}\label{eq_Dmm_Taylor}
\tilde{D}_{mm}(\varepsilon,N) &\simeq e^{-i\phi_{m}^{E}} e^{i \phi_{m}^x} \bigg[ 1 - \frac{d_{m}^{\perp 2} F_{0}^2[a_y(t_r')-a_y(t_i')]^2}{4\omega^2}\varepsilon^2 \nonumber \\
&+ \frac{d_{m}^{\perp 4} F_{0}^4[a_y(t_r')-a_y(t_i')]^4}{64\omega^4}\varepsilon^4 \bigg]
\end{align}
Our analytical formula indicates that $|\tilde{D}_{mm}(\varepsilon,N)|$ maximizes for linearly polarized fields, with $|\tilde{D}_{mm}(0,N)|^2=1$.
Only the zero order term is present for $\varepsilon=0$.
This term contains the phase accumulated due to the energy of the field-free state and to the interaction of $\varepsilon$-independent field component with the parallel component of its permanent dipole, which is the same for all molecular orientations.
Higher order terms arise as a result of the linear Stark shift induced by the interaction of the $\varepsilon$-dependent field component with the orientation-dependent dipole component along this direction.
The second order term induces cancellation of the harmonic intensity upon orientational averaging.
The degree of suppression depends on the ratio $F_0/\omega$, thus offering the possibility of control, as we show in section \ref{section_control}.
We note that, at high ellipticities, the suppression induced by the second order term could be compensated by the fourth order term, but this contribution is expected to be significantly weaker for moderate values of $F_0/\omega$.

\begin{figure}
\centering
\includegraphics[width=\linewidth, keepaspectratio=true]{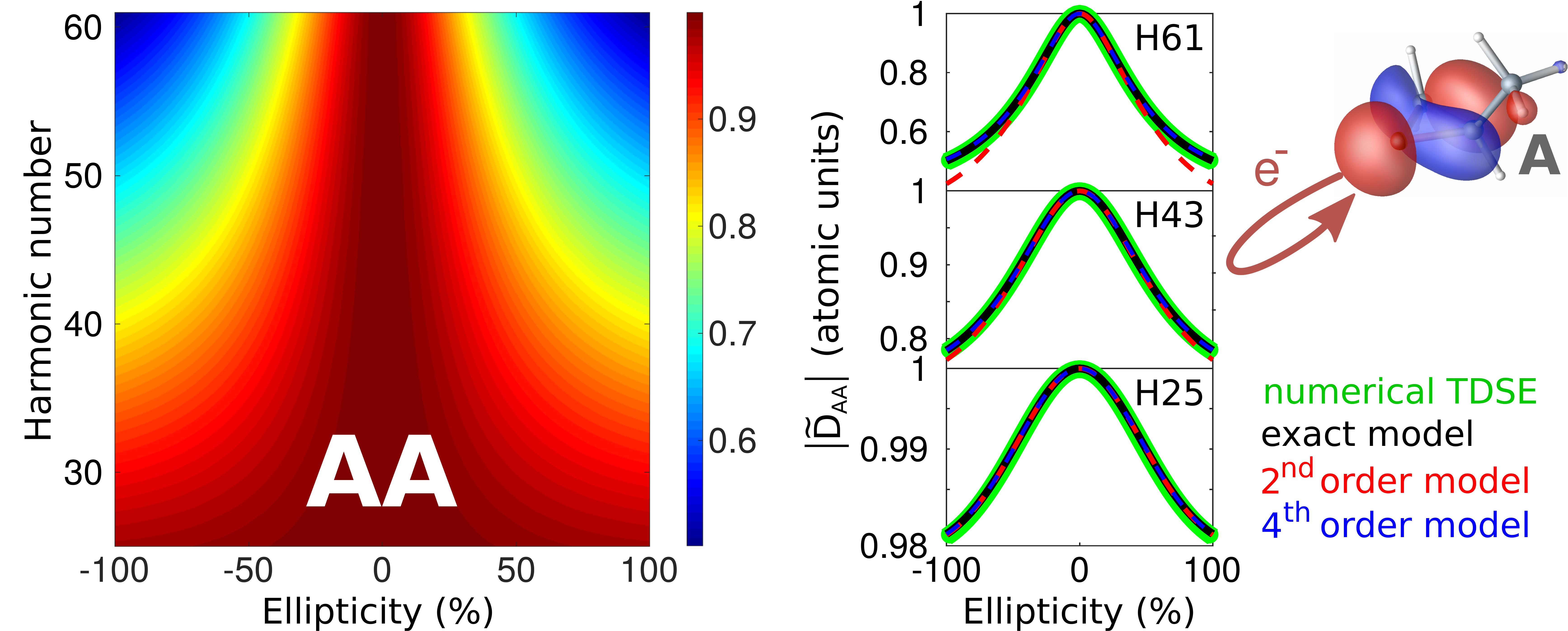}
\caption{Harmonic dipole $\tilde{D}_{mn}$ associated with the direct HHG channel AA, for laser parameters: $F_0=0.04$ a.u. $\omega=0.0224$ a.u.
Left panel: result of applying the analytical eq. \ref{eq_Dmm_Taylor}, as a function of ellipticity and harmonic number.
Right panels: comparison between numerical TDSE results (green lines), exact model solutions (black lines, eq. \ref{eq_Dmm_model}) and approximate solutions (eq. \ref{eq_Dmm_Taylor}) up to order 2 (red lines) and up to order 4 (blue lines),
as a function of ellipticity, for harmonic numbers 25 (lower panel), 43 (central panel) and 61 (upper panel).
Inset: schematic representation of the physical process. 
}
\label{fig_Dmm_model}
\end{figure}

The left panel of fig. \ref{fig_Dmm_model} contains the values of $|\tilde{D}_{mm}|$ as a function of $\varepsilon$ and $N$ resulting from applying eq. \ref{eq_Dmm_Taylor} to the direct channel AA.
These results are identical to the numerical solutions of the TDSE presented in fig. \ref{fig_Dmn_TDSE}.
For a more detailed comparison, the right panels of fig. \ref{fig_Dmm_model} contain the values of $|\tilde{D}_{mm}|$, as a function of $\varepsilon$, for $N=25$, $43$ and $61$, obtained using different approaches.
We show the numerical TDSE results (already presented in fig. \ref{fig_Dmn_TDSE}), the exact model solutions resulting from applying eq. \ref{eq_Dmm_model} and performing numerical integration in $\alpha$, and the analytical solutions of eq. \ref{eq_Dmm_Taylor}, up to order 2 and up to order 4.
The excellent agreement between the model solutions and the numerical TDSE results confirms the suppression mechanism based on the linear Stark shift.
The second order expansion (see eq. \ref{eq_Dmm_Taylor}) reproduces very well the decay of $|\tilde{D}_{mm}|$ with ellipticity in the whole range of harmonic numbers, showing that the cancellation of the amplitude of direct channels is quadratic with ellipticity.
Only for the highest-order harmonics, and for very high ellipticities, the inclusion of the fourth order term is required in order to obtain perfect agreement with the numerical TDSE results.

\begin{figure}
\centering
\includegraphics[width=\linewidth, keepaspectratio=true]{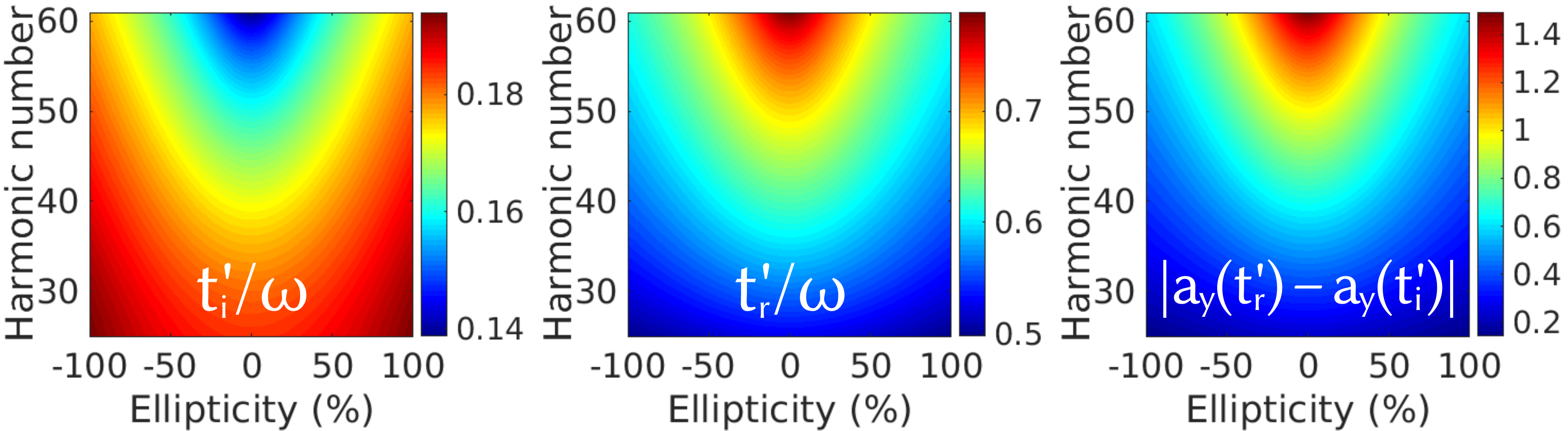}
\caption{Real part of ionization (left panel) and recombination (central panel) phases, presented in units of $\pi$, and values of the dimensionless term $|a_y(t_r')-a_y(t_i')|$, as functions of ellipticity and harmonic number, for laser parameters: $F_0=0.04$ a.u. $\omega=0.0224$ a.u.}
\label{fig_SaddlePoints}
\end{figure}

For a given non-zero ellipticity, the intensity of the direct channels drops with the harmonic number.
The reason is that higher-order harmonics are related to earlier ionization times and to later recombination times, i.e. to longer excursions in the continuum.
Therefore, the linear Stark shift-based suppression mechanism has more time to act, which leads to a stronger cancellation of achiral background in the high-energy region of the spectrum.
Fig. \ref{fig_SaddlePoints} contains the ionization and recombination times, as a function of harmonic number and ellipticity, as well as the values of the dimensionless term $|a_y(t_r')-a_y(t_i')|$.
Note that it is this term that induces the decrease of harmonic intensity with the harmonic number in the non-zero order terms of eq. \ref{eq_Dmm_Taylor}.
The values of $|a_y(t_r')-a_y(t_i')|$ decrease linearly with the harmonic number, which induces a quadratic suppression of $|\tilde{D}_{mm}|$.
Although $|a_y(t_r')-a_y(t_i')|$ also decreases with ellipticity, it does so weakly, barely distorting the quadratic decay of $|\tilde{D}_{mm}|$ with ellipticity.

The model presented in this section for the direct HHG channels reveals that $|\tilde{D}_{mm}|$ decreases quadratically with both ellipticity and harmonic number. 
The intensity associated with a given HHG channel is proportional to $|\tilde{D}_{mm}|^2$, as dictated by eq. \ref{eq_I}.
Therefore, the intensity of achiral background, resulting from the achiral contribution of direct channels to HHG, decreases with the fourth power of ellipticity and with the fourth power of harmonic number.

\subsection{Cross HHG channels}\label{section_crossChannels}

The evaluation of the contribution from cross channels is more complex because they result from the interplay between electric and magnetic interactions in the ionic core between ionization and recombination.
Assuming $|a_{mm}^{\alpha}(t_r',t_i')|^2\simeq1$, the transition amplitude for $m\neq n$ can be written as
\begin{align}\label{eq_a_mn}
a_{n\leftarrow m}^{\alpha}(t_r',t_i') &= -i \int_{t_i'}^{t_r'} e^{-i\int_{t}^{t_r'}H_{nn}^{\alpha}(t')dt'} \nonumber \\
&H_{nm}^{\alpha}(t) \text{ } e^{-i\int_{t_i'}^{t}H_{mm}^{\alpha}(t')dt'} dt
\end{align}
where the off-diagonal time-dependent matrix element $H_{nm}^{\alpha}(t)$ describing an electronic transition from the $m$ to the $n$ ionic state is given by
\begin{equation}\label{eq_H_mn}
H_{nm}^{\alpha}(t) = \bold{F}(t)\cdot \bold{d}_{nm}^{\alpha} + \bold{B}(t)\cdot \bold{m}_{nm}^{\alpha} \quad\text{(} m\neq n \text{)}
\end{equation}
and $\bold{d}_{nm}^{\alpha}$ and $\bold{m}_{nm}^{\alpha}$ are the corresponding electric and magnetic transition dipoles, which depend on the molecular orientation through the angle $\alpha$.
Inserting eqs. \ref{eq_a_mn} and \ref{eq_H_mn} into eq. \ref{eq_Dmn_tilde}, we have
\begin{align}
\tilde{D}_{mn}(\varepsilon,N) &= -\frac{i}{2\pi} \int_{t_i'}^{t_r'} dt \text{ } e^{-i[E_{n}(t_r-t)+E_{m}(t-t_i')]} \nonumber \\
&\int_{0}^{2\pi} d\alpha\text{ } [\bold{F}(t)\cdot \bold{d}_{nm}^{\alpha} + \bold{B}(t)\cdot \bold{m}_{nm}^{\alpha}] \nonumber \\
&e^{-i\int_{t}^{t_r}dt'\bold{F}(t')\cdot\bold{d}_{n}^{\alpha}} e^{-i\int_{t_i'}^{t}dt'\bold{F}(t')\cdot\bold{d}_{m}^{\alpha}}
\end{align}
By decomposing the dot products inside the exponential functions into their direction components, we obtain
\begin{align}\label{eq_Dmn_model}
\tilde{D}_{mn}(\varepsilon,N) &= -\frac{i}{2\pi} \int_{t_i'}^{t_r'} dt \text{ } e^{-i\phi_{mn}^{E}(t)} \text{ } e^{i\phi_{mn}^{x}(t)} \nonumber \\
&\int_{0}^{2\pi} d\alpha \text{ } [\bold{F}(t)\cdot \bold{d}_{nm}^{\alpha} + \bold{B}(t)\cdot \bold{m}_{nm}^{\alpha}] \text{ } e^{i\varepsilon\phi_{mn}^{y\alpha}(t)}
\end{align}
where, for the sake of clarity, we have introduced the time-dependent phases $\phi_{mn}^{E}(t)$, $\phi_{mn}^{x}(t)$ and $\phi_y(t)$:
\begin{align}
\phi_{mn}^{E}(t)&=\big[E_{n}(t_r'-t)+E_{m}(t-t_i')\big]\\
\phi_{mn}^{x}(t)&=\frac{F_0}{\omega} \big[d_{n}^{\parallel_m}\tilde{a}_x^{r}(t)+d_{m}^{\parallel_m}\tilde{a}_x^{i}(t) \big] \\
\phi_{mn}^{y\alpha}(t)&=\frac{F_0}{\omega} \big[ d_{n}^{\perp_m}\cos{(\alpha_{m,0}^{d_{n}}+\alpha)}\tilde{a}_y^{r}(t) \nonumber \\
&\quad+d_{m}^{\perp_m}\cos{(\alpha_{m,0}^{d_{n}}+\alpha)}\tilde{a}_y^{i}(t)\big]
\end{align}
and the following functions:
\begin{align}
\tilde{a}_{x,y}^{r}(t) &= a_{x,y}(t_r')-a_{x,y}(t) \\
\tilde{a}_{x,y}^{i}(t) &= a_{x,y}(t)  -a_{x,y}(t_i')
\end{align}
The term $\phi_{mn}^{E}(t)$ contains the phase accumulated due to the energy difference between the field-free ionic eigenstates, whereas $\phi_{mn}^{x}(t)$ and $\phi_{mn}^{y\alpha}(t)$ are the additional phases accumulated due to the linear Stark shift, as a result of the interaction of the ionic states with the $\varepsilon$-independent and $\varepsilon$-dependent field components, respectively.
The harmonic dipole associated with the cross channels can be split into two terms that account for electric and magnetic transitions in the ion separately, i.e.
\begin{equation}
\tilde{D}_{mn}^e(\varepsilon,N) = \tilde{D}_{mn}^e(\varepsilon,N) + \tilde{D}_{mn}^m(\varepsilon,N)
\end{equation}
with
\begin{align}
\tilde{D}_{mn}^e(\varepsilon,N) &= -\frac{i}{2\pi} \int_{t_i'}^{t_r'} dt \text{ } e^{-i\phi_{mn}^{E}(t)} e^{i\phi_{mn}^{x}(t)} \nonumber \\
&\int_{0}^{2\pi} d\alpha \text{ } \bold{F}(t)\cdot \bold{d}_{nm}^{\alpha} \text{ } e^{i\varepsilon\phi_{mn}^{y\alpha}(t)} \label{eq_Dmn^e_0}
\end{align}
and
\begin{align}
\tilde{D}_{mn}^m(\varepsilon,N) &= -\frac{i}{2\pi} \int_{t_i'}^{t_r'} dt \text{ } e^{-i\phi_{mn}^{E}(t)} e^{i\phi_{mn}^{x}(t)} \nonumber \\
&\int_{0}^{2\pi} d\alpha \text{ } \bold{B}(t)\cdot \bold{m}_{nm}^{\alpha} \text{ } e^{i\varepsilon\phi_{mn}^{y\alpha}(t)} \label{eq_Dmn^m_0}
\end{align}
In the following, we derive simple analytical expressions for these two contributions.

\subsubsection{Electric dipole transition contribution}

Purely electric dipole transitions cannot induce chiral dichroism.
Therefore, eq. \ref{eq_Dmn^e_0} can be simplified by applying Euler's formula to $e^{i\varepsilon\phi_{mn}^{y\alpha}(t)}$ and removing the terms that are odd with respect to $\varepsilon$. 
For analysis purposes, we now split $\tilde{D}_{mn}^e(\varepsilon,N)$ into its two even contributions:
\begin{equation}\label{eq_Dmn^e}
\tilde{D}_{mn}^e(\varepsilon,N) = \tilde{D}_{mn}^{e,1}(\varepsilon,N) + \tilde{D}_{mn}^{e,2}(\varepsilon,N)
\end{equation}
where
\begin{align}
\tilde{D}_{mn}^{e,1}(\varepsilon,N) &= -i \frac{F_0 d_{nm}^{\parallel_m}}{2\pi}        \int_{t_i'}^{t_r'} dt \text{ } e^{-i\phi_{mn}^{E}(t)} e^{i\phi_{mn}^{x}(t)} \nonumber \\
&\int_{0}^{2\pi} d\alpha \cos{\big(\varepsilon \phi_{\alpha}(t)\big)} \label{eq_Dmn^e1}
\end{align}
and
\begin{align}
\tilde{D}_{mn}^{e,2}(\varepsilon,N) &=    \frac{F_0 d_{nm}^{\perp_m}\varepsilon}{2\pi} \int_{t_i'}^{t_r'} dt \text{ } e^{-i\phi_{mn}^{E}(t)} e^{i\phi_{mn}^{x}(t)} \nonumber \\
&\int_{0}^{2\pi} d\alpha \cos{(\alpha_{m,0}^{d_{nm}}+\alpha)} \sin{\big(\varepsilon\phi_{\alpha}(t)\big)} \label{eq_Dmn^e2}
\end{align}
The integrals in $\alpha$ can be calculated analytically by performing a Taylor expansion of the sine and cosine functions.
In order to keep the equations simple, we truncate the expansions to the first non-constant contributions, i.e. $\sin{\big(\varepsilon \phi_{\alpha}(t)\big)}\simeq\varepsilon \phi_{\alpha}(t)$ and $\cos{\big(\varepsilon \phi_{\alpha}(t)\big)}\simeq1-\frac{1}{2}\varepsilon^2 \phi_{\alpha}^2(t)$.
We obtain:
\begin{align}
\tilde{D}_{mn}^{e,1}(\varepsilon,N) &\simeq -iF_0 d_{nm}^{\parallel_m} \int_{t_i'}^{t_r'} dt \text{ } e^{-i\phi_{mn}^{E}(t)} e^{i\phi_{mn}^{x}(t)} f_x(t) \label{eq_Dmn^e1_Taylor} \nonumber \\
& \bigg( 1 - \frac{F_0^2 \varepsilon^2}{4\omega^2} \Big| \tilde{a}_y^{r}(t)\bold{d}_{n}^{\perp_m} + \tilde{a}_y^{i}(t)\bold{d}_{m}^{\perp_m} \Big|^2 \bigg)
\end{align}
and
\begin{align}
\tilde{D}_{mn}^{e,2}(\varepsilon,N) &\simeq \frac{F_0^2\varepsilon^2}{2\omega} \int_{t_i'}^{t_r'} dt \text{ } e^{-i\phi_{mn}^{E}(t)} e^{i\phi_{mn}^{x}(t)} f_y(t) \nonumber \\
&\bold{d}_{nm}^{\perp_m} \cdot \Big( \tilde{a}_y^{r}(t)\bold{d}_{n}^{\perp_m} + \tilde{a}_y^{i}(t)\bold{d}_{m}^{\perp_m} \Big) \label{eq_Dmn^e2_Taylor}
\end{align}
where we have used the property that the dot products between the vectors $\bold{d}_{m}^{\perp_m}$, $\bold{d}_{n}^{\perp_m}$ and $\bold{d}_{nm}^{\perp_m}$ (projections of $\bold{d}_{m}^{\alpha}$, $\bold{d}_{n}^{\alpha}$ and $\bold{d}_{nm}^{\alpha}$ onto the plane perpendicular to $\bold{r}_{0}^{m}$) are invariant with respect to rotations around $\alpha$.
Both $\tilde{D}_{mn}^{e,1}$ and $\tilde{D}_{mn}^{e,2}$ (and therefore also $\tilde{D}_{mn}^{e}$) are invariant with respect to the reversal of light polarization and with respect to the exchange of enantiomer.
Indeed, the operation of exchanging the enantiomer is equivalent to reflection of the molecular system through a plane.
Let us consider reflection through a plane that contains $\bold{r}_{0}^{m}$, which points along the $\hat{\bold{x}}$ direction in the laboratory frame.
The value of $d_{nm}^{\parallel_m}$ will remain unaffected.
Although this reflection can modify the direction of the vectors $\bold{d}_{m}^{\perp_m}$, $\bold{d}_{n}^{\perp_m}$ and $\bold{d}_{nm}^{\perp_m}$, the dot products between them will not change.
Therefore, $\tilde{D}_{mn}^{e,1}$ and $\tilde{D}_{mn}^{e,1}$ will remain unaltered.

We note that the behaviour of $\tilde{D}_{mn}^{e,1}$ is very similar to that of the dipole associated with the direct channels ($\tilde{D}_{mm}$, see eq. \ref{eq_Dmm_Taylor}).
The zero order term provides an $\varepsilon$-independent background that is proportional to $F_0$ and to $d_{nm}^{\parallel_m}$, and the second order term induces a quadratic suppression with ellipticity that can be modulated by tuning field parameters, in particular the ratio $F_{0}/\omega$.
Thus, orientational averaging also induces suppression of the achiral background associated with the cross channels, which results from the interaction of the transition dipole with the $\varepsilon$-dependent field component.
However, this suppression can be compensated by $\tilde{D}_{mn}^{e,2}$, which increases quadratically with ellipticity.
The relative strengths of $\tilde{D}_{mn}^{e,1}$ and $\tilde{D}_{mn}^{e,2}$ depend on the orientation of the transition dipole $d_{nm}$ with respect to the direction that maximizes strong-field ionization $\bold{r}_{0}^{m}$.

\begin{figure}
\centering
\includegraphics[width=\linewidth, keepaspectratio=true]{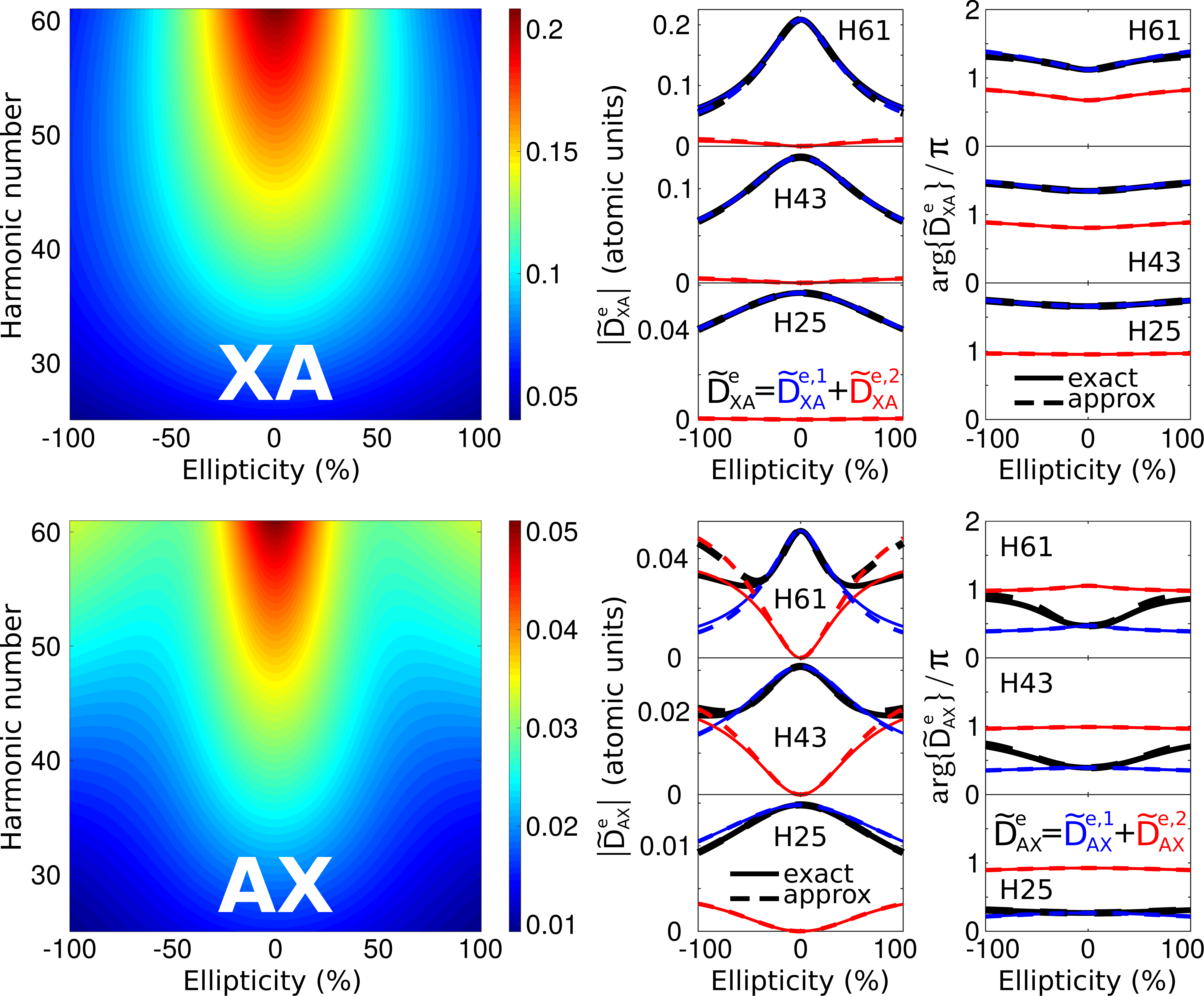}
\caption{Electric contribution to the harmonic dipole $\tilde{D}_{mn}^e$ associated with the cross HHG channels XA (upper panels) and AX (lower panels), for laser parameters $F_0=0.04$ a.u. and $\omega=0.0224$ a.u.
Left panels: exact model solutions of $|\tilde{D}_{mn}^e|$ (eq. \ref{eq_Dmn^e}) as a function of ellipticity and harmonic number.
Right panels: model solutions of $\tilde{D}_{mn}^e$ (black lines) and of its two contributions: $\tilde{D}_{mn}^{e,1}$ (blue lines) and $\tilde{D}_{mn}^{e,2}$ (red lines),
as a function of ellipticity, for harmonic numbers 25 (lower panel), 43 (central panel) and 61 (upper panel).
The first column shows the absolute values $|\tilde{D}_{mn}^{e}|$, $|\tilde{D}_{mn}^{e,1}|$ and $|\tilde{D}_{mn}^{e,2}|$, while the second column shows their phases $\arg(\tilde{D}_{mn}^{e})$, $\arg(\tilde{D}_{mn}^{e,1})$ and $\arg(\tilde{D}_{mn}^{e,2})$.
Full lines: exact solutions (eqs. \ref{eq_Dmn^e1} and \ref{eq_Dmn^e2}); dashed lines: approximate analytical solutions (eqs. \ref{eq_Dmn^e1_Taylor} and \ref{eq_Dmn^e2_Taylor}).
}
\label{fig_De_mn_model}
\end{figure}

The electric contributions to the harmonic dipoles associated with the cross channels XA and AX are presented in fig. \ref{fig_De_mn_model}.
The left panels show the absolute values $|\tilde{D}_{mn}^{e}|$ as a function of ellipticity and harmonic number, evaluated using eq. \ref{eq_Dmn^e_0}.
The agreement with numerical TDSE solutions (not shown) is excellent for both channels.
The right panels of fig. \ref{fig_De_mn_model} show the absolute values and phases of $\tilde{D}_{mn}^{e}$ and of its two contributions, $\tilde{D}_{mn}^{e,1}$ and $\tilde{D}_{mn}^{e,2}$, for harmonic numbers 25, 43 and 61, as a function of ellipticity,
evaluated using the exact eqs. \ref{eq_Dmn^e1} and \ref{eq_Dmn^e2} and the approximate analytical eqs. \ref{eq_Dmn^e1_Taylor} and \ref{eq_Dmn^e2_Taylor}.
The agreement is very good in all cases, which validates the use of the approximations described above.
Only near the cutoff, and for high ellipticities, we find poorer agreement between the exact values of $\tilde{D}_{mn}^{e,2}$ and the analytical solutions provided by eq. \ref{eq_Dmn^e2_Taylor},
which indicates that higher order expansion terms of the sine function in eq. \ref{eq_Dmn^e2} are not negligible in this region of the spectrum. 

The modulation of the two cross channels with ellipticity is very different.
The reason is the different orientation of the transition dipole $\bold{d}_{XA}$ ($=\bold{d}_{AX}$) with respect to the direction that maximizes strong-field ionization from the two ionic states ($\bold{r}_{0}^{X}$ and $\bold{r}_{0}^{A}$), as already pointed out.
Indeed, whereas $\bold{d}_{XA}$ and $\bold{r}_{0}^{X}$ are close to being parallel, $\bold{d}_{XA}$ is essentially orthogonal to $\bold{r}_{0}^{A}$.
In the case of the XA channel, $d_{XA}^{\parallel_X}>|d_{XA}^{\perp_X}|$ (see table \ref{table_DME_projections}) leads to $|\tilde{D}_{XA}^{e,1}|>>|\tilde{D}_{XA}^{e,2}|$, and therefore $\tilde{D}_{XA}^{e}$ decays quadratically decay with ellipticity, like the dipole associated to the direct channels.
In contrast, $\tilde{D}_{AX}^{e,1}$ and $\tilde{D}_{AX}^{e,2}$ have comparable strength at high ellipticities and, as a result, the AX channel exhibits a more complex behaviour.
We note that the ratio $|\tilde{D}_{mn}^{e,2}|/|\tilde{D}_{mn}^{e,1}|$ increases with the harmonic number in both HHG channels, as longer excursion times lead to stronger suppression of $\tilde{D}_{mn}^{e,1}$ and to larger enhancement of $\tilde{D}_{mn}^{e,2}$.
Of course, for weak ellipticities the term $\tilde{D}_{XA}^{e,1}$ is always dominant, and the reason why the amplitude of the XA channel is approximately four times stronger than that of the AX channel is simply that $|d_{XA}^{\parallel_X}| \simeq 4 |d_{XA}^{\parallel_A}|$.

\subsubsection{Magnetic dipole transition contribution}

The term accounting for the magnetic dipole transitions in the ion (eq. $\ref{eq_Dmn^e}$) can also be simplified by applying Euler's formula to $e^{i\varepsilon\phi_{mn}^{y\alpha}(t)}$ and removing the contributions that cancel due to symmetry.
The resulting expression can be written as
\begin{equation}\label{eq_Dmn^m}
\tilde{D}_{mn}^{m}(\varepsilon,N) = \tilde{D}_{mn}^{m,1}(\varepsilon,N) + \tilde{D}_{mn}^{m,2}(\varepsilon,N)
\end{equation}
where the two non-vanishing odd contributions are given by
\begin{align}
\tilde{D}_{mn}^{m,1}(\varepsilon,N) &= i\frac{\varepsilon F_0 m_{nm}^{\parallel_m}}{2\pi c} \int_{t_i'}^{t_r'} dt \text{ } e^{-i\phi_{mn}^{E}(t)} e^{i\phi_{mn}^{x}(t)} f_y(t) \nonumber \\
&\int_{0}^{2\pi} d\alpha \cos{\big(\varepsilon \phi_{\alpha}(t)\big)} \label{eq_Dmn^m1}
\end{align}
and
\begin{align}
\tilde{D}_{mn}^{m,2}(\varepsilon,N) &=  \frac{F_0 m_{nm}^{\perp_m}}{2\pi c} \int_{t_i'}^{t_r'} dt \text{ } e^{-i\phi_{mn}^{E}(t)} e^{i\phi_{mn}^{x}(t)} f_x(t) \nonumber \\
&\int_{0}^{2\pi} d\alpha \cos{(\alpha_{m,0}^{m_{nm}}+\alpha)} \sin{\big(\varepsilon\phi_{\alpha}(t)\big)} \label{eq_Dmn^m2}
\end{align}
We can perform the integration over molecular orientations by expanding the sine and cosine $\varepsilon$-dependent functions in Taylor series up to first non-constant contributions.
Using equivalent arguments as in the previous section, we obtain the following expressions:
\begin{align}
\tilde{D}_{mn}^{m,1}(\varepsilon,N) &\simeq i\frac{F_0 m_{nm}^{\parallel_m}\varepsilon}{c} \int_{t_i'}^{t_r'} dt \text{ } e^{-i\phi_{mn}^{E}(t)} e^{i\phi_{mn}^{x}(t)} f_y(t) \label{eq_Dmn^m1_Taylor} \nonumber \\
& \bigg( 1 -\frac{F_0^2 \varepsilon^2}{4\omega^2} \Big| \tilde{a}_y^{r}(t)\bold{d}_{n}^{\perp_m} + \tilde{a}_y^{i}(t)\bold{d}_{m}^{\perp_m} \Big|^2 \bigg)
\end{align}
and
\begin{align}
\tilde{D}_{mn}^{m,2}(\varepsilon,N) &= \frac{F_0^2\varepsilon}{2c\omega} \int_{t_i'}^{t_r'} dt \text{ } e^{-i\phi_{mn}^{E}(t)} e^{i\phi_{mn}^{x}(t)} f_x(t) \nonumber \\
&\bold{m}_{nm}^{\perp_m} \cdot \Big( \tilde{a}_y^{r}(t)\bold{d}_{n}^{\perp_m} + \tilde{a}_y^{i}(t)\bold{d}_{m}^{\perp_m} \Big) \label{eq_Dmn^m2_Taylor}
\end{align}
These analytical equations show that $\tilde{D}_{mn}^{m,1}$ and $\tilde{D}_{mn}^{m,2}$ are pseudoscalars that change sign with the reversal of light polarization and with the exchange of enantiomer.
This can be easily understood by considering the effect of reflecting the molecular system through a plane that contains $\bold{r}_{0}^{m}$, which is equivalent to exchanging the enantiomer, as we did in the previous section.
The dot products between $\bold{d}_{m}^{\perp_m}$ and $\bold{d}_{n}^{\perp_m}$ will not be affected by this operation, but the sign of $m_{nm}^{\parallel_m}$ will change.
As a result, $\tilde{D}_{mn}^{m,1}$ will flip sign.
Reflection through this plane will also change the sign of the dot products $\bold{m}_{nm}^{\perp_m} \cdot \bold{d}_{m}^{\perp_m}$ and $\bold{m}_{nm}^{\perp_m} \cdot \bold{d}_{n}^{\perp_m}$ because these quantities are molecular pseudoscalars, and therefore $\tilde{D}_{mn}^{m,2}$ will flip sign too.
The full magnetic dipole contribution $\tilde{D}_{mn}^{m}$ is also a pseudoscalar because it is the sum of two pseudoscalars.

\begin{figure}
\centering
\includegraphics[width=\linewidth, keepaspectratio=true]{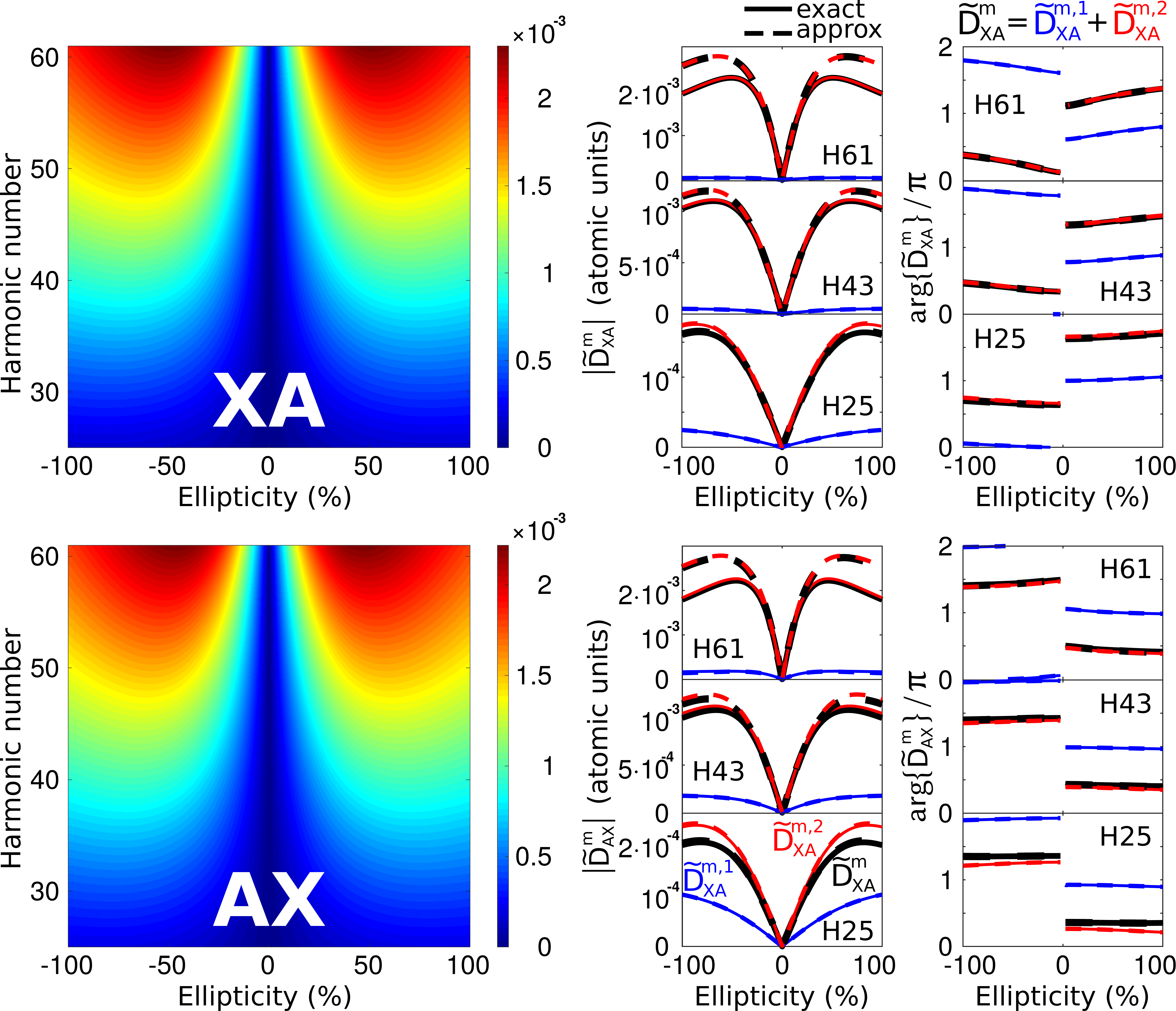}
\caption{Magnetic contribution to the harmonic dipole $\tilde{D}_{mn}^m$ associated with the cross HHG channels XA (upper panels) and AX (lower panels), for laser parameters $F_0=0.04$ a.u. and $\omega=0.0224$ a.u.
Left panels: exact model solutions of $|\tilde{D}_{mn}^m|$ (eq. \ref{eq_Dmn^m}) as a function of ellipticity and harmonic number.
Right panels: model solutions of $\tilde{D}_{mn}^m$ (black lines) and of its two contributions: $\tilde{D}_{mn}^{m,1}$ (blue lines) and $\tilde{D}_{mn}^{m,2}$ (red lines),
as a function of ellipticity, for harmonic numbers 25 (lower panel), 43 (central panel) and 61 (upper panel).
The first column shows the absolute values $|\tilde{D}_{mn}^{m}|$, $|\tilde{D}_{mn}^{m,1}|$ and $|\tilde{D}_{mn}^{m,2}|$, while the second column shows their phases $\arg(\tilde{D}_{mn}^{m})$, $\arg(\tilde{D}_{mn}^{m,1})$ and $\arg(\tilde{D}_{mn}^{m,2})$.
Full lines: exact solutions (eqs. \ref{eq_Dmn^m1} and \ref{eq_Dmn^m2}); dashed lines: approximate analytical solutions (eqs. \ref{eq_Dmn^m1_Taylor} and \ref{eq_Dmn^m2_Taylor}).
}
\label{fig_Dm_mn_model}
\end{figure}

The values of $\tilde{D}_{XA}^{m}$ and $\tilde{D}_{AX}^{m}$ are presented in fig. \ref{fig_Dm_mn_model}.
The left panels contain the absolute values $|\tilde{D}_{XA}^{m}|$ and $|\tilde{D}_{AX}^{m}|$ resulting from applying eq. \ref{eq_Dmn^m_0}, as a function of ellipticity and harmonic number.
The two channels exhibit a very similar behaviour.
In both cases, $|\tilde{D}_{mn}^{m}|$ increases with the harmonic number.
The reason is that higher-order harmonics are associated with longer excursion times, as already discussed, and thus the probability of magnetic transition between ionic states is higher.
For linearly polarized fields, $\tilde{D}_{mn}^{m}$ vanishes as a result of coherent orientational averaging.

The right panels of fig. \ref{fig_Dm_mn_model} contain the absolute values and phases of $\tilde{D}_{mn}^{m}$, together with those of its two contributions, $\tilde{D}_{mn}^{m,1}$ and $\tilde{D}_{mn}^{m,2}$,
evaluated using the exact eqs. \ref{eq_Dmn^m1} and \ref{eq_Dmn^m2} and the approximate analytical eqs. \ref{eq_Dmn^m1_Taylor} and \ref{eq_Dmn^m2_Taylor}, for harmonic numbers 25, 43 and 61, as a function of ellipticity.
All magnetic contributions are odd functions with respect to ellipticity because they are pseudoscalars.
We find that $\tilde{D}_{mn}^{m,2}$ is significantly stronger than $\tilde{D}_{mn}^{m,1}$ in both channels.
The reason is that the magnetic transition dipole $\bold{m}_{XA}$ ($=-\bold{m}_{XA}$) is essentially orthogonal to both $\bold{r}_{0}^{X}$ and $\bold{r}_{0}^{A}$, and therefore $m_{XA}^{\parallel_X}$ and $m_{XA}^{\parallel_A}$ are very small.
This is especially dramatic in the case of the XA channel, where $|m_{XA}^{\perp_X}| \simeq 45 |m_{XA}^{\parallel_X}|$, and thus $\tilde{D}_{AX}^{m,1}$ is negligible in the whole range of harmonic numbers and ellipticities.
We note that there is excellent agreement between the exact eqs. \ref{eq_Dmn^m1} and \ref{eq_Dmn^m2} and the approximate analytical eqs. \ref{eq_Dmn^m1_Taylor} and \ref{eq_Dmn^m2_Taylor} in most regions of the spectra.
Only near the cutoff, and for high ellipticities, the agreement is not so good, which reveals that higher-order terms in the Taylor expansions play some role in this region of the HHG spectra. 

The absolute values $|\tilde{D}_{mn}^{e}|$ and $|\tilde{D}_{mn}^{m}|$ are symmetric with respect to ellipticity and remain unaltered with the exchange of enantiomer.
Chiral dichroism arises when $\tilde{D}_{mn}^{e}$ and $\tilde{D}_{mn}^{m}$ are added coherently because, whereas the phase of $\tilde{D}_{mn}^{e}$ remains unchanged with the reversal of light polarization and with the exchange of enantiomer,
$\tilde{D}_{mn}^{m}$ flips sign, it is a pseudoscalar.
Therefore, in order to have strong chiral response in a given HHG channel, there needs to be a good balance between electric and magnetic contributions.
If one of the two terms is significantly stronger than the other, $|\tilde{D}_{mn}(\varepsilon,N)|$ and $|\tilde{D}_{mn}(-\varepsilon,N)|$ will be very similar and therefore chiral dichroism will be weak.

\subsection{Ellipticity dependence of recombination amplitudes}

Finally, we note that within our primitive model, there is one more $\varepsilon$-dependent contribution that should be given a consideration, because it has the same order in $\varepsilon$ as the other terms considered here.
This contribution is associated with the $\varepsilon$-dependent phase of the recombination matrix element introduced in eq. \ref{eq_a_rec}.
The phase of this matrix element can be written as
\begin{equation}
\Psi_{mn}^{\alpha}(\varepsilon,N) \simeq \Psi_{mn}^{\alpha(0)} + \varepsilon \Psi_{mn}^{\alpha(1)}(N)
\end{equation}
where $\Psi_{mn}^{\alpha(0)}$ is the phase of the recombination dipole in the direction of the $\hat{\bold{x}}$ axis, and the weak dependence on ellipticity is given by
\begin{equation}
\Psi_{mn}^{\alpha(1)}(N) = \frac{\partial\theta}{\partial\varepsilon} \frac{\partial\Psi_{mn}^{\alpha}}{\partial\theta}(0,N)
\end{equation}
where $\theta$ is the recombination angle with respect to the $\hat{\bold{x}}$ axis:
\begin{equation}
\theta \simeq \frac{k_y}{k_x} \simeq \varepsilon \underbrace{ \frac{\tilde{p}_y+a_y(t_r')}{\tilde{p}_x+a_x(t_r')} }_{=q(N)}
\end{equation}
with $k_x$ and $k_y$ being the projections of the recombination velocity onto the $\hat{\bold{x}}$ and $\hat{\bold{y}}$ axes in the laboratory frame,
and $\tilde{p}_x$ and $\tilde{p}_y$ are given by \cite{bookChapter_SmirnovaIvanov_AttosecondAndXUVPhysics}:
\begin{align}
\tilde{p}_x &= \frac{1}{t_i'-t_r'} \int_{t_i'}^{t_r'} a_x(\tau) \text{ } d\tau \\
\tilde{p}_y &= \frac{1}{t_i'-t_r'} \int_{t_i'}^{t_r'} a_y(\tau) \text{ } d\tau
\end{align}
Thus, the linear $\varepsilon$-dependence of $\Psi_{mn}^{\alpha}$ is characterized by
\begin{equation}
\Psi_{mn}^{\alpha(1)}(N) = \frac{\partial\Psi_{mn}^{\alpha}}{\partial\theta}(0,N) \text{ } q(N)
\end{equation}
Including this term into the integrals over $\alpha$ in eqs. \ref{eq_Dmn^e1}, \ref{eq_Dmn^e2}, \ref{eq_Dmn^m1} and \ref{eq_Dmn^m2} we obtain expressions that have the following general structure:
\begin{align}
I_1 &= \int_{0}^{2\pi} d\alpha \underbrace{ \cos{\big(\varepsilon \phi_{\alpha}(t)\big)} }_{g_1^{\alpha}} \text{ } e^{i\varepsilon\Psi_{mn}^{\alpha(1)}(N)} \label{eq_I1} \\
I_2 &= \int_{0}^{2\pi} d\alpha \underbrace{ \cos{(\alpha_{m,0}+\alpha)} \sin{\big(\varepsilon\phi_{\alpha}(t)\big)}  }_{g_2^{\alpha}} \text{ } e^{i\varepsilon \Psi_{mn}^{\alpha(1)}(N)} \label{eq_I2}
\end{align}
One can estimate the outcome of angular integration for small values of $\varepsilon$ as follows.
\begin{align}
I_i &\simeq \int_{0}^{2\pi} g_i^{\alpha} \big[ \cos{\big(\varepsilon\Psi_{mn}^{\alpha(1)}(N)\big)} + i \sin{\big(\varepsilon\Psi_{mn}^{\alpha(1)}(N)\big)} \big] d\alpha \nonumber \\
    &\simeq \int_{0}^{2\pi} g_i^{\alpha} \big[ 1 + i \varepsilon\Psi_{mn}^{\alpha(1)}(N) \big] d\alpha \simeq e^{i\varepsilon G_i} \int_{0}^{2\pi} g_i^{\alpha} d\alpha
\end{align}
where
\begin{equation}
G_i = \frac{\int_{0}^{2\pi} d\alpha \text{ } g_i^{\alpha} \text{ } \Psi_{mn}^{\alpha(1)}(N) }{\int_{0}^{2\pi} d\alpha \text{ } g_i^{\alpha}}
\end{equation}
is a molecular-specific constant.
The expansion is justified for weakly elliptical fields, such as those used in \cite{Cireasa2015NatPhys}, and also for bi-elliptical fields, since the electron returns to the core with nearly straight trajectory.
We do not consider this molecular specific contribution in this paper.

\section{Enhancing chiral response in HHG}\label{section_control}

The HHG emission in a medium of randomly oriented molecules results from the coherent addition of radiation emitted from all the centers in the macroscopic sample.
The model presented in the previous section shows that the linear Stark effect can induce cancellation of the achiral signal associated with the direct HHG channels while preserving the chiral response of the cross channels.
In this section, we illustrate how to exploit this suppression mechanism to enhance chiral response in HHG.

\begin{figure}
\centering
\includegraphics[width=\linewidth, keepaspectratio=true]{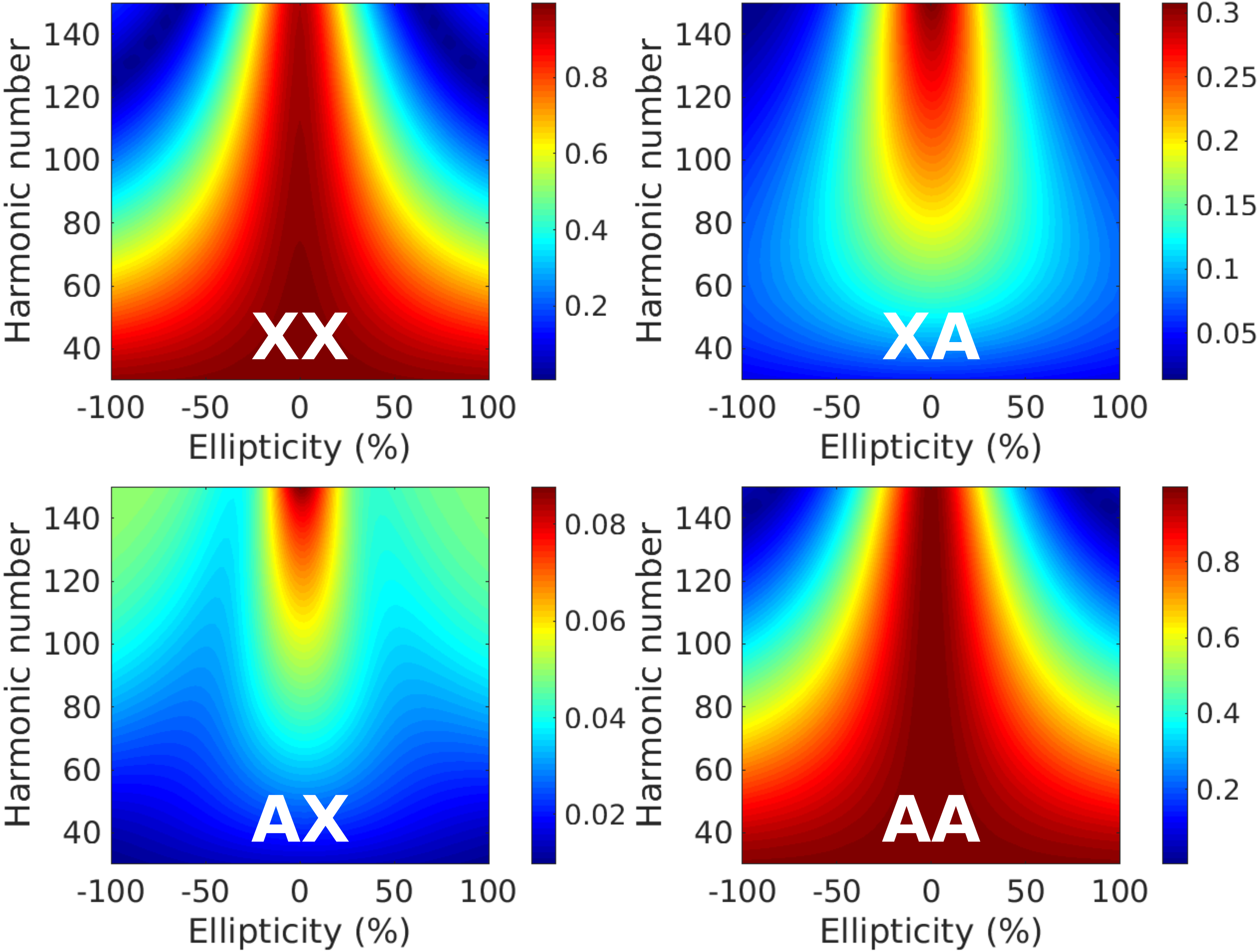}
\caption{Same as fig. \ref{fig_Dmn_TDSE} for laser parameters $F_0=0.05$ a.u. and $\omega=0.018$ a.u.}
\label{fig_Dmn_TDSE_opt}
\end{figure}

\begin{figure}
\centering
\includegraphics[width=\linewidth, keepaspectratio=true]{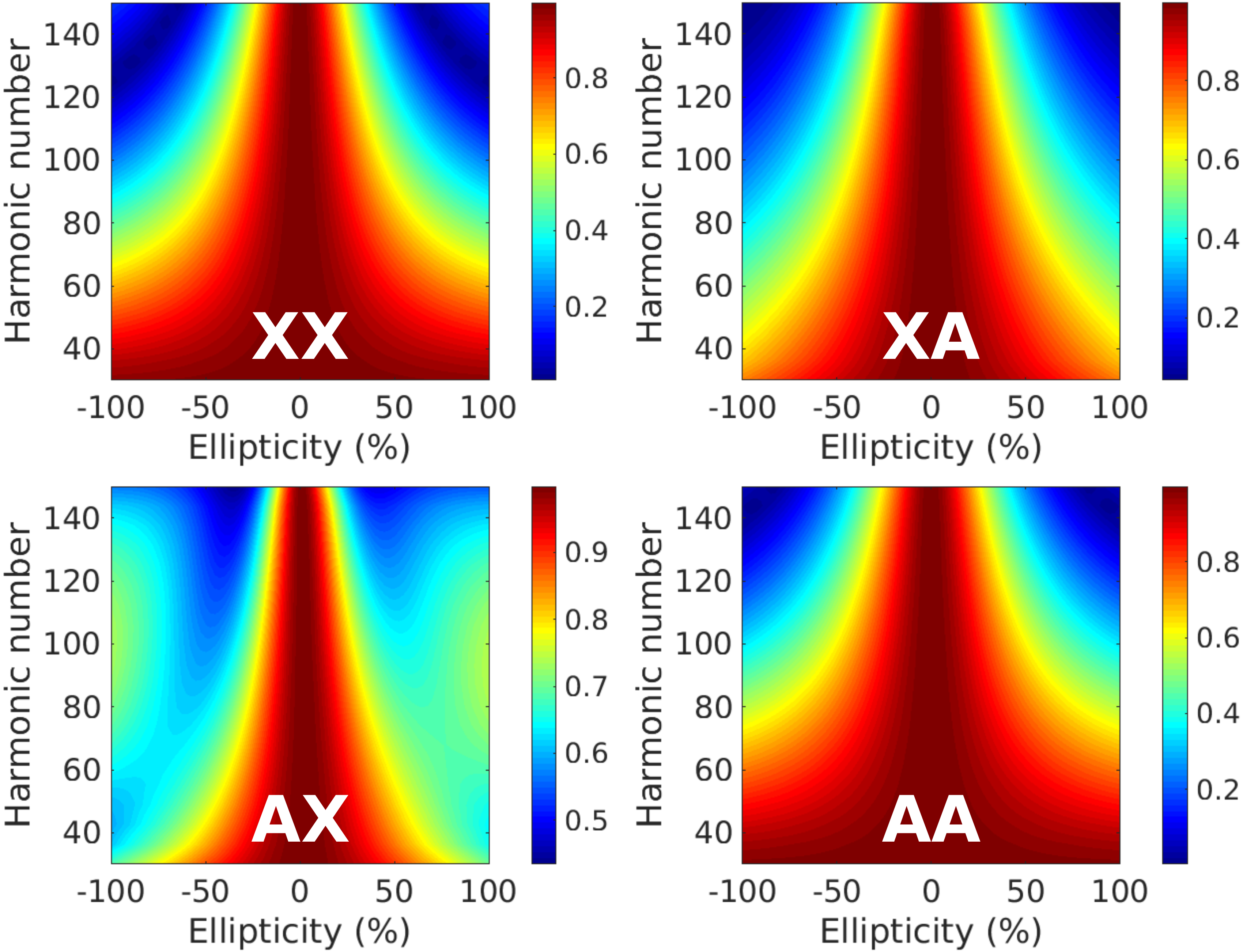}
\caption{Same as fig. \ref{fig_Dmn_norm_TDSE} for laser parameters $F_0=0.05$ a.u. and $\omega=0.018$ a.u.
The absolute values of $|\tilde{D}_{mn}|$ are presented in fig. \ref{fig_Dmn_TDSE_opt}.}
\label{fig_Dmn_TDSE_norm_opt}
\end{figure}

Our analytical model reveals that the harmonic dipole associated with the direct HHG channels decreases quadratically with ellipticity, and that the degree of suppression is proportional to $F_0^2/\omega^2$ (see eq. \ref{eq_Dmm_Taylor}).
Thus, one can induce stronger cancellation of achiral background by increasing the field intensity and/or using longer wavelength radiation.
In order to illustrate this possibility, we have evaluated the different channel contributions to HHG driven by bi-elliptical fields with amplitude $F_0=0.05$ a.u., frequency $\omega=0.018$ a.u. and ellipticity $\varepsilon\in[-1,1]$.
As a result of increasing the ratio $F_0/\omega$, the cutoff value increases from H60 to H150, for $\varepsilon=0$.
The values of $|\tilde{D}_{mn}|$ that result from numerical solution of the TDSE are presented in fig. \ref{fig_Dmn_TDSE_opt}, as a function of harmonic number and ellipticity.
For a better analysis, we show in fig. \ref{fig_Dmn_TDSE_norm_opt} the result of normalizing $|\tilde{D}_{mn}|$ to its maximum value, for each harmonic number.

The values of $|\tilde{D}_{mn}|$ in figs. \ref{fig_Dmn_TDSE_opt} and \ref{fig_Dmn_TDSE_norm_opt} exhibit similar overall trends to those presented in the previous section (figs. \ref{fig_Dmn_TDSE} and \ref{fig_Dmn_norm_TDSE}), but we find some important differences.
As a result using higher field amplitude and longer wavelength, the intensity of the cross channels is now stronger.
A more intense field interacting with the ion during longer excursion times induces larger population transfer between ionic states.
Nonetheless, direct channels are still dominant if the ellipticity of the driving field is weak.
But, for large ellipticities, their intensity drops due to linear Stark shift.
Increasing the ratio $F_0/\omega$ enhances this suppression.
Indeed, our calculations show that $\tilde{D}_{XX}$ and $\tilde{D}_{AA}$ vanish completely in the region of ellipticities $75\%-100\%$ and harmonics numbers $120-150$, showing that the use of tailored fields can lead to complete cancellation of achiral background.
The enantiosensitive cross channels dominate this region of the spectrum.

\begin{figure}
\centering
\includegraphics[width=\linewidth, keepaspectratio=true]{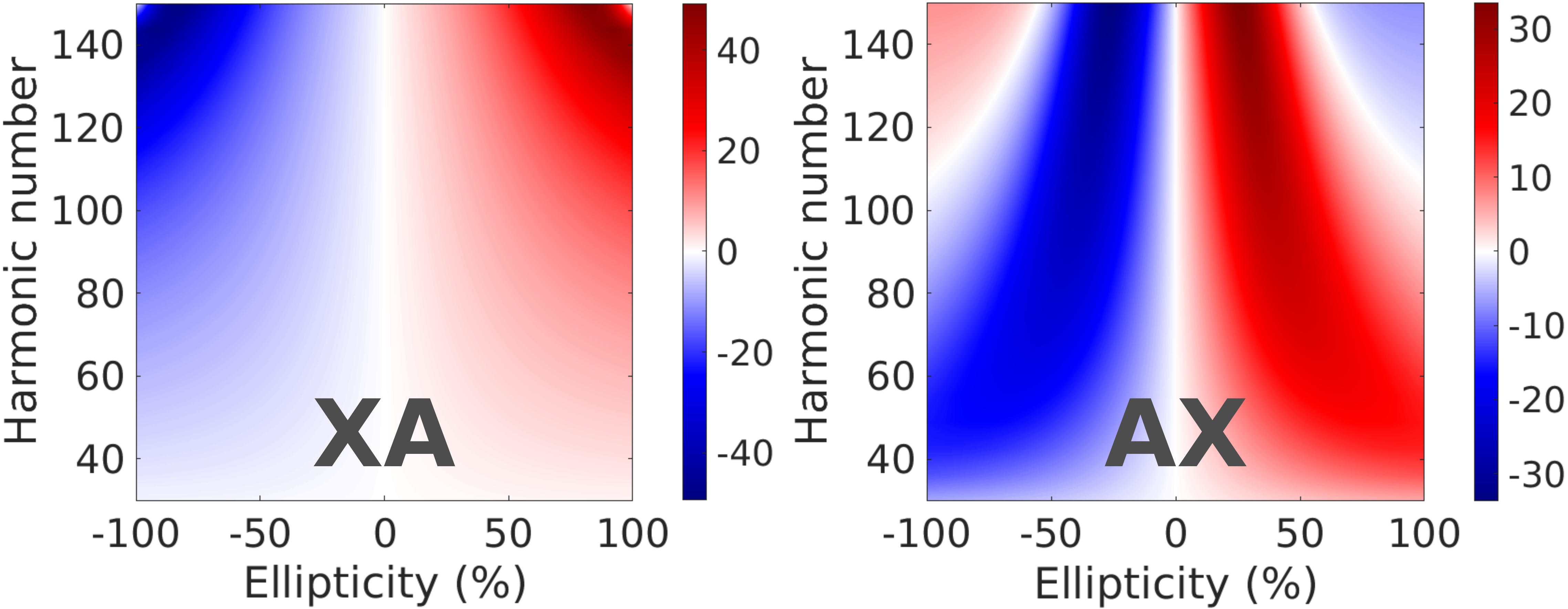}
\caption{Same as fig. \ref{fig_CD_TDSE} for laser parameters $F_0=0.05$ a.u. and $\omega=0.018$ a.u.}
\label{fig_CD_TDSE_opt}
\end{figure}

The chiral dichroism (eq. \ref{eq_CD}) associated with the cross channels XA and AX is presented in fig. \ref{fig_CD_TDSE_opt}.
We obtain values that are significantly larger than those presented in the previous section (fig. \ref{fig_CD_TDSE}), calculated using a less intense driving field with higher frequency.
These results show that increasing the ratio $F_0/\omega$ not only leads to stronger cancellation of the achiral signal of the direct channels, but also to an amplification of the chiral response in the cross channels.
The reason for this enhancement can be fully explained using the model presented in the previous section.
As already stated, in order to obtain strong chiral dichroism in a cross HHG channel, the amplitudes of $\tilde{D}_{mn}^{e}$ and $\tilde{D}_{mn}^{m}$ need to be comparable.
Electric dipole transitions between ionic states are usually more intense than magnetic transitions and, in general, $|\tilde{D}_{mn}^{e}|>>|\tilde{D}_{mn}^{m}|$, as shown in figs. \ref{fig_De_mn_model} and \ref{fig_Dm_mn_model}.
One can significantly reduce the amplitude of $\tilde{D}_{mn}^{e}$ at high ellipticities by enhancing the ratio $F_0/\omega$, because $\tilde{D}_{mn}^{e,1}$ drops with ellipticity in a similar way to the direct channels.
In addition, increasing $F_0$ and reducing $\omega$ leads to an enhancement of the magnetic contribution $\tilde{D}_{mn}^{m}$.
The combination of these two effects leads to a better balance between $\tilde{D}_{mn}^{e}$ and $\tilde{D}_{mn}^{m}$ and, as a result, to stronger chiral response in the cross channels.

\section{Conclusions}

Chiral dichroism in HHG results from the interplay between electric and magnetic interactions in the multi-electron dynamics driven in the ion between ionization and recombination.
In order to observe strong chiral response in the harmonic spectrum of chiral molecules, the achiral background associated with the direct HHG channels needs to be suppressed.
One suppression mechanism consists of destructive interference between different channels.
This mechanism was exploited in the first cHHG experiments \cite{Cireasa2015NatPhys}, that used weakly-elliptical drivers to record different harmonic signals from opposite enantiomers of propylene oxide in the dynamical region of destructive interference between the direct channels XX and AA.

Here we have demonstrated an alternative strategy to suppress achiral background in cHHG that does not require destructive interference between different channels.
Our analytical model reveals that the use of bi-elliptical fields can induce destructive interference at the single-channel level, based on a fundamentally different mechanism: the linear Stark effect.
HHG channels accumulate an additional phase due to the interaction of the ionic states with the strong field.
This extra phase depends on the relative orientation of the molecule with respect to the field.
Although partial orientation is induced by strong-field ionization, we have shown that, as long as the permanent dipole of the ionic state is not parallel to the main ionization direction,
the coherent addition of harmonic radiation emitted from different molecules in the macroscopic sample leads to cancellation of the achiral background associated with the direct HHG channels.
Our model reveals that this suppression mechanism can be controlled by tuning the parameters of the driving field, in particular its frequency, intensity and ellipticity.

We stress once again that our analytical model provides a simple vision for cHHG driven by chiral light pulses:
(1) it quantifies the Stark suppression of achiral signal associated with the direct HHG channels,
(2) it shows that the delicate interplay between electric and magnetic interactions stops suppression of the enantiosensitive cross HHG channels,
(3) it derives the rotationally invariant molecular pseudoscalars responsible for cHHG, and
(4) it shows that one can control and enhance the chiral response in cHHG by tuning the parameters of the applied radiation.
We expect that the recipes proposed here can be exploited in the design of future experiments for observing strong chiral response in cHHG using tailored driving fields.

\section*{Acknowledgements}
The authors acknowledge fruitful discussions with Misha Ivanov.
DA and OS acknowledge support from the DFG SPP 1840 ``Quantum Dynamics in Tailored Intense Fields'' and DFG grant SM 292/5-1; SP and OS acknowledge support MEDEA.
The MEDEA project has received funding from the European Union's Horizon 2020 research and innovation programme under the Marie Sk\l{}odowska-Curie grant agreement No 641789.

\bibliography{Bibliography}

\begin{thebibliography}{74}%
\makeatletter
\providecommand \@ifxundefined [1]{%
 \@ifx{#1\undefined}
}%
\providecommand \@ifnum [1]{%
 \ifnum #1\expandafter \@firstoftwo
 \else \expandafter \@secondoftwo
 \fi
}%
\providecommand \@ifx [1]{%
 \ifx #1\expandafter \@firstoftwo
 \else \expandafter \@secondoftwo
 \fi
}%
\providecommand \natexlab [1]{#1}%
\providecommand \enquote  [1]{``#1''}%
\providecommand \bibnamefont  [1]{#1}%
\providecommand \bibfnamefont [1]{#1}%
\providecommand \citenamefont [1]{#1}%
\providecommand \href@noop [0]{\@secondoftwo}%
\providecommand \href [0]{\begingroup \@sanitize@url \@href}%
\providecommand \@href[1]{\@@startlink{#1}\@@href}%
\providecommand \@@href[1]{\endgroup#1\@@endlink}%
\providecommand \@sanitize@url [0]{\catcode `\\12\catcode `\$12\catcode
  `\&12\catcode `\#12\catcode `\^12\catcode `\_12\catcode `\%12\relax}%
\providecommand \@@startlink[1]{}%
\providecommand \@@endlink[0]{}%
\providecommand \url  [0]{\begingroup\@sanitize@url \@url }%
\providecommand \@url [1]{\endgroup\@href {#1}{\urlprefix }}%
\providecommand \urlprefix  [0]{URL }%
\providecommand \Eprint [0]{\href }%
\providecommand \doibase [0]{http://dx.doi.org/}%
\providecommand \selectlanguage [0]{\@gobble}%
\providecommand \bibinfo  [0]{\@secondoftwo}%
\providecommand \bibfield  [0]{\@secondoftwo}%
\providecommand \translation [1]{[#1]}%
\providecommand \BibitemOpen [0]{}%
\providecommand \bibitemStop [0]{}%
\providecommand \bibitemNoStop [0]{.\EOS\space}%
\providecommand \EOS [0]{\spacefactor3000\relax}%
\providecommand \BibitemShut  [1]{\csname bibitem#1\endcsname}%
\let\auto@bib@innerbib\@empty
\bibitem [{\citenamefont {Ferray}\ \emph {et~al.}(1988)\citenamefont {Ferray},
  \citenamefont {L'Huillier}, \citenamefont {Li}, \citenamefont {Lompre},
  \citenamefont {Mainfray},\ and\ \citenamefont {Manus}}]{Ferray1998JPB}%
  \BibitemOpen
  \bibfield  {author} {\bibinfo {author} {\bibfnamefont {M.}~\bibnamefont
  {Ferray}}, \bibinfo {author} {\bibfnamefont {A.}~\bibnamefont {L'Huillier}},
  \bibinfo {author} {\bibfnamefont {X.~F.}\ \bibnamefont {Li}}, \bibinfo
  {author} {\bibfnamefont {L.~A.}\ \bibnamefont {Lompre}}, \bibinfo {author}
  {\bibfnamefont {G.}~\bibnamefont {Mainfray}}, \ and\ \bibinfo {author}
  {\bibfnamefont {C.}~\bibnamefont {Manus}},\ }\href
  {http://stacks.iop.org/0953-4075/21/i=3/a=001} {\bibfield  {journal}
  {\bibinfo  {journal} {Journal of Physics B: Atomic, Molecular and Optical
  Physics}\ }\textbf {\bibinfo {volume} {21}},\ \bibinfo {pages} {L31}
  (\bibinfo {year} {1988})}\BibitemShut {NoStop}%
\bibitem [{\citenamefont {Krausz}\ and\ \citenamefont
  {Ivanov}(2009)}]{Krausz2009RevModPhys}%
  \BibitemOpen
  \bibfield  {author} {\bibinfo {author} {\bibfnamefont {F.}~\bibnamefont
  {Krausz}}\ and\ \bibinfo {author} {\bibfnamefont {M.}~\bibnamefont
  {Ivanov}},\ }\href {\doibase 10.1103/RevModPhys.81.163} {\bibfield  {journal}
  {\bibinfo  {journal} {Rev. Mod. Phys.}\ }\textbf {\bibinfo {volume} {81}},\
  \bibinfo {pages} {163} (\bibinfo {year} {2009})}\BibitemShut {NoStop}%
\bibitem [{\citenamefont {Corkum}(1993)}]{Corkum1993PRL}%
  \BibitemOpen
  \bibfield  {author} {\bibinfo {author} {\bibfnamefont {P.~B.}\ \bibnamefont
  {Corkum}},\ }\href {\doibase 10.1103/PhysRevLett.71.1994} {\bibfield
  {journal} {\bibinfo  {journal} {Phys. Rev. Lett.}\ }\textbf {\bibinfo
  {volume} {71}},\ \bibinfo {pages} {1994} (\bibinfo {year}
  {1993})}\BibitemShut {NoStop}%
\bibitem [{\citenamefont {Uiberacker}\ \emph {et~al.}(2007)\citenamefont
  {Uiberacker}, \citenamefont {Uphues}, \citenamefont {Schultze}, \citenamefont
  {Verhoef}, \citenamefont {Yakovlev}, \citenamefont {Kling}, \citenamefont
  {Rauschenberger}, \citenamefont {Kabachnik}, \citenamefont {Schr{\"o}der},
  \citenamefont {Lezius}, \citenamefont {Kompa}, \citenamefont {Muller},
  \citenamefont {Vrakking}, \citenamefont {Hendel}, \citenamefont {Kleineberg},
  \citenamefont {Heinzmann}, \citenamefont {Drescher},\ and\ \citenamefont
  {Krausz}}]{Uiberacker2007}%
  \BibitemOpen
  \bibfield  {author} {\bibinfo {author} {\bibfnamefont {M.}~\bibnamefont
  {Uiberacker}}, \bibinfo {author} {\bibfnamefont {T.}~\bibnamefont {Uphues}},
  \bibinfo {author} {\bibfnamefont {M.}~\bibnamefont {Schultze}}, \bibinfo
  {author} {\bibfnamefont {A.~J.}\ \bibnamefont {Verhoef}}, \bibinfo {author}
  {\bibfnamefont {V.}~\bibnamefont {Yakovlev}}, \bibinfo {author}
  {\bibfnamefont {M.~F.}\ \bibnamefont {Kling}}, \bibinfo {author}
  {\bibfnamefont {J.}~\bibnamefont {Rauschenberger}}, \bibinfo {author}
  {\bibfnamefont {N.~M.}\ \bibnamefont {Kabachnik}}, \bibinfo {author}
  {\bibfnamefont {H.}~\bibnamefont {Schr{\"o}der}}, \bibinfo {author}
  {\bibfnamefont {M.}~\bibnamefont {Lezius}}, \bibinfo {author} {\bibfnamefont
  {K.~L.}\ \bibnamefont {Kompa}}, \bibinfo {author} {\bibfnamefont {H.-G.}\
  \bibnamefont {Muller}}, \bibinfo {author} {\bibfnamefont {M.~J.~J.}\
  \bibnamefont {Vrakking}}, \bibinfo {author} {\bibfnamefont {S.}~\bibnamefont
  {Hendel}}, \bibinfo {author} {\bibfnamefont {U.}~\bibnamefont {Kleineberg}},
  \bibinfo {author} {\bibfnamefont {U.}~\bibnamefont {Heinzmann}}, \bibinfo
  {author} {\bibfnamefont {M.}~\bibnamefont {Drescher}}, \ and\ \bibinfo
  {author} {\bibfnamefont {F.}~\bibnamefont {Krausz}},\ }\href
  {http://dx.doi.org/10.1038/nature05648} {\bibfield  {journal} {\bibinfo
  {journal} {Nature}\ }\textbf {\bibinfo {volume} {446}},\ \bibinfo {pages}
  {627 EP } (\bibinfo {year} {2007})},\ \bibinfo {note} {article}\BibitemShut
  {NoStop}%
\bibitem [{\citenamefont {Schultze}\ \emph {et~al.}(2010)\citenamefont
  {Schultze}, \citenamefont {Fie{\ss}}, \citenamefont {Karpowicz},
  \citenamefont {Gagnon}, \citenamefont {Korbman}, \citenamefont {Hofstetter},
  \citenamefont {Neppl}, \citenamefont {Cavalieri}, \citenamefont {Komninos},
  \citenamefont {Mercouris}, \citenamefont {Nicolaides}, \citenamefont
  {Pazourek}, \citenamefont {Nagele}, \citenamefont {Feist}, \citenamefont
  {Burgd{\"o}rfer}, \citenamefont {Azzeer}, \citenamefont {Ernstorfer},
  \citenamefont {Kienberger}, \citenamefont {Kleineberg}, \citenamefont
  {Goulielmakis}, \citenamefont {Krausz},\ and\ \citenamefont
  {Yakovlev}}]{Schultze2010Science}%
  \BibitemOpen
  \bibfield  {author} {\bibinfo {author} {\bibfnamefont {M.}~\bibnamefont
  {Schultze}}, \bibinfo {author} {\bibfnamefont {M.}~\bibnamefont {Fie{\ss}}},
  \bibinfo {author} {\bibfnamefont {N.}~\bibnamefont {Karpowicz}}, \bibinfo
  {author} {\bibfnamefont {J.}~\bibnamefont {Gagnon}}, \bibinfo {author}
  {\bibfnamefont {M.}~\bibnamefont {Korbman}}, \bibinfo {author} {\bibfnamefont
  {M.}~\bibnamefont {Hofstetter}}, \bibinfo {author} {\bibfnamefont
  {S.}~\bibnamefont {Neppl}}, \bibinfo {author} {\bibfnamefont {A.~L.}\
  \bibnamefont {Cavalieri}}, \bibinfo {author} {\bibfnamefont {Y.}~\bibnamefont
  {Komninos}}, \bibinfo {author} {\bibfnamefont {T.}~\bibnamefont {Mercouris}},
  \bibinfo {author} {\bibfnamefont {C.~A.}\ \bibnamefont {Nicolaides}},
  \bibinfo {author} {\bibfnamefont {R.}~\bibnamefont {Pazourek}}, \bibinfo
  {author} {\bibfnamefont {S.}~\bibnamefont {Nagele}}, \bibinfo {author}
  {\bibfnamefont {J.}~\bibnamefont {Feist}}, \bibinfo {author} {\bibfnamefont
  {J.}~\bibnamefont {Burgd{\"o}rfer}}, \bibinfo {author} {\bibfnamefont
  {A.~M.}\ \bibnamefont {Azzeer}}, \bibinfo {author} {\bibfnamefont
  {R.}~\bibnamefont {Ernstorfer}}, \bibinfo {author} {\bibfnamefont
  {R.}~\bibnamefont {Kienberger}}, \bibinfo {author} {\bibfnamefont
  {U.}~\bibnamefont {Kleineberg}}, \bibinfo {author} {\bibfnamefont
  {E.}~\bibnamefont {Goulielmakis}}, \bibinfo {author} {\bibfnamefont
  {F.}~\bibnamefont {Krausz}}, \ and\ \bibinfo {author} {\bibfnamefont {V.~S.}\
  \bibnamefont {Yakovlev}},\ }\href {\doibase 10.1126/science.1189401}
  {\bibfield  {journal} {\bibinfo  {journal} {Science}\ }\textbf {\bibinfo
  {volume} {328}},\ \bibinfo {pages} {1658} (\bibinfo {year}
  {2010})}\BibitemShut {NoStop}%
\bibitem [{\citenamefont {Goulielmakis}\ \emph {et~al.}(2010)\citenamefont
  {Goulielmakis}, \citenamefont {Loh}, \citenamefont {Wirth}, \citenamefont
  {Santra}, \citenamefont {Rohringer}, \citenamefont {Yakovlev}, \citenamefont
  {Zherebtsov}, \citenamefont {Pfeifer}, \citenamefont {Azzeer}, \citenamefont
  {Kling}, \citenamefont {Leone},\ and\ \citenamefont
  {Krausz}}]{Goulielmakis2010Nature}%
  \BibitemOpen
  \bibfield  {author} {\bibinfo {author} {\bibfnamefont {E.}~\bibnamefont
  {Goulielmakis}}, \bibinfo {author} {\bibfnamefont {Z.-H.}\ \bibnamefont
  {Loh}}, \bibinfo {author} {\bibfnamefont {A.}~\bibnamefont {Wirth}}, \bibinfo
  {author} {\bibfnamefont {R.}~\bibnamefont {Santra}}, \bibinfo {author}
  {\bibfnamefont {N.}~\bibnamefont {Rohringer}}, \bibinfo {author}
  {\bibfnamefont {V.~S.}\ \bibnamefont {Yakovlev}}, \bibinfo {author}
  {\bibfnamefont {S.}~\bibnamefont {Zherebtsov}}, \bibinfo {author}
  {\bibfnamefont {T.}~\bibnamefont {Pfeifer}}, \bibinfo {author} {\bibfnamefont
  {A.~M.}\ \bibnamefont {Azzeer}}, \bibinfo {author} {\bibfnamefont {M.~F.}\
  \bibnamefont {Kling}}, \bibinfo {author} {\bibfnamefont {S.~R.}\ \bibnamefont
  {Leone}}, \ and\ \bibinfo {author} {\bibfnamefont {F.}~\bibnamefont
  {Krausz}},\ }\href {http://dx.doi.org/10.1038/nature09212} {\bibfield
  {journal} {\bibinfo  {journal} {Nature}\ }\textbf {\bibinfo {volume} {466}},\
  \bibinfo {pages} {739 EP } (\bibinfo {year} {2010})}\BibitemShut {NoStop}%
\bibitem [{\citenamefont {Ott}\ \emph {et~al.}(2014)\citenamefont {Ott},
  \citenamefont {Kaldun}, \citenamefont {Argenti}, \citenamefont {Raith},
  \citenamefont {Meyer}, \citenamefont {Laux}, \citenamefont {Zhang},
  \citenamefont {Bl{\"a}ttermann}, \citenamefont {Hagstotz}, \citenamefont
  {Ding}, \citenamefont {Heck}, \citenamefont {Madro{\~n}ero}, \citenamefont
  {Mart{\'i}n},\ and\ \citenamefont {Pfeifer}}]{Ott2014Nature}%
  \BibitemOpen
  \bibfield  {author} {\bibinfo {author} {\bibfnamefont {C.}~\bibnamefont
  {Ott}}, \bibinfo {author} {\bibfnamefont {A.}~\bibnamefont {Kaldun}},
  \bibinfo {author} {\bibfnamefont {L.}~\bibnamefont {Argenti}}, \bibinfo
  {author} {\bibfnamefont {P.}~\bibnamefont {Raith}}, \bibinfo {author}
  {\bibfnamefont {K.}~\bibnamefont {Meyer}}, \bibinfo {author} {\bibfnamefont
  {M.}~\bibnamefont {Laux}}, \bibinfo {author} {\bibfnamefont {Y.}~\bibnamefont
  {Zhang}}, \bibinfo {author} {\bibfnamefont {A.}~\bibnamefont
  {Bl{\"a}ttermann}}, \bibinfo {author} {\bibfnamefont {S.}~\bibnamefont
  {Hagstotz}}, \bibinfo {author} {\bibfnamefont {T.}~\bibnamefont {Ding}},
  \bibinfo {author} {\bibfnamefont {R.}~\bibnamefont {Heck}}, \bibinfo {author}
  {\bibfnamefont {J.}~\bibnamefont {Madro{\~n}ero}}, \bibinfo {author}
  {\bibfnamefont {F.}~\bibnamefont {Mart{\'i}n}}, \ and\ \bibinfo {author}
  {\bibfnamefont {T.}~\bibnamefont {Pfeifer}},\ }\href
  {http://dx.doi.org/10.1038/nature14026} {\bibfield  {journal} {\bibinfo
  {journal} {Nature}\ }\textbf {\bibinfo {volume} {516}},\ \bibinfo {pages}
  {374 EP } (\bibinfo {year} {2014})}\BibitemShut {NoStop}%
\bibitem [{\citenamefont {Gruson}\ \emph {et~al.}(2016)\citenamefont {Gruson},
  \citenamefont {Barreau}, \citenamefont {Jim{\'e}nez-Galan}, \citenamefont
  {Risoud}, \citenamefont {Caillat}, \citenamefont {Maquet}, \citenamefont
  {Carr{\'e}}, \citenamefont {Lepetit}, \citenamefont {Hergott}, \citenamefont
  {Ruchon}, \citenamefont {Argenti}, \citenamefont {Ta{\"\i}eb}, \citenamefont
  {Mart{\'\i}n},\ and\ \citenamefont {Sali{\`e}res}}]{Gruson2016Science}%
  \BibitemOpen
  \bibfield  {author} {\bibinfo {author} {\bibfnamefont {V.}~\bibnamefont
  {Gruson}}, \bibinfo {author} {\bibfnamefont {L.}~\bibnamefont {Barreau}},
  \bibinfo {author} {\bibfnamefont {{\'A}.}~\bibnamefont {Jim{\'e}nez-Galan}},
  \bibinfo {author} {\bibfnamefont {F.}~\bibnamefont {Risoud}}, \bibinfo
  {author} {\bibfnamefont {J.}~\bibnamefont {Caillat}}, \bibinfo {author}
  {\bibfnamefont {A.}~\bibnamefont {Maquet}}, \bibinfo {author} {\bibfnamefont
  {B.}~\bibnamefont {Carr{\'e}}}, \bibinfo {author} {\bibfnamefont
  {F.}~\bibnamefont {Lepetit}}, \bibinfo {author} {\bibfnamefont {J.-F.}\
  \bibnamefont {Hergott}}, \bibinfo {author} {\bibfnamefont {T.}~\bibnamefont
  {Ruchon}}, \bibinfo {author} {\bibfnamefont {L.}~\bibnamefont {Argenti}},
  \bibinfo {author} {\bibfnamefont {R.}~\bibnamefont {Ta{\"\i}eb}}, \bibinfo
  {author} {\bibfnamefont {F.}~\bibnamefont {Mart{\'\i}n}}, \ and\ \bibinfo
  {author} {\bibfnamefont {P.}~\bibnamefont {Sali{\`e}res}},\ }\href {\doibase
  10.1126/science.aah5188} {\bibfield  {journal} {\bibinfo  {journal}
  {Science}\ }\textbf {\bibinfo {volume} {354}},\ \bibinfo {pages} {734}
  (\bibinfo {year} {2016})}\BibitemShut {NoStop}%
\bibitem [{\citenamefont {Cirelli}\ \emph {et~al.}(2018)\citenamefont
  {Cirelli}, \citenamefont {Marante}, \citenamefont {Heuser}, \citenamefont
  {Petersson}, \citenamefont {Gal{\'a}n}, \citenamefont {Argenti},
  \citenamefont {Zhong}, \citenamefont {Busto}, \citenamefont {Isinger},
  \citenamefont {Nandi}, \citenamefont {Maclot}, \citenamefont {Rading},
  \citenamefont {Johnsson}, \citenamefont {Gisselbrecht}, \citenamefont
  {Lucchini}, \citenamefont {Gallmann}, \citenamefont {Dahlstr{\"o}m},
  \citenamefont {Lindroth}, \citenamefont {L'Huillier}, \citenamefont
  {Mart{\'i}n},\ and\ \citenamefont {Keller}}]{Cirelli2018NatComm}%
  \BibitemOpen
  \bibfield  {author} {\bibinfo {author} {\bibfnamefont {C.}~\bibnamefont
  {Cirelli}}, \bibinfo {author} {\bibfnamefont {C.}~\bibnamefont {Marante}},
  \bibinfo {author} {\bibfnamefont {S.}~\bibnamefont {Heuser}}, \bibinfo
  {author} {\bibfnamefont {C.~L.~M.}\ \bibnamefont {Petersson}}, \bibinfo
  {author} {\bibfnamefont {{\'A}.~J.}\ \bibnamefont {Gal{\'a}n}}, \bibinfo
  {author} {\bibfnamefont {L.}~\bibnamefont {Argenti}}, \bibinfo {author}
  {\bibfnamefont {S.}~\bibnamefont {Zhong}}, \bibinfo {author} {\bibfnamefont
  {D.}~\bibnamefont {Busto}}, \bibinfo {author} {\bibfnamefont
  {M.}~\bibnamefont {Isinger}}, \bibinfo {author} {\bibfnamefont
  {S.}~\bibnamefont {Nandi}}, \bibinfo {author} {\bibfnamefont
  {S.}~\bibnamefont {Maclot}}, \bibinfo {author} {\bibfnamefont
  {L.}~\bibnamefont {Rading}}, \bibinfo {author} {\bibfnamefont
  {P.}~\bibnamefont {Johnsson}}, \bibinfo {author} {\bibfnamefont
  {M.}~\bibnamefont {Gisselbrecht}}, \bibinfo {author} {\bibfnamefont
  {M.}~\bibnamefont {Lucchini}}, \bibinfo {author} {\bibfnamefont
  {L.}~\bibnamefont {Gallmann}}, \bibinfo {author} {\bibfnamefont {J.~M.}\
  \bibnamefont {Dahlstr{\"o}m}}, \bibinfo {author} {\bibfnamefont
  {E.}~\bibnamefont {Lindroth}}, \bibinfo {author} {\bibfnamefont
  {A.}~\bibnamefont {L'Huillier}}, \bibinfo {author} {\bibfnamefont
  {F.}~\bibnamefont {Mart{\'i}n}}, \ and\ \bibinfo {author} {\bibfnamefont
  {U.}~\bibnamefont {Keller}},\ }\href {\doibase 10.1038/s41467-018-03009-1}
  {\bibfield  {journal} {\bibinfo  {journal} {Nature Communications}\ }\textbf
  {\bibinfo {volume} {9}},\ \bibinfo {pages} {955} (\bibinfo {year}
  {2018})}\BibitemShut {NoStop}%
\bibitem [{\citenamefont {Sansone}\ \emph {et~al.}(2010)\citenamefont
  {Sansone}, \citenamefont {Kelkensberg}, \citenamefont {P{\'e}rez-Torres},
  \citenamefont {Morales}, \citenamefont {Kling}, \citenamefont {Siu},
  \citenamefont {Ghafur}, \citenamefont {Johnsson}, \citenamefont {Swoboda},
  \citenamefont {Benedetti}, \citenamefont {Ferrari}, \citenamefont
  {L{\'e}pine}, \citenamefont {Sanz-Vicario}, \citenamefont {Zherebtsov},
  \citenamefont {Znakovskaya}, \citenamefont {L'Huillier}, \citenamefont
  {Ivanov}, \citenamefont {Nisoli}, \citenamefont {Mart{\'i}n},\ and\
  \citenamefont {Vrakking}}]{Sansone2010}%
  \BibitemOpen
  \bibfield  {author} {\bibinfo {author} {\bibfnamefont {G.}~\bibnamefont
  {Sansone}}, \bibinfo {author} {\bibfnamefont {F.}~\bibnamefont
  {Kelkensberg}}, \bibinfo {author} {\bibfnamefont {J.~F.}\ \bibnamefont
  {P{\'e}rez-Torres}}, \bibinfo {author} {\bibfnamefont {F.}~\bibnamefont
  {Morales}}, \bibinfo {author} {\bibfnamefont {M.~F.}\ \bibnamefont {Kling}},
  \bibinfo {author} {\bibfnamefont {W.}~\bibnamefont {Siu}}, \bibinfo {author}
  {\bibfnamefont {O.}~\bibnamefont {Ghafur}}, \bibinfo {author} {\bibfnamefont
  {P.}~\bibnamefont {Johnsson}}, \bibinfo {author} {\bibfnamefont
  {M.}~\bibnamefont {Swoboda}}, \bibinfo {author} {\bibfnamefont
  {E.}~\bibnamefont {Benedetti}}, \bibinfo {author} {\bibfnamefont
  {F.}~\bibnamefont {Ferrari}}, \bibinfo {author} {\bibfnamefont
  {F.}~\bibnamefont {L{\'e}pine}}, \bibinfo {author} {\bibfnamefont {J.~L.}\
  \bibnamefont {Sanz-Vicario}}, \bibinfo {author} {\bibfnamefont
  {S.}~\bibnamefont {Zherebtsov}}, \bibinfo {author} {\bibfnamefont
  {I.}~\bibnamefont {Znakovskaya}}, \bibinfo {author} {\bibfnamefont
  {A.}~\bibnamefont {L'Huillier}}, \bibinfo {author} {\bibfnamefont {M.~Y.}\
  \bibnamefont {Ivanov}}, \bibinfo {author} {\bibfnamefont {M.}~\bibnamefont
  {Nisoli}}, \bibinfo {author} {\bibfnamefont {F.}~\bibnamefont {Mart{\'i}n}},
  \ and\ \bibinfo {author} {\bibfnamefont {M.~J.~J.}\ \bibnamefont
  {Vrakking}},\ }\href {http://dx.doi.org/10.1038/nature09084} {\bibfield
  {journal} {\bibinfo  {journal} {Nature}\ }\textbf {\bibinfo {volume} {465}},\
  \bibinfo {pages} {763 EP } (\bibinfo {year} {2010})}\BibitemShut {NoStop}%
\bibitem [{\citenamefont {Ranitovic}\ \emph {et~al.}(2014)\citenamefont
  {Ranitovic}, \citenamefont {Hogle}, \citenamefont {Rivi{\`e}re},
  \citenamefont {Palacios}, \citenamefont {Tong}, \citenamefont {Toshima},
  \citenamefont {Gonz{\'a}lez-Castrillo}, \citenamefont {Martin}, \citenamefont
  {Mart{\'\i}n}, \citenamefont {Murnane},\ and\ \citenamefont
  {Kapteyn}}]{Ranitovic2014PNAS}%
  \BibitemOpen
  \bibfield  {author} {\bibinfo {author} {\bibfnamefont {P.}~\bibnamefont
  {Ranitovic}}, \bibinfo {author} {\bibfnamefont {C.~W.}\ \bibnamefont
  {Hogle}}, \bibinfo {author} {\bibfnamefont {P.}~\bibnamefont {Rivi{\`e}re}},
  \bibinfo {author} {\bibfnamefont {A.}~\bibnamefont {Palacios}}, \bibinfo
  {author} {\bibfnamefont {X.-M.}\ \bibnamefont {Tong}}, \bibinfo {author}
  {\bibfnamefont {N.}~\bibnamefont {Toshima}}, \bibinfo {author} {\bibfnamefont
  {A.}~\bibnamefont {Gonz{\'a}lez-Castrillo}}, \bibinfo {author} {\bibfnamefont
  {L.}~\bibnamefont {Martin}}, \bibinfo {author} {\bibfnamefont
  {F.}~\bibnamefont {Mart{\'\i}n}}, \bibinfo {author} {\bibfnamefont {M.~M.}\
  \bibnamefont {Murnane}}, \ and\ \bibinfo {author} {\bibfnamefont
  {H.}~\bibnamefont {Kapteyn}},\ }\href {\doibase 10.1073/pnas.1321999111}
  {\bibfield  {journal} {\bibinfo  {journal} {Proceedings of the National
  Academy of Sciences}\ }\textbf {\bibinfo {volume} {111}},\ \bibinfo {pages}
  {912} (\bibinfo {year} {2014})}\BibitemShut {NoStop}%
\bibitem [{\citenamefont {Calegari}\ \emph {et~al.}(2014)\citenamefont
  {Calegari}, \citenamefont {Ayuso}, \citenamefont {Trabattoni}, \citenamefont
  {Belshaw}, \citenamefont {De~Camillis}, \citenamefont {Anumula},
  \citenamefont {Frassetto}, \citenamefont {Poletto}, \citenamefont {Palacios},
  \citenamefont {Decleva}, \citenamefont {Greenwood}, \citenamefont
  {Mart{\'\i}n},\ and\ \citenamefont {Nisoli}}]{Calegari2014Science}%
  \BibitemOpen
  \bibfield  {author} {\bibinfo {author} {\bibfnamefont {F.}~\bibnamefont
  {Calegari}}, \bibinfo {author} {\bibfnamefont {D.}~\bibnamefont {Ayuso}},
  \bibinfo {author} {\bibfnamefont {A.}~\bibnamefont {Trabattoni}}, \bibinfo
  {author} {\bibfnamefont {L.}~\bibnamefont {Belshaw}}, \bibinfo {author}
  {\bibfnamefont {S.}~\bibnamefont {De~Camillis}}, \bibinfo {author}
  {\bibfnamefont {S.}~\bibnamefont {Anumula}}, \bibinfo {author} {\bibfnamefont
  {F.}~\bibnamefont {Frassetto}}, \bibinfo {author} {\bibfnamefont
  {L.}~\bibnamefont {Poletto}}, \bibinfo {author} {\bibfnamefont
  {A.}~\bibnamefont {Palacios}}, \bibinfo {author} {\bibfnamefont
  {P.}~\bibnamefont {Decleva}}, \bibinfo {author} {\bibfnamefont {J.~B.}\
  \bibnamefont {Greenwood}}, \bibinfo {author} {\bibfnamefont {F.}~\bibnamefont
  {Mart{\'\i}n}}, \ and\ \bibinfo {author} {\bibfnamefont {M.}~\bibnamefont
  {Nisoli}},\ }\href {\doibase 10.1126/science.1254061} {\bibfield  {journal}
  {\bibinfo  {journal} {Science}\ }\textbf {\bibinfo {volume} {346}},\ \bibinfo
  {pages} {336} (\bibinfo {year} {2014})}\BibitemShut {NoStop}%
\bibitem [{\citenamefont {Trabattoni}\ \emph {et~al.}(2015)\citenamefont
  {Trabattoni}, \citenamefont {Klinker}, \citenamefont {Gonz\'alez-V\'azquez},
  \citenamefont {Liu}, \citenamefont {Sansone}, \citenamefont {Linguerri},
  \citenamefont {Hochlaf}, \citenamefont {Klei}, \citenamefont {Vrakking},
  \citenamefont {Mart\'{\i}n}, \citenamefont {Nisoli},\ and\ \citenamefont
  {Calegari}}]{Trabattoni2015PRX}%
  \BibitemOpen
  \bibfield  {author} {\bibinfo {author} {\bibfnamefont {A.}~\bibnamefont
  {Trabattoni}}, \bibinfo {author} {\bibfnamefont {M.}~\bibnamefont {Klinker}},
  \bibinfo {author} {\bibfnamefont {J.}~\bibnamefont {Gonz\'alez-V\'azquez}},
  \bibinfo {author} {\bibfnamefont {C.}~\bibnamefont {Liu}}, \bibinfo {author}
  {\bibfnamefont {G.}~\bibnamefont {Sansone}}, \bibinfo {author} {\bibfnamefont
  {R.}~\bibnamefont {Linguerri}}, \bibinfo {author} {\bibfnamefont
  {M.}~\bibnamefont {Hochlaf}}, \bibinfo {author} {\bibfnamefont
  {J.}~\bibnamefont {Klei}}, \bibinfo {author} {\bibfnamefont {M.~J.~J.}\
  \bibnamefont {Vrakking}}, \bibinfo {author} {\bibfnamefont {F.}~\bibnamefont
  {Mart\'{\i}n}}, \bibinfo {author} {\bibfnamefont {M.}~\bibnamefont {Nisoli}},
  \ and\ \bibinfo {author} {\bibfnamefont {F.}~\bibnamefont {Calegari}},\
  }\href {\doibase 10.1103/PhysRevX.5.041053} {\bibfield  {journal} {\bibinfo
  {journal} {Phys. Rev. X}\ }\textbf {\bibinfo {volume} {5}},\ \bibinfo {pages}
  {041053} (\bibinfo {year} {2015})}\BibitemShut {NoStop}%
\bibitem [{\citenamefont {Calegari}\ \emph {et~al.}(2015)\citenamefont
  {Calegari}, \citenamefont {Ayuso}, \citenamefont {Trabattoni}, \citenamefont
  {Belshaw}, \citenamefont {Camillis}, \citenamefont {Frassetto}, \citenamefont
  {Poletto}, \citenamefont {Palacios}, \citenamefont {Decleva}, \citenamefont
  {Greenwood}, \citenamefont {Mart\'in},\ and\ \citenamefont
  {Nisoli}}]{Calegari2015JSTQE}%
  \BibitemOpen
  \bibfield  {author} {\bibinfo {author} {\bibfnamefont {F.}~\bibnamefont
  {Calegari}}, \bibinfo {author} {\bibfnamefont {D.}~\bibnamefont {Ayuso}},
  \bibinfo {author} {\bibfnamefont {A.}~\bibnamefont {Trabattoni}}, \bibinfo
  {author} {\bibfnamefont {L.}~\bibnamefont {Belshaw}}, \bibinfo {author}
  {\bibfnamefont {S.~D.}\ \bibnamefont {Camillis}}, \bibinfo {author}
  {\bibfnamefont {F.}~\bibnamefont {Frassetto}}, \bibinfo {author}
  {\bibfnamefont {L.}~\bibnamefont {Poletto}}, \bibinfo {author} {\bibfnamefont
  {A.}~\bibnamefont {Palacios}}, \bibinfo {author} {\bibfnamefont
  {P.}~\bibnamefont {Decleva}}, \bibinfo {author} {\bibfnamefont {J.~B.}\
  \bibnamefont {Greenwood}}, \bibinfo {author} {\bibfnamefont {F.}~\bibnamefont
  {Mart\'in}}, \ and\ \bibinfo {author} {\bibfnamefont {M.}~\bibnamefont
  {Nisoli}},\ }\href {\doibase 10.1109/JSTQE.2015.2419218} {\bibfield
  {journal} {\bibinfo  {journal} {IEEE Journal of Selected Topics in Quantum
  Electronics}\ }\textbf {\bibinfo {volume} {21}},\ \bibinfo {pages} {1}
  (\bibinfo {year} {2015})}\BibitemShut {NoStop}%
\bibitem [{\citenamefont {Cavalieri}\ \emph {et~al.}(2007)\citenamefont
  {Cavalieri}, \citenamefont {M{\"u}ller}, \citenamefont {Uphues},
  \citenamefont {Yakovlev}, \citenamefont {Baltuska}, \citenamefont {Horvath},
  \citenamefont {Schmidt}, \citenamefont {Bl{\"u}mel}, \citenamefont
  {Holzwarth}, \citenamefont {Hendel}, \citenamefont {Drescher}, \citenamefont
  {Kleineberg}, \citenamefont {Echenique}, \citenamefont {Kienberger},
  \citenamefont {Krausz},\ and\ \citenamefont {Heinzmann}}]{Cavalieri2007}%
  \BibitemOpen
  \bibfield  {author} {\bibinfo {author} {\bibfnamefont {A.~L.}\ \bibnamefont
  {Cavalieri}}, \bibinfo {author} {\bibfnamefont {N.}~\bibnamefont
  {M{\"u}ller}}, \bibinfo {author} {\bibfnamefont {T.}~\bibnamefont {Uphues}},
  \bibinfo {author} {\bibfnamefont {V.~S.}\ \bibnamefont {Yakovlev}}, \bibinfo
  {author} {\bibfnamefont {A.}~\bibnamefont {Baltuska}}, \bibinfo {author}
  {\bibfnamefont {B.}~\bibnamefont {Horvath}}, \bibinfo {author} {\bibfnamefont
  {B.}~\bibnamefont {Schmidt}}, \bibinfo {author} {\bibfnamefont
  {L.}~\bibnamefont {Bl{\"u}mel}}, \bibinfo {author} {\bibfnamefont
  {R.}~\bibnamefont {Holzwarth}}, \bibinfo {author} {\bibfnamefont
  {S.}~\bibnamefont {Hendel}}, \bibinfo {author} {\bibfnamefont
  {M.}~\bibnamefont {Drescher}}, \bibinfo {author} {\bibfnamefont
  {U.}~\bibnamefont {Kleineberg}}, \bibinfo {author} {\bibfnamefont {P.~M.}\
  \bibnamefont {Echenique}}, \bibinfo {author} {\bibfnamefont {R.}~\bibnamefont
  {Kienberger}}, \bibinfo {author} {\bibfnamefont {F.}~\bibnamefont {Krausz}},
  \ and\ \bibinfo {author} {\bibfnamefont {U.}~\bibnamefont {Heinzmann}},\
  }\href {http://dx.doi.org/10.1038/nature06229} {\bibfield  {journal}
  {\bibinfo  {journal} {Nature}\ }\textbf {\bibinfo {volume} {449}},\ \bibinfo
  {pages} {1029 EP } (\bibinfo {year} {2007})}\BibitemShut {NoStop}%
\bibitem [{\citenamefont {Lein}(2005)}]{Lein2005PRL}%
  \BibitemOpen
  \bibfield  {author} {\bibinfo {author} {\bibfnamefont {M.}~\bibnamefont
  {Lein}},\ }\href {\doibase 10.1103/PhysRevLett.94.053004} {\bibfield
  {journal} {\bibinfo  {journal} {Phys. Rev. Lett.}\ }\textbf {\bibinfo
  {volume} {94}},\ \bibinfo {pages} {053004} (\bibinfo {year}
  {2005})}\BibitemShut {NoStop}%
\bibitem [{\citenamefont {Baker}\ \emph {et~al.}(2006)\citenamefont {Baker},
  \citenamefont {Robinson}, \citenamefont {Haworth}, \citenamefont {Teng},
  \citenamefont {Smith}, \citenamefont {Chiril{\u a}}, \citenamefont {Lein},
  \citenamefont {Tisch},\ and\ \citenamefont {Marangos}}]{Baker2006Science}%
  \BibitemOpen
  \bibfield  {author} {\bibinfo {author} {\bibfnamefont {S.}~\bibnamefont
  {Baker}}, \bibinfo {author} {\bibfnamefont {J.~S.}\ \bibnamefont {Robinson}},
  \bibinfo {author} {\bibfnamefont {C.~A.}\ \bibnamefont {Haworth}}, \bibinfo
  {author} {\bibfnamefont {H.}~\bibnamefont {Teng}}, \bibinfo {author}
  {\bibfnamefont {R.~A.}\ \bibnamefont {Smith}}, \bibinfo {author}
  {\bibfnamefont {C.~C.}\ \bibnamefont {Chiril{\u a}}}, \bibinfo {author}
  {\bibfnamefont {M.}~\bibnamefont {Lein}}, \bibinfo {author} {\bibfnamefont
  {J.~W.~G.}\ \bibnamefont {Tisch}}, \ and\ \bibinfo {author} {\bibfnamefont
  {J.~P.}\ \bibnamefont {Marangos}},\ }\href {\doibase 10.1126/science.1123904}
  {\bibfield  {journal} {\bibinfo  {journal} {Science}\ }\textbf {\bibinfo
  {volume} {312}},\ \bibinfo {pages} {424} (\bibinfo {year}
  {2006})}\BibitemShut {NoStop}%
\bibitem [{\citenamefont {Baker}\ \emph {et~al.}(2008)\citenamefont {Baker},
  \citenamefont {Robinson}, \citenamefont {Lein}, \citenamefont
  {Chiril\ifmmode~\u{a}\else \u{a}\fi{}}, \citenamefont {Torres}, \citenamefont
  {Bandulet}, \citenamefont {Comtois}, \citenamefont {Kieffer}, \citenamefont
  {Villeneuve}, \citenamefont {Tisch},\ and\ \citenamefont
  {Marangos}}]{Baker2008PRL}%
  \BibitemOpen
  \bibfield  {author} {\bibinfo {author} {\bibfnamefont {S.}~\bibnamefont
  {Baker}}, \bibinfo {author} {\bibfnamefont {J.~S.}\ \bibnamefont {Robinson}},
  \bibinfo {author} {\bibfnamefont {M.}~\bibnamefont {Lein}}, \bibinfo {author}
  {\bibfnamefont {C.~C.}\ \bibnamefont {Chiril\ifmmode~\u{a}\else \u{a}\fi{}}},
  \bibinfo {author} {\bibfnamefont {R.}~\bibnamefont {Torres}}, \bibinfo
  {author} {\bibfnamefont {H.~C.}\ \bibnamefont {Bandulet}}, \bibinfo {author}
  {\bibfnamefont {D.}~\bibnamefont {Comtois}}, \bibinfo {author} {\bibfnamefont
  {J.~C.}\ \bibnamefont {Kieffer}}, \bibinfo {author} {\bibfnamefont {D.~M.}\
  \bibnamefont {Villeneuve}}, \bibinfo {author} {\bibfnamefont {J.~W.~G.}\
  \bibnamefont {Tisch}}, \ and\ \bibinfo {author} {\bibfnamefont {J.~P.}\
  \bibnamefont {Marangos}},\ }\href {\doibase 10.1103/PhysRevLett.101.053901}
  {\bibfield  {journal} {\bibinfo  {journal} {Phys. Rev. Lett.}\ }\textbf
  {\bibinfo {volume} {101}},\ \bibinfo {pages} {053901} (\bibinfo {year}
  {2008})}\BibitemShut {NoStop}%
\bibitem [{\citenamefont {Smirnova}\ \emph
  {et~al.}(2009{\natexlab{a}})\citenamefont {Smirnova}, \citenamefont
  {Mairesse}, \citenamefont {Patchkovskii}, \citenamefont {Dudovich},
  \citenamefont {Villeneuve}, \citenamefont {Corkum},\ and\ \citenamefont
  {Ivanov}}]{Smirnova2009Nature}%
  \BibitemOpen
  \bibfield  {author} {\bibinfo {author} {\bibfnamefont {O.}~\bibnamefont
  {Smirnova}}, \bibinfo {author} {\bibfnamefont {Y.}~\bibnamefont {Mairesse}},
  \bibinfo {author} {\bibfnamefont {S.}~\bibnamefont {Patchkovskii}}, \bibinfo
  {author} {\bibfnamefont {N.}~\bibnamefont {Dudovich}}, \bibinfo {author}
  {\bibfnamefont {D.}~\bibnamefont {Villeneuve}}, \bibinfo {author}
  {\bibfnamefont {P.}~\bibnamefont {Corkum}}, \ and\ \bibinfo {author}
  {\bibfnamefont {M.~Y.}\ \bibnamefont {Ivanov}},\ }\href {\doibase
  10.1038/nature08253} {\bibfield  {journal} {\bibinfo  {journal} {Nature}\
  }\textbf {\bibinfo {volume} {460}},\ \bibinfo {pages} {972} (\bibinfo {year}
  {2009}{\natexlab{a}})}\BibitemShut {NoStop}%
\bibitem [{\citenamefont {Smirnova}\ \emph
  {et~al.}(2009{\natexlab{b}})\citenamefont {Smirnova}, \citenamefont
  {Patchkovskii}, \citenamefont {Mairesse}, \citenamefont {Dudovich},\ and\
  \citenamefont {Ivanov}}]{Smirnova2009PNAS}%
  \BibitemOpen
  \bibfield  {author} {\bibinfo {author} {\bibfnamefont {O.}~\bibnamefont
  {Smirnova}}, \bibinfo {author} {\bibfnamefont {S.}~\bibnamefont
  {Patchkovskii}}, \bibinfo {author} {\bibfnamefont {Y.}~\bibnamefont
  {Mairesse}}, \bibinfo {author} {\bibfnamefont {N.}~\bibnamefont {Dudovich}},
  \ and\ \bibinfo {author} {\bibfnamefont {M.~Y.}\ \bibnamefont {Ivanov}},\
  }\href {\doibase 10.1073/pnas.0907434106} {\bibfield  {journal} {\bibinfo
  {journal} {Proceedings of the National Academy of Sciences}\ }\textbf
  {\bibinfo {volume} {106}},\ \bibinfo {pages} {16556} (\bibinfo {year}
  {2009}{\natexlab{b}})}\BibitemShut {NoStop}%
\bibitem [{\citenamefont {Shafir}\ \emph {et~al.}(2012)\citenamefont {Shafir},
  \citenamefont {Soifer}, \citenamefont {Bruner}, \citenamefont {Dagan},
  \citenamefont {Mairesse}, \citenamefont {Patchkovskii}, \citenamefont
  {Ivanov}, \citenamefont {Smirnova},\ and\ \citenamefont
  {Dudovich}}]{Shafir2012}%
  \BibitemOpen
  \bibfield  {author} {\bibinfo {author} {\bibfnamefont {D.}~\bibnamefont
  {Shafir}}, \bibinfo {author} {\bibfnamefont {H.}~\bibnamefont {Soifer}},
  \bibinfo {author} {\bibfnamefont {B.~D.}\ \bibnamefont {Bruner}}, \bibinfo
  {author} {\bibfnamefont {M.}~\bibnamefont {Dagan}}, \bibinfo {author}
  {\bibfnamefont {Y.}~\bibnamefont {Mairesse}}, \bibinfo {author}
  {\bibfnamefont {S.}~\bibnamefont {Patchkovskii}}, \bibinfo {author}
  {\bibfnamefont {M.~Y.}\ \bibnamefont {Ivanov}}, \bibinfo {author}
  {\bibfnamefont {O.}~\bibnamefont {Smirnova}}, \ and\ \bibinfo {author}
  {\bibfnamefont {N.}~\bibnamefont {Dudovich}},\ }\href {\doibase
  10.1038/nature11025} {\bibfield  {journal} {\bibinfo  {journal} {Nature}\
  }\textbf {\bibinfo {volume} {485}},\ \bibinfo {pages} {343} (\bibinfo {year}
  {2012})}\BibitemShut {NoStop}%
\bibitem [{\citenamefont {Pedatzur}\ \emph {et~al.}(2015)\citenamefont
  {Pedatzur}, \citenamefont {Orenstein}, \citenamefont {Serbinenko},
  \citenamefont {Soifer}, \citenamefont {Bruner}, \citenamefont {Uzan},
  \citenamefont {Brambila}, \citenamefont {Harvey}, \citenamefont {Torlina},
  \citenamefont {Morales}, \citenamefont {Smirnova},\ and\ \citenamefont
  {Dudovich}}]{Pedatzur2015}%
  \BibitemOpen
  \bibfield  {author} {\bibinfo {author} {\bibfnamefont {O.}~\bibnamefont
  {Pedatzur}}, \bibinfo {author} {\bibfnamefont {G.}~\bibnamefont {Orenstein}},
  \bibinfo {author} {\bibfnamefont {V.}~\bibnamefont {Serbinenko}}, \bibinfo
  {author} {\bibfnamefont {H.}~\bibnamefont {Soifer}}, \bibinfo {author}
  {\bibfnamefont {B.~D.}\ \bibnamefont {Bruner}}, \bibinfo {author}
  {\bibfnamefont {A.~J.}\ \bibnamefont {Uzan}}, \bibinfo {author}
  {\bibfnamefont {D.~S.}\ \bibnamefont {Brambila}}, \bibinfo {author}
  {\bibfnamefont {A.~G.}\ \bibnamefont {Harvey}}, \bibinfo {author}
  {\bibfnamefont {L.}~\bibnamefont {Torlina}}, \bibinfo {author} {\bibfnamefont
  {F.}~\bibnamefont {Morales}}, \bibinfo {author} {\bibfnamefont
  {O.}~\bibnamefont {Smirnova}}, \ and\ \bibinfo {author} {\bibfnamefont
  {N.}~\bibnamefont {Dudovich}},\ }\href {http://dx.doi.org/10.1038/nphys3436}
  {\bibfield  {journal} {\bibinfo  {journal} {Nat Phys}\ }\textbf {\bibinfo
  {volume} {11}},\ \bibinfo {pages} {815} (\bibinfo {year} {2015})},\ \bibinfo
  {note} {letter}\BibitemShut {NoStop}%
\bibitem [{\citenamefont {Bruner}\ \emph {et~al.}(2016)\citenamefont {Bruner},
  \citenamefont {Masin}, \citenamefont {Negro}, \citenamefont {Morales},
  \citenamefont {Brambila}, \citenamefont {Devetta}, \citenamefont {Facciala},
  \citenamefont {Harvey}, \citenamefont {Ivanov}, \citenamefont {Mairesse},
  \citenamefont {Patchkovskii}, \citenamefont {Serbinenko}, \citenamefont
  {Soifer}, \citenamefont {Stagira}, \citenamefont {Vozzi}, \citenamefont
  {Dudovich},\ and\ \citenamefont {Smirnova}}]{Bruner2016FD}%
  \BibitemOpen
  \bibfield  {author} {\bibinfo {author} {\bibfnamefont {B.~D.}\ \bibnamefont
  {Bruner}}, \bibinfo {author} {\bibfnamefont {Z.}~\bibnamefont {Masin}},
  \bibinfo {author} {\bibfnamefont {M.}~\bibnamefont {Negro}}, \bibinfo
  {author} {\bibfnamefont {F.}~\bibnamefont {Morales}}, \bibinfo {author}
  {\bibfnamefont {D.}~\bibnamefont {Brambila}}, \bibinfo {author}
  {\bibfnamefont {M.}~\bibnamefont {Devetta}}, \bibinfo {author} {\bibfnamefont
  {D.}~\bibnamefont {Facciala}}, \bibinfo {author} {\bibfnamefont {A.~G.}\
  \bibnamefont {Harvey}}, \bibinfo {author} {\bibfnamefont {M.}~\bibnamefont
  {Ivanov}}, \bibinfo {author} {\bibfnamefont {Y.}~\bibnamefont {Mairesse}},
  \bibinfo {author} {\bibfnamefont {S.}~\bibnamefont {Patchkovskii}}, \bibinfo
  {author} {\bibfnamefont {V.}~\bibnamefont {Serbinenko}}, \bibinfo {author}
  {\bibfnamefont {H.}~\bibnamefont {Soifer}}, \bibinfo {author} {\bibfnamefont
  {S.}~\bibnamefont {Stagira}}, \bibinfo {author} {\bibfnamefont
  {C.}~\bibnamefont {Vozzi}}, \bibinfo {author} {\bibfnamefont
  {N.}~\bibnamefont {Dudovich}}, \ and\ \bibinfo {author} {\bibfnamefont
  {O.}~\bibnamefont {Smirnova}},\ }\href {\doibase 10.1039/C6FD00130K}
  {\bibfield  {journal} {\bibinfo  {journal} {Faraday Discuss.}\ }\textbf
  {\bibinfo {volume} {194}},\ \bibinfo {pages} {369} (\bibinfo {year}
  {2016})}\BibitemShut {NoStop}%
\bibitem [{\citenamefont {Cireasa}\ \emph {et~al.}(2015)\citenamefont
  {Cireasa}, \citenamefont {Boguslavskiy}, \citenamefont {Pons}, \citenamefont
  {Wong}, \citenamefont {Descamps}, \citenamefont {Petit}, \citenamefont {Ruf},
  \citenamefont {Thire}, \citenamefont {Ferre}, \citenamefont {Suarez},
  \citenamefont {Higuet}, \citenamefont {Schmidt}, \citenamefont {Alharbi},
  \citenamefont {Legare}, \citenamefont {Blanchet}, \citenamefont {Fabre},
  \citenamefont {Patchkovskii}, \citenamefont {Smirnova}, \citenamefont
  {Mairesse},\ and\ \citenamefont {Bhardwaj}}]{Cireasa2015NatPhys}%
  \BibitemOpen
  \bibfield  {author} {\bibinfo {author} {\bibfnamefont {R.}~\bibnamefont
  {Cireasa}}, \bibinfo {author} {\bibfnamefont {A.~E.}\ \bibnamefont
  {Boguslavskiy}}, \bibinfo {author} {\bibfnamefont {B.}~\bibnamefont {Pons}},
  \bibinfo {author} {\bibfnamefont {M.~C.~H.}\ \bibnamefont {Wong}}, \bibinfo
  {author} {\bibfnamefont {D.}~\bibnamefont {Descamps}}, \bibinfo {author}
  {\bibfnamefont {S.}~\bibnamefont {Petit}}, \bibinfo {author} {\bibfnamefont
  {H.}~\bibnamefont {Ruf}}, \bibinfo {author} {\bibfnamefont {N.}~\bibnamefont
  {Thire}}, \bibinfo {author} {\bibfnamefont {A.}~\bibnamefont {Ferre}},
  \bibinfo {author} {\bibfnamefont {J.}~\bibnamefont {Suarez}}, \bibinfo
  {author} {\bibfnamefont {J.}~\bibnamefont {Higuet}}, \bibinfo {author}
  {\bibfnamefont {B.~E.}\ \bibnamefont {Schmidt}}, \bibinfo {author}
  {\bibfnamefont {A.~F.}\ \bibnamefont {Alharbi}}, \bibinfo {author}
  {\bibfnamefont {F.}~\bibnamefont {Legare}}, \bibinfo {author} {\bibfnamefont
  {V.}~\bibnamefont {Blanchet}}, \bibinfo {author} {\bibfnamefont
  {B.}~\bibnamefont {Fabre}}, \bibinfo {author} {\bibfnamefont
  {S.}~\bibnamefont {Patchkovskii}}, \bibinfo {author} {\bibfnamefont
  {O.}~\bibnamefont {Smirnova}}, \bibinfo {author} {\bibfnamefont
  {Y.}~\bibnamefont {Mairesse}}, \ and\ \bibinfo {author} {\bibfnamefont
  {V.~R.}\ \bibnamefont {Bhardwaj}},\ }\href
  {http://dx.doi.org/10.1038/nphys3369} {\bibfield  {journal} {\bibinfo
  {journal} {Nature Physics}\ }\textbf {\bibinfo {volume} {11}},\ \bibinfo
  {pages} {654} (\bibinfo {year} {2015})},\ \bibinfo {note}
  {letter}\BibitemShut {NoStop}%
\bibitem [{\citenamefont {Wade}(2003)}]{book_Wade_OrganicChemistry}%
  \BibitemOpen
  \bibfield  {author} {\bibinfo {author} {\bibfnamefont {L.~G.}\ \bibnamefont
  {Wade}},\ }\href@noop {} {\emph {\bibinfo {title} {Organic Chemistry}}}\
  (\bibinfo  {publisher} {Prentice Hall},\ \bibinfo {year} {2003})\BibitemShut
  {NoStop}%
\bibitem [{\citenamefont {Carreira}\ and\ \citenamefont
  {Yamamoto}(2012)}]{book_ComprehensiveChirality}%
  \BibitemOpen
  \bibfield  {author} {\bibinfo {author} {\bibfnamefont {E.~M.}\ \bibnamefont
  {Carreira}}\ and\ \bibinfo {author} {\bibfnamefont {H.}~\bibnamefont
  {Yamamoto}},\ }\href@noop {} {\emph {\bibinfo {title} {Comprehensive
  Chirality}}}\ (\bibinfo  {publisher} {Elsevier Ltd.},\ \bibinfo {year}
  {2012})\BibitemShut {NoStop}%
\bibitem [{\citenamefont {Smirnova}\ \emph {et~al.}(2015)\citenamefont
  {Smirnova}, \citenamefont {Mairesse},\ and\ \citenamefont
  {Patchkovskii}}]{Smirnova2015JPB}%
  \BibitemOpen
  \bibfield  {author} {\bibinfo {author} {\bibfnamefont {O.}~\bibnamefont
  {Smirnova}}, \bibinfo {author} {\bibfnamefont {Y.}~\bibnamefont {Mairesse}},
  \ and\ \bibinfo {author} {\bibfnamefont {S.}~\bibnamefont {Patchkovskii}},\
  }\href {http://stacks.iop.org/0953-4075/48/i=23/a=234005} {\bibfield
  {journal} {\bibinfo  {journal} {Journal of Physics B: Atomic, Molecular and
  Optical Physics}\ }\textbf {\bibinfo {volume} {48}},\ \bibinfo {pages}
  {234005} (\bibinfo {year} {2015})}\BibitemShut {NoStop}%
\bibitem [{\citenamefont {Ayuso}\ \emph {et~al.}(2018)\citenamefont {Ayuso},
  \citenamefont {Decleva}, \citenamefont {Patchkovskii},\ and\ \citenamefont
  {Smirnova}}]{Ayuso2018JPB}%
  \BibitemOpen
  \bibfield  {author} {\bibinfo {author} {\bibfnamefont {D.}~\bibnamefont
  {Ayuso}}, \bibinfo {author} {\bibfnamefont {P.}~\bibnamefont {Decleva}},
  \bibinfo {author} {\bibfnamefont {S.}~\bibnamefont {Patchkovskii}}, \ and\
  \bibinfo {author} {\bibfnamefont {O.}~\bibnamefont {Smirnova}},\ }\href
  {http://stacks.iop.org/0953-4075/51/i=6/a=06LT01} {\bibfield  {journal}
  {\bibinfo  {journal} {Journal of Physics B: Atomic, Molecular and Optical
  Physics}\ }\textbf {\bibinfo {volume} {51}},\ \bibinfo {pages} {06LT01}
  (\bibinfo {year} {2018})}\BibitemShut {NoStop}%
\bibitem [{\citenamefont {Berova}\ \emph {et~al.}(2013)\citenamefont {Berova},
  \citenamefont {Polavarapu}, \citenamefont {Nakanishi},\ and\ \citenamefont
  {Woody}}]{book_ComprehensiveChiropticalSpectroscopy}%
  \BibitemOpen
  \bibfield  {author} {\bibinfo {author} {\bibfnamefont {N.}~\bibnamefont
  {Berova}}, \bibinfo {author} {\bibfnamefont {P.~L.}\ \bibnamefont
  {Polavarapu}}, \bibinfo {author} {\bibfnamefont {K.}~\bibnamefont
  {Nakanishi}}, \ and\ \bibinfo {author} {\bibfnamefont {R.~W.}\ \bibnamefont
  {Woody}},\ }\href@noop {} {\emph {\bibinfo {title} {Comprehensive Chiroptical
  Spectroscopy}}}\ (\bibinfo  {publisher} {Wiley},\ \bibinfo {year}
  {2013})\BibitemShut {NoStop}%
\bibitem [{\citenamefont {Tinoco}\ and\ \citenamefont
  {Turner}(1976)}]{Tinoco1976JACS}%
  \BibitemOpen
  \bibfield  {author} {\bibinfo {author} {\bibfnamefont {I.}~\bibnamefont
  {Tinoco}}\ and\ \bibinfo {author} {\bibfnamefont {D.~H.}\ \bibnamefont
  {Turner}},\ }\href {\doibase 10.1021/ja00437a003} {\bibfield  {journal}
  {\bibinfo  {journal} {Journal of the American Chemical Society}\ }\textbf
  {\bibinfo {volume} {98}},\ \bibinfo {pages} {6453} (\bibinfo {year}
  {1976})},\ \Eprint
  {http://arxiv.org/abs/http://dx.doi.org/10.1021/ja00437a003}
  {http://dx.doi.org/10.1021/ja00437a003} \BibitemShut {NoStop}%
\bibitem [{\citenamefont {Castiglioni}\ \emph {et~al.}(2014)\citenamefont
  {Castiglioni}, \citenamefont {Abbate}, \citenamefont {Lebon},\ and\
  \citenamefont {Longhi}}]{Castiglioni2014}%
  \BibitemOpen
  \bibfield  {author} {\bibinfo {author} {\bibfnamefont {E.}~\bibnamefont
  {Castiglioni}}, \bibinfo {author} {\bibfnamefont {S.}~\bibnamefont {Abbate}},
  \bibinfo {author} {\bibfnamefont {F.}~\bibnamefont {Lebon}}, \ and\ \bibinfo
  {author} {\bibfnamefont {G.}~\bibnamefont {Longhi}},\ }\href
  {http://stacks.iop.org/2050-6120/2/i=2/a=024006} {\bibfield  {journal}
  {\bibinfo  {journal} {Methods and Applications in Fluorescence}\ }\textbf
  {\bibinfo {volume} {2}},\ \bibinfo {pages} {024006} (\bibinfo {year}
  {2014})}\BibitemShut {NoStop}%
\bibitem [{\citenamefont {Parchansky}\ \emph {et~al.}(2014)\citenamefont
  {Parchansky}, \citenamefont {Kapitan},\ and\ \citenamefont
  {Bour}}]{Parchansky2014RSC}%
  \BibitemOpen
  \bibfield  {author} {\bibinfo {author} {\bibfnamefont {V.}~\bibnamefont
  {Parchansky}}, \bibinfo {author} {\bibfnamefont {J.}~\bibnamefont {Kapitan}},
  \ and\ \bibinfo {author} {\bibfnamefont {P.}~\bibnamefont {Bour}},\ }\href
  {\doibase 10.1039/C4RA10416A} {\bibfield  {journal} {\bibinfo  {journal} {RSC
  Adv.}\ }\textbf {\bibinfo {volume} {4}},\ \bibinfo {pages} {57125} (\bibinfo
  {year} {2014})}\BibitemShut {NoStop}%
\bibitem [{\citenamefont {Ritchie}(1976)}]{Ritchie1976PRA}%
  \BibitemOpen
  \bibfield  {author} {\bibinfo {author} {\bibfnamefont {B.}~\bibnamefont
  {Ritchie}},\ }\href {\doibase 10.1103/PhysRevA.13.1411} {\bibfield  {journal}
  {\bibinfo  {journal} {Phys. Rev. A}\ }\textbf {\bibinfo {volume} {13}},\
  \bibinfo {pages} {1411} (\bibinfo {year} {1976})}\BibitemShut {NoStop}%
\bibitem [{\citenamefont {Powis}(2000)}]{Powis2000JCP}%
  \BibitemOpen
  \bibfield  {author} {\bibinfo {author} {\bibfnamefont {I.}~\bibnamefont
  {Powis}},\ }\href {\doibase 10.1063/1.480581} {\bibfield  {journal} {\bibinfo
   {journal} {The Journal of Chemical Physics}\ }\textbf {\bibinfo {volume}
  {112}},\ \bibinfo {pages} {301} (\bibinfo {year} {2000})}\BibitemShut
  {NoStop}%
\bibitem [{\citenamefont {B\"owering}\ \emph {et~al.}(2001)\citenamefont
  {B\"owering}, \citenamefont {Lischke}, \citenamefont {Schmidtke},
  \citenamefont {M\"uller}, \citenamefont {Khalil},\ and\ \citenamefont
  {Heinzmann}}]{Bowering2001PRL}%
  \BibitemOpen
  \bibfield  {author} {\bibinfo {author} {\bibfnamefont {N.}~\bibnamefont
  {B\"owering}}, \bibinfo {author} {\bibfnamefont {T.}~\bibnamefont {Lischke}},
  \bibinfo {author} {\bibfnamefont {B.}~\bibnamefont {Schmidtke}}, \bibinfo
  {author} {\bibfnamefont {N.}~\bibnamefont {M\"uller}}, \bibinfo {author}
  {\bibfnamefont {T.}~\bibnamefont {Khalil}}, \ and\ \bibinfo {author}
  {\bibfnamefont {U.}~\bibnamefont {Heinzmann}},\ }\href {\doibase
  10.1103/PhysRevLett.86.1187} {\bibfield  {journal} {\bibinfo  {journal}
  {Phys. Rev. Lett.}\ }\textbf {\bibinfo {volume} {86}},\ \bibinfo {pages}
  {1187} (\bibinfo {year} {2001})}\BibitemShut {NoStop}%
\bibitem [{\citenamefont {Garcia}\ \emph {et~al.}(2003)\citenamefont {Garcia},
  \citenamefont {Nahon}, \citenamefont {Lebech}, \citenamefont {Houver},
  \citenamefont {Dowek},\ and\ \citenamefont {Powis}}]{Garcia2003JCP}%
  \BibitemOpen
  \bibfield  {author} {\bibinfo {author} {\bibfnamefont {G.~A.}\ \bibnamefont
  {Garcia}}, \bibinfo {author} {\bibfnamefont {L.}~\bibnamefont {Nahon}},
  \bibinfo {author} {\bibfnamefont {M.}~\bibnamefont {Lebech}}, \bibinfo
  {author} {\bibfnamefont {J.-C.}\ \bibnamefont {Houver}}, \bibinfo {author}
  {\bibfnamefont {D.}~\bibnamefont {Dowek}}, \ and\ \bibinfo {author}
  {\bibfnamefont {I.}~\bibnamefont {Powis}},\ }\href {\doibase
  10.1063/1.1621379} {\bibfield  {journal} {\bibinfo  {journal} {The Journal of
  Chemical Physics}\ }\textbf {\bibinfo {volume} {119}},\ \bibinfo {pages}
  {8781} (\bibinfo {year} {2003})}\BibitemShut {NoStop}%
\bibitem [{\citenamefont {Lux}\ \emph {et~al.}(2012)\citenamefont {Lux},
  \citenamefont {Wollenhaupt}, \citenamefont {Bolze}, \citenamefont {Liang},
  \citenamefont {Köhler}, \citenamefont {Sarpe},\ and\ \citenamefont
  {Baumert}}]{Lux2012Angewandte}%
  \BibitemOpen
  \bibfield  {author} {\bibinfo {author} {\bibfnamefont {C.}~\bibnamefont
  {Lux}}, \bibinfo {author} {\bibfnamefont {M.}~\bibnamefont {Wollenhaupt}},
  \bibinfo {author} {\bibfnamefont {T.}~\bibnamefont {Bolze}}, \bibinfo
  {author} {\bibfnamefont {Q.}~\bibnamefont {Liang}}, \bibinfo {author}
  {\bibfnamefont {J.}~\bibnamefont {Köhler}}, \bibinfo {author} {\bibfnamefont
  {C.}~\bibnamefont {Sarpe}}, \ and\ \bibinfo {author} {\bibfnamefont
  {T.}~\bibnamefont {Baumert}},\ }\href {\doibase 10.1002/anie.201109035}
  {\bibfield  {journal} {\bibinfo  {journal} {Angewandte Chemie International
  Edition}\ }\textbf {\bibinfo {volume} {51}},\ \bibinfo {pages} {5001}
  (\bibinfo {year} {2012})}\BibitemShut {NoStop}%
\bibitem [{\citenamefont {Garcia}\ \emph {et~al.}(2013)\citenamefont {Garcia},
  \citenamefont {Nahon}, \citenamefont {Daly},\ and\ \citenamefont
  {Powis}}]{Garcia2013NatComm}%
  \BibitemOpen
  \bibfield  {author} {\bibinfo {author} {\bibfnamefont {G.~A.}\ \bibnamefont
  {Garcia}}, \bibinfo {author} {\bibfnamefont {L.}~\bibnamefont {Nahon}},
  \bibinfo {author} {\bibfnamefont {S.}~\bibnamefont {Daly}}, \ and\ \bibinfo
  {author} {\bibfnamefont {I.}~\bibnamefont {Powis}},\ }\href
  {http://dx.doi.org/10.1038/ncomms3132} {\ \textbf {\bibinfo {volume} {4}},\
  \bibinfo {pages} {2132 EP } (\bibinfo {year} {2013})},\ \bibinfo {note}
  {article}\BibitemShut {NoStop}%
\bibitem [{\citenamefont {Lehmann}\ \emph {et~al.}(2013)\citenamefont
  {Lehmann}, \citenamefont {Ram}, \citenamefont {Powis},\ and\ \citenamefont
  {Janssen}}]{Stefan2013JCP}%
  \BibitemOpen
  \bibfield  {author} {\bibinfo {author} {\bibfnamefont {C.~S.}\ \bibnamefont
  {Lehmann}}, \bibinfo {author} {\bibfnamefont {N.~B.}\ \bibnamefont {Ram}},
  \bibinfo {author} {\bibfnamefont {I.}~\bibnamefont {Powis}}, \ and\ \bibinfo
  {author} {\bibfnamefont {M.~H.~M.}\ \bibnamefont {Janssen}},\ }\href
  {\doibase 10.1063/1.4844295} {\bibfield  {journal} {\bibinfo  {journal} {The
  Journal of Chemical Physics}\ }\textbf {\bibinfo {volume} {139}},\ \bibinfo
  {pages} {234307} (\bibinfo {year} {2013})}\BibitemShut {NoStop}%
\bibitem [{\citenamefont {Janssen}\ and\ \citenamefont
  {Powis}(2014)}]{Janssen2014PCCP}%
  \BibitemOpen
  \bibfield  {author} {\bibinfo {author} {\bibfnamefont {M.~H.~M.}\
  \bibnamefont {Janssen}}\ and\ \bibinfo {author} {\bibfnamefont
  {I.}~\bibnamefont {Powis}},\ }\href {\doibase 10.1039/C3CP53741B} {\bibfield
  {journal} {\bibinfo  {journal} {Phys. Chem. Chem. Phys.}\ }\textbf {\bibinfo
  {volume} {16}},\ \bibinfo {pages} {856} (\bibinfo {year} {2014})}\BibitemShut
  {NoStop}%
\bibitem [{\citenamefont {Lux}\ \emph {et~al.}(2015)\citenamefont {Lux},
  \citenamefont {Wollenhaupt}, \citenamefont {Sarpe},\ and\ \citenamefont
  {Baumert}}]{Lux2015ChemPhysChem}%
  \BibitemOpen
  \bibfield  {author} {\bibinfo {author} {\bibfnamefont {C.}~\bibnamefont
  {Lux}}, \bibinfo {author} {\bibfnamefont {M.}~\bibnamefont {Wollenhaupt}},
  \bibinfo {author} {\bibfnamefont {C.}~\bibnamefont {Sarpe}}, \ and\ \bibinfo
  {author} {\bibfnamefont {T.}~\bibnamefont {Baumert}},\ }\href {\doibase
  10.1002/cphc.201402643} {\bibfield  {journal} {\bibinfo  {journal}
  {ChemPhysChem}\ }\textbf {\bibinfo {volume} {16}},\ \bibinfo {pages} {115}
  (\bibinfo {year} {2015})}\BibitemShut {NoStop}%
\bibitem [{\citenamefont {Rhee}\ \emph {et~al.}(2009)\citenamefont {Rhee},
  \citenamefont {June}, \citenamefont {Lee}, \citenamefont {Lee}, \citenamefont
  {Ha}, \citenamefont {Kim}, \citenamefont {Jeon},\ and\ \citenamefont
  {Cho}}]{Rhee2009Nature}%
  \BibitemOpen
  \bibfield  {author} {\bibinfo {author} {\bibfnamefont {H.}~\bibnamefont
  {Rhee}}, \bibinfo {author} {\bibfnamefont {Y.-G.}\ \bibnamefont {June}},
  \bibinfo {author} {\bibfnamefont {J.-S.}\ \bibnamefont {Lee}}, \bibinfo
  {author} {\bibfnamefont {K.-K.}\ \bibnamefont {Lee}}, \bibinfo {author}
  {\bibfnamefont {J.-H.}\ \bibnamefont {Ha}}, \bibinfo {author} {\bibfnamefont
  {Z.~H.}\ \bibnamefont {Kim}}, \bibinfo {author} {\bibfnamefont {S.-J.}\
  \bibnamefont {Jeon}}, \ and\ \bibinfo {author} {\bibfnamefont
  {M.}~\bibnamefont {Cho}},\ }\href {\doibase 10.1038/nature07846} {\bibfield
  {journal} {\bibinfo  {journal} {Nature}\ }\textbf {\bibinfo {volume} {458}},\
  \bibinfo {pages} {310} (\bibinfo {year} {2009})}\BibitemShut {NoStop}%
\bibitem [{\citenamefont {Pitzer}\ \emph {et~al.}(2013)\citenamefont {Pitzer},
  \citenamefont {Kunitski}, \citenamefont {Johnson}, \citenamefont {Jahnke},
  \citenamefont {Sann}, \citenamefont {Sturm}, \citenamefont {Schmidt},
  \citenamefont {Schmidt-B{\"o}cking}, \citenamefont {D{\"o}rner},
  \citenamefont {Stohner}, \citenamefont {Kiedrowski}, \citenamefont
  {Reggelin}, \citenamefont {Marquardt}, \citenamefont {Schie{\ss}er},
  \citenamefont {Berger},\ and\ \citenamefont
  {Sch{\"o}ffler}}]{Pitzer2013Science}%
  \BibitemOpen
  \bibfield  {author} {\bibinfo {author} {\bibfnamefont {M.}~\bibnamefont
  {Pitzer}}, \bibinfo {author} {\bibfnamefont {M.}~\bibnamefont {Kunitski}},
  \bibinfo {author} {\bibfnamefont {A.~S.}\ \bibnamefont {Johnson}}, \bibinfo
  {author} {\bibfnamefont {T.}~\bibnamefont {Jahnke}}, \bibinfo {author}
  {\bibfnamefont {H.}~\bibnamefont {Sann}}, \bibinfo {author} {\bibfnamefont
  {F.}~\bibnamefont {Sturm}}, \bibinfo {author} {\bibfnamefont {L.~P.~H.}\
  \bibnamefont {Schmidt}}, \bibinfo {author} {\bibfnamefont {H.}~\bibnamefont
  {Schmidt-B{\"o}cking}}, \bibinfo {author} {\bibfnamefont {R.}~\bibnamefont
  {D{\"o}rner}}, \bibinfo {author} {\bibfnamefont {J.}~\bibnamefont {Stohner}},
  \bibinfo {author} {\bibfnamefont {J.}~\bibnamefont {Kiedrowski}}, \bibinfo
  {author} {\bibfnamefont {M.}~\bibnamefont {Reggelin}}, \bibinfo {author}
  {\bibfnamefont {S.}~\bibnamefont {Marquardt}}, \bibinfo {author}
  {\bibfnamefont {A.}~\bibnamefont {Schie{\ss}er}}, \bibinfo {author}
  {\bibfnamefont {R.}~\bibnamefont {Berger}}, \ and\ \bibinfo {author}
  {\bibfnamefont {M.~S.}\ \bibnamefont {Sch{\"o}ffler}},\ }\href {\doibase
  10.1126/science.1240362} {\bibfield  {journal} {\bibinfo  {journal}
  {Science}\ }\textbf {\bibinfo {volume} {341}},\ \bibinfo {pages} {1096}
  (\bibinfo {year} {2013})}\BibitemShut {NoStop}%
\bibitem [{\citenamefont {Herwig}\ \emph {et~al.}(2013)\citenamefont {Herwig},
  \citenamefont {Zawatzky}, \citenamefont {Grieser}, \citenamefont {Heber},
  \citenamefont {Jordon-Thaden}, \citenamefont {Krantz}, \citenamefont
  {Novotn{\'y}}, \citenamefont {Repnow}, \citenamefont {Schurig}, \citenamefont
  {Schwalm}, \citenamefont {Vager}, \citenamefont {Wolf}, \citenamefont
  {Trapp},\ and\ \citenamefont {Kreckel}}]{Herwig2013Science}%
  \BibitemOpen
  \bibfield  {author} {\bibinfo {author} {\bibfnamefont {P.}~\bibnamefont
  {Herwig}}, \bibinfo {author} {\bibfnamefont {K.}~\bibnamefont {Zawatzky}},
  \bibinfo {author} {\bibfnamefont {M.}~\bibnamefont {Grieser}}, \bibinfo
  {author} {\bibfnamefont {O.}~\bibnamefont {Heber}}, \bibinfo {author}
  {\bibfnamefont {B.}~\bibnamefont {Jordon-Thaden}}, \bibinfo {author}
  {\bibfnamefont {C.}~\bibnamefont {Krantz}}, \bibinfo {author} {\bibfnamefont
  {O.}~\bibnamefont {Novotn{\'y}}}, \bibinfo {author} {\bibfnamefont
  {R.}~\bibnamefont {Repnow}}, \bibinfo {author} {\bibfnamefont
  {V.}~\bibnamefont {Schurig}}, \bibinfo {author} {\bibfnamefont
  {D.}~\bibnamefont {Schwalm}}, \bibinfo {author} {\bibfnamefont
  {Z.}~\bibnamefont {Vager}}, \bibinfo {author} {\bibfnamefont
  {A.}~\bibnamefont {Wolf}}, \bibinfo {author} {\bibfnamefont {O.}~\bibnamefont
  {Trapp}}, \ and\ \bibinfo {author} {\bibfnamefont {H.}~\bibnamefont
  {Kreckel}},\ }\href {\doibase 10.1126/science.1246549} {\bibfield  {journal}
  {\bibinfo  {journal} {Science}\ }\textbf {\bibinfo {volume} {342}},\ \bibinfo
  {pages} {1084} (\bibinfo {year} {2013})}\BibitemShut {NoStop}%
\bibitem [{\citenamefont {Patterson}\ \emph {et~al.}(2013)\citenamefont
  {Patterson}, \citenamefont {Schnell},\ and\ \citenamefont
  {Doyle}}]{Patterson2013Nature}%
  \BibitemOpen
  \bibfield  {author} {\bibinfo {author} {\bibfnamefont {D.}~\bibnamefont
  {Patterson}}, \bibinfo {author} {\bibfnamefont {M.}~\bibnamefont {Schnell}},
  \ and\ \bibinfo {author} {\bibfnamefont {J.~M.}\ \bibnamefont {Doyle}},\
  }\href {http://dx.doi.org/10.1038/nature12150} {\bibfield  {journal}
  {\bibinfo  {journal} {Nature}\ }\textbf {\bibinfo {volume} {497}},\ \bibinfo
  {pages} {475} (\bibinfo {year} {2013})},\ \bibinfo {note}
  {letter}\BibitemShut {NoStop}%
\bibitem [{\citenamefont {Fidler}\ \emph {et~al.}(2014)\citenamefont {Fidler},
  \citenamefont {Singh}, \citenamefont {Long}, \citenamefont {Dahlberg},\ and\
  \citenamefont {Engel}}]{Fidler2014NatComm}%
  \BibitemOpen
  \bibfield  {author} {\bibinfo {author} {\bibfnamefont {A.~F.}\ \bibnamefont
  {Fidler}}, \bibinfo {author} {\bibfnamefont {V.~P.}\ \bibnamefont {Singh}},
  \bibinfo {author} {\bibfnamefont {P.~D.}\ \bibnamefont {Long}}, \bibinfo
  {author} {\bibfnamefont {P.~D.}\ \bibnamefont {Dahlberg}}, \ and\ \bibinfo
  {author} {\bibfnamefont {G.~S.}\ \bibnamefont {Engel}},\ }\href
  {http://dx.doi.org/10.1038/ncomms4286} {\ \textbf {\bibinfo {volume} {5}},\
  \bibinfo {pages} {3286 EP } (\bibinfo {year} {2014})},\ \bibinfo {note}
  {article}\BibitemShut {NoStop}%
\bibitem [{\citenamefont {Rouxel}\ \emph {et~al.}(2017)\citenamefont {Rouxel},
  \citenamefont {Kowalewski},\ and\ \citenamefont {Mukamel}}]{Rouxel2017SD}%
  \BibitemOpen
  \bibfield  {author} {\bibinfo {author} {\bibfnamefont {J.~R.}\ \bibnamefont
  {Rouxel}}, \bibinfo {author} {\bibfnamefont {M.}~\bibnamefont {Kowalewski}},
  \ and\ \bibinfo {author} {\bibfnamefont {S.}~\bibnamefont {Mukamel}},\ }\href
  {\doibase 10.1063/1.4974260} {\bibfield  {journal} {\bibinfo  {journal}
  {Structural Dynamics}\ }\textbf {\bibinfo {volume} {4}},\ \bibinfo {pages}
  {044006} (\bibinfo {year} {2017})}\BibitemShut {NoStop}%
\bibitem [{\citenamefont {Zhang}\ \emph {et~al.}(2017)\citenamefont {Zhang},
  \citenamefont {Rouxel}, \citenamefont {Autschbach}, \citenamefont {Govind},\
  and\ \citenamefont {Mukamel}}]{Zhang2017CS}%
  \BibitemOpen
  \bibfield  {author} {\bibinfo {author} {\bibfnamefont {Y.}~\bibnamefont
  {Zhang}}, \bibinfo {author} {\bibfnamefont {J.~R.}\ \bibnamefont {Rouxel}},
  \bibinfo {author} {\bibfnamefont {J.}~\bibnamefont {Autschbach}}, \bibinfo
  {author} {\bibfnamefont {N.}~\bibnamefont {Govind}}, \ and\ \bibinfo {author}
  {\bibfnamefont {S.}~\bibnamefont {Mukamel}},\ }\href {\doibase
  10.1039/C7SC01347G} {\bibfield  {journal} {\bibinfo  {journal} {Chem. Sci.}\
  }\textbf {\bibinfo {volume} {8}},\ \bibinfo {pages} {5969} (\bibinfo {year}
  {2017})}\BibitemShut {NoStop}%
\bibitem [{\citenamefont {Beaulieu}\ \emph
  {et~al.}(2016{\natexlab{a}})\citenamefont {Beaulieu}, \citenamefont {Comby},
  \citenamefont {Fabre}, \citenamefont {Descamps}, \citenamefont {Ferre},
  \citenamefont {Garcia}, \citenamefont {Geneaux}, \citenamefont {Legare},
  \citenamefont {Nahon}, \citenamefont {Petit}, \citenamefont {Ruchon},
  \citenamefont {Pons}, \citenamefont {Blanchet},\ and\ \citenamefont
  {Mairesse}}]{Beaulieu2016FD}%
  \BibitemOpen
  \bibfield  {author} {\bibinfo {author} {\bibfnamefont {S.}~\bibnamefont
  {Beaulieu}}, \bibinfo {author} {\bibfnamefont {A.}~\bibnamefont {Comby}},
  \bibinfo {author} {\bibfnamefont {B.}~\bibnamefont {Fabre}}, \bibinfo
  {author} {\bibfnamefont {D.}~\bibnamefont {Descamps}}, \bibinfo {author}
  {\bibfnamefont {A.}~\bibnamefont {Ferre}}, \bibinfo {author} {\bibfnamefont
  {G.}~\bibnamefont {Garcia}}, \bibinfo {author} {\bibfnamefont
  {R.}~\bibnamefont {Geneaux}}, \bibinfo {author} {\bibfnamefont
  {F.}~\bibnamefont {Legare}}, \bibinfo {author} {\bibfnamefont
  {L.}~\bibnamefont {Nahon}}, \bibinfo {author} {\bibfnamefont
  {S.}~\bibnamefont {Petit}}, \bibinfo {author} {\bibfnamefont
  {T.}~\bibnamefont {Ruchon}}, \bibinfo {author} {\bibfnamefont
  {B.}~\bibnamefont {Pons}}, \bibinfo {author} {\bibfnamefont {V.}~\bibnamefont
  {Blanchet}}, \ and\ \bibinfo {author} {\bibfnamefont {Y.}~\bibnamefont
  {Mairesse}},\ }\href {\doibase 10.1039/C6FD00113K} {\bibfield  {journal}
  {\bibinfo  {journal} {Faraday Discuss.}\ }\textbf {\bibinfo {volume} {194}},\
  \bibinfo {pages} {325} (\bibinfo {year} {2016}{\natexlab{a}})}\BibitemShut
  {NoStop}%
\bibitem [{\citenamefont {Beaulieu}\ \emph
  {et~al.}(2016{\natexlab{b}})\citenamefont {Beaulieu}, \citenamefont {Comby},
  \citenamefont {Descamps}, \citenamefont {Fabre}, \citenamefont {Garcia},
  \citenamefont {Geneaux}, \citenamefont {Harvey}, \citenamefont {Legare},
  \citenamefont {Masin}, \citenamefont {Nahon}, \citenamefont {Ordonez},
  \citenamefont {Petit}, \citenamefont {Pons}, \citenamefont {Mairesse},
  \citenamefont {Smirnova},\ and\ \citenamefont
  {Blanchet}}]{Beaulieu2016arXiv}%
  \BibitemOpen
  \bibfield  {author} {\bibinfo {author} {\bibfnamefont {S.}~\bibnamefont
  {Beaulieu}}, \bibinfo {author} {\bibfnamefont {A.}~\bibnamefont {Comby}},
  \bibinfo {author} {\bibfnamefont {D.}~\bibnamefont {Descamps}}, \bibinfo
  {author} {\bibfnamefont {B.}~\bibnamefont {Fabre}}, \bibinfo {author}
  {\bibfnamefont {G.~A.}\ \bibnamefont {Garcia}}, \bibinfo {author}
  {\bibfnamefont {R.}~\bibnamefont {Geneaux}}, \bibinfo {author} {\bibfnamefont
  {A.~G.}\ \bibnamefont {Harvey}}, \bibinfo {author} {\bibfnamefont
  {F.}~\bibnamefont {Legare}}, \bibinfo {author} {\bibfnamefont
  {Z.}~\bibnamefont {Masin}}, \bibinfo {author} {\bibfnamefont
  {L.}~\bibnamefont {Nahon}}, \bibinfo {author} {\bibfnamefont {A.~F.}\
  \bibnamefont {Ordonez}}, \bibinfo {author} {\bibfnamefont {S.}~\bibnamefont
  {Petit}}, \bibinfo {author} {\bibfnamefont {B.}~\bibnamefont {Pons}},
  \bibinfo {author} {\bibfnamefont {Y.}~\bibnamefont {Mairesse}}, \bibinfo
  {author} {\bibfnamefont {O.}~\bibnamefont {Smirnova}}, \ and\ \bibinfo
  {author} {\bibfnamefont {V.}~\bibnamefont {Blanchet}},\ }\href@noop {}
  {\enquote {\bibinfo {title} {Photoexcitation circular dichroism in chiral
  molecules},}\ } (\bibinfo {year} {2016}{\natexlab{b}}),\ \Eprint
  {http://arxiv.org/abs/arXiv:1612.08764} {arXiv:1612.08764} \BibitemShut
  {NoStop}%
\bibitem [{\citenamefont {Eichmann}\ \emph {et~al.}(1995)\citenamefont
  {Eichmann}, \citenamefont {Egbert}, \citenamefont {Nolte}, \citenamefont
  {Momma}, \citenamefont {Wellegehausen}, \citenamefont {Becker}, \citenamefont
  {Long},\ and\ \citenamefont {McIver}}]{Eichmann1995PRA}%
  \BibitemOpen
  \bibfield  {author} {\bibinfo {author} {\bibfnamefont {H.}~\bibnamefont
  {Eichmann}}, \bibinfo {author} {\bibfnamefont {A.}~\bibnamefont {Egbert}},
  \bibinfo {author} {\bibfnamefont {S.}~\bibnamefont {Nolte}}, \bibinfo
  {author} {\bibfnamefont {C.}~\bibnamefont {Momma}}, \bibinfo {author}
  {\bibfnamefont {B.}~\bibnamefont {Wellegehausen}}, \bibinfo {author}
  {\bibfnamefont {W.}~\bibnamefont {Becker}}, \bibinfo {author} {\bibfnamefont
  {S.}~\bibnamefont {Long}}, \ and\ \bibinfo {author} {\bibfnamefont {J.~K.}\
  \bibnamefont {McIver}},\ }\href {\doibase 10.1103/PhysRevA.51.R3414}
  {\bibfield  {journal} {\bibinfo  {journal} {Physical Review A}\ }\textbf
  {\bibinfo {volume} {51}},\ \bibinfo {pages} {R3414} (\bibinfo {year}
  {1995})}\BibitemShut {NoStop}%
\bibitem [{\citenamefont {Long}\ \emph {et~al.}(1995)\citenamefont {Long},
  \citenamefont {Becker},\ and\ \citenamefont {McIver}}]{Long1995PRA}%
  \BibitemOpen
  \bibfield  {author} {\bibinfo {author} {\bibfnamefont {S.}~\bibnamefont
  {Long}}, \bibinfo {author} {\bibfnamefont {W.}~\bibnamefont {Becker}}, \ and\
  \bibinfo {author} {\bibfnamefont {J.~K.}\ \bibnamefont {McIver}},\ }\href
  {\doibase 10.1103/PhysRevA.52.2262} {\bibfield  {journal} {\bibinfo
  {journal} {Physical Review A}\ }\textbf {\bibinfo {volume} {52}},\ \bibinfo
  {pages} {2262} (\bibinfo {year} {1995})}\BibitemShut {NoStop}%
\bibitem [{\citenamefont {Milo\ifmmode \check{s}\else
  \v{s}\fi{}evi\ifmmode~\acute{c}\else \'{c}\fi{}}\ \emph
  {et~al.}(2000)\citenamefont {Milo\ifmmode \check{s}\else
  \v{s}\fi{}evi\ifmmode~\acute{c}\else \'{c}\fi{}}, \citenamefont {Becker},\
  and\ \citenamefont {Kopold}}]{Milosevic2000PRA}%
  \BibitemOpen
  \bibfield  {author} {\bibinfo {author} {\bibfnamefont {D.~B.}\ \bibnamefont
  {Milo\ifmmode \check{s}\else \v{s}\fi{}evi\ifmmode~\acute{c}\else
  \'{c}\fi{}}}, \bibinfo {author} {\bibfnamefont {W.}~\bibnamefont {Becker}}, \
  and\ \bibinfo {author} {\bibfnamefont {R.}~\bibnamefont {Kopold}},\ }\href
  {\doibase 10.1103/PhysRevA.61.063403} {\bibfield  {journal} {\bibinfo
  {journal} {Physical Review A}\ }\textbf {\bibinfo {volume} {61}},\ \bibinfo
  {pages} {063403} (\bibinfo {year} {2000})}\BibitemShut {NoStop}%
\bibitem [{\citenamefont {Milo\ifmmode \check{s}\else
  \v{s}\fi{}evi\ifmmode~\acute{c}\else \'{c}\fi{}}\ and\ \citenamefont
  {Becker}(2000)}]{Milosevic2000PRA_2}%
  \BibitemOpen
  \bibfield  {author} {\bibinfo {author} {\bibfnamefont {D.~B.}\ \bibnamefont
  {Milo\ifmmode \check{s}\else \v{s}\fi{}evi\ifmmode~\acute{c}\else
  \'{c}\fi{}}}\ and\ \bibinfo {author} {\bibfnamefont {W.}~\bibnamefont
  {Becker}},\ }\href {\doibase 10.1103/PhysRevA.62.011403} {\bibfield
  {journal} {\bibinfo  {journal} {Physical Review A}\ }\textbf {\bibinfo
  {volume} {62}},\ \bibinfo {pages} {011403} (\bibinfo {year}
  {2000})}\BibitemShut {NoStop}%
\bibitem [{\citenamefont {Zuo}\ and\ \citenamefont
  {Bandrauk}(1995)}]{Zuo1995JNOPM}%
  \BibitemOpen
  \bibfield  {author} {\bibinfo {author} {\bibfnamefont {T.}~\bibnamefont
  {Zuo}}\ and\ \bibinfo {author} {\bibfnamefont {A.~D.}\ \bibnamefont
  {Bandrauk}},\ }\href {\doibase 10.1142/S0218863595000227} {\bibfield
  {journal} {\bibinfo  {journal} {J. Nonlinear Optic. Phys. Mat. 6}\ }\textbf
  {\bibinfo {volume} {04}},\ \bibinfo {pages} {533} (\bibinfo {year}
  {1995})}\BibitemShut {NoStop}%
\bibitem [{\citenamefont {Ivanov}\ and\ \citenamefont
  {Pisanty}(2014)}]{Ivanov2014NatPhot}%
  \BibitemOpen
  \bibfield  {author} {\bibinfo {author} {\bibfnamefont {M.}~\bibnamefont
  {Ivanov}}\ and\ \bibinfo {author} {\bibfnamefont {E.}~\bibnamefont
  {Pisanty}},\ }\href {http://dx.doi.org/10.1038/nphoton.2014.141} {\bibfield
  {journal} {\bibinfo  {journal} {Nature Photonics}\ }\textbf {\bibinfo
  {volume} {8}},\ \bibinfo {pages} {501} (\bibinfo {year} {2014})},\ \bibinfo
  {note} {news and Views}\BibitemShut {NoStop}%
\bibitem [{\citenamefont {Fleischer}\ \emph {et~al.}(2014)\citenamefont
  {Fleischer}, \citenamefont {Kfir}, \citenamefont {Diskin}, \citenamefont
  {Sidorenko},\ and\ \citenamefont {Cohen}}]{Fleischer2014NatPhot}%
  \BibitemOpen
  \bibfield  {author} {\bibinfo {author} {\bibfnamefont {A.}~\bibnamefont
  {Fleischer}}, \bibinfo {author} {\bibfnamefont {O.}~\bibnamefont {Kfir}},
  \bibinfo {author} {\bibfnamefont {T.}~\bibnamefont {Diskin}}, \bibinfo
  {author} {\bibfnamefont {P.}~\bibnamefont {Sidorenko}}, \ and\ \bibinfo
  {author} {\bibfnamefont {O.}~\bibnamefont {Cohen}},\ }\href
  {http://dx.doi.org/10.1038/nphoton.2014.108} {\bibfield  {journal} {\bibinfo
  {journal} {Nature Photonics}\ }\textbf {\bibinfo {volume} {8}},\ \bibinfo
  {pages} {543} (\bibinfo {year} {2014})},\ \bibinfo {note}
  {article}\BibitemShut {NoStop}%
\bibitem [{\citenamefont {Pisanty}\ \emph {et~al.}(2014)\citenamefont
  {Pisanty}, \citenamefont {Sukiasyan},\ and\ \citenamefont
  {Ivanov}}]{Pisanty2014PRA}%
  \BibitemOpen
  \bibfield  {author} {\bibinfo {author} {\bibfnamefont {E.}~\bibnamefont
  {Pisanty}}, \bibinfo {author} {\bibfnamefont {S.}~\bibnamefont {Sukiasyan}},
  \ and\ \bibinfo {author} {\bibfnamefont {M.}~\bibnamefont {Ivanov}},\ }\href
  {\doibase 10.1103/PhysRevA.90.043829} {\bibfield  {journal} {\bibinfo
  {journal} {Phys. Rev. A}\ }\textbf {\bibinfo {volume} {90}},\ \bibinfo
  {pages} {043829} (\bibinfo {year} {2014})}\BibitemShut {NoStop}%
\bibitem [{\citenamefont {Kfir}\ \emph {et~al.}(2015)\citenamefont {Kfir},
  \citenamefont {Grychtol}, \citenamefont {Turgut}, \citenamefont {Knut},
  \citenamefont {Zusin}, \citenamefont {Popmintchev}, \citenamefont
  {Popmintchev}, \citenamefont {Nembach}, \citenamefont {Shaw}, \citenamefont
  {Fleischer}, \citenamefont {Kapteyn}, \citenamefont {Murnane},\ and\
  \citenamefont {Cohen}}]{Kfir2015NatPhot}%
  \BibitemOpen
  \bibfield  {author} {\bibinfo {author} {\bibfnamefont {O.}~\bibnamefont
  {Kfir}}, \bibinfo {author} {\bibfnamefont {P.}~\bibnamefont {Grychtol}},
  \bibinfo {author} {\bibfnamefont {E.}~\bibnamefont {Turgut}}, \bibinfo
  {author} {\bibfnamefont {R.}~\bibnamefont {Knut}}, \bibinfo {author}
  {\bibfnamefont {D.}~\bibnamefont {Zusin}}, \bibinfo {author} {\bibfnamefont
  {D.}~\bibnamefont {Popmintchev}}, \bibinfo {author} {\bibfnamefont
  {T.}~\bibnamefont {Popmintchev}}, \bibinfo {author} {\bibfnamefont
  {H.}~\bibnamefont {Nembach}}, \bibinfo {author} {\bibfnamefont {J.~M.}\
  \bibnamefont {Shaw}}, \bibinfo {author} {\bibfnamefont {A.}~\bibnamefont
  {Fleischer}}, \bibinfo {author} {\bibfnamefont {H.}~\bibnamefont {Kapteyn}},
  \bibinfo {author} {\bibfnamefont {M.}~\bibnamefont {Murnane}}, \ and\
  \bibinfo {author} {\bibfnamefont {O.}~\bibnamefont {Cohen}},\ }\href
  {http://dx.doi.org/10.1038/nphoton.2014.293} {\bibfield  {journal} {\bibinfo
  {journal} {Nature Photonics}\ }\textbf {\bibinfo {volume} {9}},\ \bibinfo
  {pages} {99} (\bibinfo {year} {2015})},\ \bibinfo {note}
  {article}\BibitemShut {NoStop}%
\bibitem [{\citenamefont {Milo\v{s}evi\'{c}}(2015)}]{Milosevic2015OptLett}%
  \BibitemOpen
  \bibfield  {author} {\bibinfo {author} {\bibfnamefont {D.~B.}\ \bibnamefont
  {Milo\v{s}evi\'{c}}},\ }\href {\doibase 10.1364/OL.40.002381} {\bibfield
  {journal} {\bibinfo  {journal} {Opt. Lett.}\ }\textbf {\bibinfo {volume}
  {40}},\ \bibinfo {pages} {2381} (\bibinfo {year} {2015})}\BibitemShut
  {NoStop}%
\bibitem [{\citenamefont {Medi\ifmmode~\check{s}\else \v{s}\fi{}auskas}\ \emph
  {et~al.}(2015)\citenamefont {Medi\ifmmode~\check{s}\else \v{s}\fi{}auskas},
  \citenamefont {Wragg}, \citenamefont {van~der Hart},\ and\ \citenamefont
  {Ivanov}}]{Medisauskas2015PRL}%
  \BibitemOpen
  \bibfield  {author} {\bibinfo {author} {\bibfnamefont {L.}~\bibnamefont
  {Medi\ifmmode~\check{s}\else \v{s}\fi{}auskas}}, \bibinfo {author}
  {\bibfnamefont {J.}~\bibnamefont {Wragg}}, \bibinfo {author} {\bibfnamefont
  {H.}~\bibnamefont {van~der Hart}}, \ and\ \bibinfo {author} {\bibfnamefont
  {M.~Y.}\ \bibnamefont {Ivanov}},\ }\href {\doibase
  10.1103/PhysRevLett.115.153001} {\bibfield  {journal} {\bibinfo  {journal}
  {Physical Review Letters}\ }\textbf {\bibinfo {volume} {115}},\ \bibinfo
  {pages} {153001} (\bibinfo {year} {2015})}\BibitemShut {NoStop}%
\bibitem [{\citenamefont {Mauger}\ \emph {et~al.}(2016)\citenamefont {Mauger},
  \citenamefont {Bandrauk},\ and\ \citenamefont {Uzer}}]{Mauger2016JPB}%
  \BibitemOpen
  \bibfield  {author} {\bibinfo {author} {\bibfnamefont {F.}~\bibnamefont
  {Mauger}}, \bibinfo {author} {\bibfnamefont {A.~D.}\ \bibnamefont
  {Bandrauk}}, \ and\ \bibinfo {author} {\bibfnamefont {T.}~\bibnamefont
  {Uzer}},\ }\href {http://stacks.iop.org/0953-4075/49/i=10/a=10LT01}
  {\bibfield  {journal} {\bibinfo  {journal} {Journal of Physics B: Atomic,
  Molecular and Optical Physics}\ }\textbf {\bibinfo {volume} {49}},\ \bibinfo
  {pages} {10LT01} (\bibinfo {year} {2016})}\BibitemShut {NoStop}%
\bibitem [{\citenamefont {Bandrauk}\ \emph {et~al.}(2016)\citenamefont
  {Bandrauk}, \citenamefont {Mauger},\ and\ \citenamefont
  {Yuan}}]{Bandrauk2016JPB}%
  \BibitemOpen
  \bibfield  {author} {\bibinfo {author} {\bibfnamefont {A.~D.}\ \bibnamefont
  {Bandrauk}}, \bibinfo {author} {\bibfnamefont {F.}~\bibnamefont {Mauger}}, \
  and\ \bibinfo {author} {\bibfnamefont {K.-J.}\ \bibnamefont {Yuan}},\ }\href
  {http://stacks.iop.org/0953-4075/49/i=23/a=23LT01} {\bibfield  {journal}
  {\bibinfo  {journal} {Journal of Physics B: Atomic, Molecular and Optical
  Physics}\ }\textbf {\bibinfo {volume} {49}},\ \bibinfo {pages} {23LT01}
  (\bibinfo {year} {2016})}\BibitemShut {NoStop}%
\bibitem [{\citenamefont {Pisanty}\ and\ \citenamefont
  {Jim\'enez-Gal\'an}(2017)}]{Pisanty2017PRA}%
  \BibitemOpen
  \bibfield  {author} {\bibinfo {author} {\bibfnamefont {E.}~\bibnamefont
  {Pisanty}}\ and\ \bibinfo {author} {\bibfnamefont {A.}~\bibnamefont
  {Jim\'enez-Gal\'an}},\ }\href {\doibase 10.1103/PhysRevA.96.063401}
  {\bibfield  {journal} {\bibinfo  {journal} {Phys. Rev. A}\ }\textbf {\bibinfo
  {volume} {96}},\ \bibinfo {pages} {063401} (\bibinfo {year}
  {2017})}\BibitemShut {NoStop}%
\bibitem [{\citenamefont {Dorney}\ \emph {et~al.}(2017)\citenamefont {Dorney},
  \citenamefont {Ellis}, \citenamefont {Hern\'andez-Garc\'{\i}a}, \citenamefont
  {Hickstein}, \citenamefont {Mancuso}, \citenamefont {Brooks}, \citenamefont
  {Fan}, \citenamefont {Fan}, \citenamefont {Zusin}, \citenamefont {Gentry},
  \citenamefont {Grychtol}, \citenamefont {Kapteyn},\ and\ \citenamefont
  {Murnane}}]{Dorney2017PRL}%
  \BibitemOpen
  \bibfield  {author} {\bibinfo {author} {\bibfnamefont {K.~M.}\ \bibnamefont
  {Dorney}}, \bibinfo {author} {\bibfnamefont {J.~L.}\ \bibnamefont {Ellis}},
  \bibinfo {author} {\bibfnamefont {C.}~\bibnamefont
  {Hern\'andez-Garc\'{\i}a}}, \bibinfo {author} {\bibfnamefont {D.~D.}\
  \bibnamefont {Hickstein}}, \bibinfo {author} {\bibfnamefont {C.~A.}\
  \bibnamefont {Mancuso}}, \bibinfo {author} {\bibfnamefont {N.}~\bibnamefont
  {Brooks}}, \bibinfo {author} {\bibfnamefont {T.}~\bibnamefont {Fan}},
  \bibinfo {author} {\bibfnamefont {G.}~\bibnamefont {Fan}}, \bibinfo {author}
  {\bibfnamefont {D.}~\bibnamefont {Zusin}}, \bibinfo {author} {\bibfnamefont
  {C.}~\bibnamefont {Gentry}}, \bibinfo {author} {\bibfnamefont
  {P.}~\bibnamefont {Grychtol}}, \bibinfo {author} {\bibfnamefont {H.~C.}\
  \bibnamefont {Kapteyn}}, \ and\ \bibinfo {author} {\bibfnamefont {M.~M.}\
  \bibnamefont {Murnane}},\ }\href {\doibase 10.1103/PhysRevLett.119.063201}
  {\bibfield  {journal} {\bibinfo  {journal} {Phys. Rev. Lett.}\ }\textbf
  {\bibinfo {volume} {119}},\ \bibinfo {pages} {063201} (\bibinfo {year}
  {2017})}\BibitemShut {NoStop}%
\bibitem [{\citenamefont {Jim\'enez-Gal\'an}\ \emph {et~al.}(2018)\citenamefont
  {Jim\'enez-Gal\'an}, \citenamefont {Zhavoronkov}, \citenamefont {Ayuso},
  \citenamefont {Morales}, \citenamefont {Patchkovskii}, \citenamefont
  {Schloz}, \citenamefont {Pisanty}, \citenamefont {Smirnova},\ and\
  \citenamefont {Ivanov}}]{Jimenez2018PRA}%
  \BibitemOpen
  \bibfield  {author} {\bibinfo {author} {\bibfnamefont {A.}~\bibnamefont
  {Jim\'enez-Gal\'an}}, \bibinfo {author} {\bibfnamefont {N.}~\bibnamefont
  {Zhavoronkov}}, \bibinfo {author} {\bibfnamefont {D.}~\bibnamefont {Ayuso}},
  \bibinfo {author} {\bibfnamefont {F.}~\bibnamefont {Morales}}, \bibinfo
  {author} {\bibfnamefont {S.}~\bibnamefont {Patchkovskii}}, \bibinfo {author}
  {\bibfnamefont {M.}~\bibnamefont {Schloz}}, \bibinfo {author} {\bibfnamefont
  {E.}~\bibnamefont {Pisanty}}, \bibinfo {author} {\bibfnamefont
  {O.}~\bibnamefont {Smirnova}}, \ and\ \bibinfo {author} {\bibfnamefont
  {M.}~\bibnamefont {Ivanov}},\ }\href {\doibase 10.1103/PhysRevA.97.023409}
  {\bibfield  {journal} {\bibinfo  {journal} {Phys. Rev. A}\ }\textbf {\bibinfo
  {volume} {97}},\ \bibinfo {pages} {023409} (\bibinfo {year}
  {2018})}\BibitemShut {NoStop}%
\bibitem [{\citenamefont {Yuan}\ and\ \citenamefont
  {Bandrauk}(2018)}]{Bandrauk2018PRA}%
  \BibitemOpen
  \bibfield  {author} {\bibinfo {author} {\bibfnamefont {K.-J.}\ \bibnamefont
  {Yuan}}\ and\ \bibinfo {author} {\bibfnamefont {A.~D.}\ \bibnamefont
  {Bandrauk}},\ }\href {\doibase 10.1103/PhysRevA.97.023408} {\bibfield
  {journal} {\bibinfo  {journal} {Phys. Rev. A}\ }\textbf {\bibinfo {volume}
  {97}},\ \bibinfo {pages} {023408} (\bibinfo {year} {2018})}\BibitemShut
  {NoStop}%
\bibitem [{\citenamefont {Milo\ifmmode \check{s}\else
  \v{s}\fi{}evi\ifmmode~\acute{c}\else \'{c}\fi{}}(2016)}]{MilosevicPRA2016}%
  \BibitemOpen
  \bibfield  {author} {\bibinfo {author} {\bibfnamefont {D.~B.}\ \bibnamefont
  {Milo\ifmmode \check{s}\else \v{s}\fi{}evi\ifmmode~\acute{c}\else
  \'{c}\fi{}}},\ }\href {\doibase 10.1103/PhysRevA.93.051402} {\bibfield
  {journal} {\bibinfo  {journal} {Physical Review A}\ }\textbf {\bibinfo
  {volume} {93}},\ \bibinfo {pages} {051402} (\bibinfo {year}
  {2016})}\BibitemShut {NoStop}%
\bibitem [{\citenamefont {Ayuso}\ \emph {et~al.}(2017)\citenamefont {Ayuso},
  \citenamefont {Jim\'enez-Gal\'an}, \citenamefont {Morales}, \citenamefont
  {Ivanov},\ and\ \citenamefont {Smirnova}}]{Ayuso2017NJP}%
  \BibitemOpen
  \bibfield  {author} {\bibinfo {author} {\bibfnamefont {D.}~\bibnamefont
  {Ayuso}}, \bibinfo {author} {\bibfnamefont {A.}~\bibnamefont
  {Jim\'enez-Gal\'an}}, \bibinfo {author} {\bibfnamefont {F.}~\bibnamefont
  {Morales}}, \bibinfo {author} {\bibfnamefont {M.}~\bibnamefont {Ivanov}}, \
  and\ \bibinfo {author} {\bibfnamefont {O.}~\bibnamefont {Smirnova}},\ }\href
  {http://stacks.iop.org/1367-2630/19/i=7/a=073007} {\bibfield  {journal}
  {\bibinfo  {journal} {New Journal of Physics}\ }\textbf {\bibinfo {volume}
  {19}},\ \bibinfo {pages} {073007} (\bibinfo {year} {2017})}\BibitemShut
  {NoStop}%
\bibitem [{\citenamefont {Baykusheva}\ \emph {et~al.}(2016)\citenamefont
  {Baykusheva}, \citenamefont {Ahsan}, \citenamefont {Lin},\ and\ \citenamefont
  {W\"orner}}]{Baykusheva2016PRL}%
  \BibitemOpen
  \bibfield  {author} {\bibinfo {author} {\bibfnamefont {D.}~\bibnamefont
  {Baykusheva}}, \bibinfo {author} {\bibfnamefont {M.~S.}\ \bibnamefont
  {Ahsan}}, \bibinfo {author} {\bibfnamefont {N.}~\bibnamefont {Lin}}, \ and\
  \bibinfo {author} {\bibfnamefont {H.~J.}\ \bibnamefont {W\"orner}},\ }\href
  {\doibase 10.1103/PhysRevLett.116.123001} {\bibfield  {journal} {\bibinfo
  {journal} {Physical Review Letters}\ }\textbf {\bibinfo {volume} {116}},\
  \bibinfo {pages} {123001} (\bibinfo {year} {2016})}\BibitemShut {NoStop}%
\bibitem [{\citenamefont {\'{A}lvaro Jim\'{e}nez-Gal\'{a}n}\ \emph
  {et~al.}(2017)\citenamefont {\'{A}lvaro Jim\'{e}nez-Gal\'{a}n}, \citenamefont
  {Zhavoronkov}, \citenamefont {Schloz}, \citenamefont {Morales},\ and\
  \citenamefont {Ivanov}}]{Jimenez2017OptExpress}%
  \BibitemOpen
  \bibfield  {author} {\bibinfo {author} {\bibnamefont {\'{A}lvaro
  Jim\'{e}nez-Gal\'{a}n}}, \bibinfo {author} {\bibfnamefont {N.}~\bibnamefont
  {Zhavoronkov}}, \bibinfo {author} {\bibfnamefont {M.}~\bibnamefont {Schloz}},
  \bibinfo {author} {\bibfnamefont {F.}~\bibnamefont {Morales}}, \ and\
  \bibinfo {author} {\bibfnamefont {M.}~\bibnamefont {Ivanov}},\ }\href
  {\doibase 10.1364/OE.25.022880} {\bibfield  {journal} {\bibinfo  {journal}
  {Opt. Express}\ }\textbf {\bibinfo {volume} {25}},\ \bibinfo {pages} {22880}
  (\bibinfo {year} {2017})}\BibitemShut {NoStop}%
\bibitem [{\citenamefont {Spanner}\ and\ \citenamefont
  {Patchkovskii}(2009)}]{Spanner2009PRA}%
  \BibitemOpen
  \bibfield  {author} {\bibinfo {author} {\bibfnamefont {M.}~\bibnamefont
  {Spanner}}\ and\ \bibinfo {author} {\bibfnamefont {S.}~\bibnamefont
  {Patchkovskii}},\ }\href {\doibase 10.1103/PhysRevA.80.063411} {\bibfield
  {journal} {\bibinfo  {journal} {Phys. Rev. A}\ }\textbf {\bibinfo {volume}
  {80}},\ \bibinfo {pages} {063411} (\bibinfo {year} {2009})}\BibitemShut
  {NoStop}%
\bibitem [{\citenamefont {Spanner}\ \emph {et~al.}(2012)\citenamefont
  {Spanner}, \citenamefont {Patchkovskii}, \citenamefont {Zhou}, \citenamefont
  {Matsika}, \citenamefont {Kotur},\ and\ \citenamefont
  {Weinacht}}]{Spanner2012PRA}%
  \BibitemOpen
  \bibfield  {author} {\bibinfo {author} {\bibfnamefont {M.}~\bibnamefont
  {Spanner}}, \bibinfo {author} {\bibfnamefont {S.}~\bibnamefont
  {Patchkovskii}}, \bibinfo {author} {\bibfnamefont {C.}~\bibnamefont {Zhou}},
  \bibinfo {author} {\bibfnamefont {S.}~\bibnamefont {Matsika}}, \bibinfo
  {author} {\bibfnamefont {M.}~\bibnamefont {Kotur}}, \ and\ \bibinfo {author}
  {\bibfnamefont {T.~C.}\ \bibnamefont {Weinacht}},\ }\href {\doibase
  10.1103/PhysRevA.86.053406} {\bibfield  {journal} {\bibinfo  {journal} {Phys.
  Rev. A}\ }\textbf {\bibinfo {volume} {86}},\ \bibinfo {pages} {053406}
  (\bibinfo {year} {2012})}\BibitemShut {NoStop}%
\bibitem [{\citenamefont {Smirnova}\ and\ \citenamefont
  {Ivanov}(2014)}]{bookChapter_SmirnovaIvanov_AttosecondAndXUVPhysics}%
  \BibitemOpen
  \bibfield  {author} {\bibinfo {author} {\bibfnamefont {O.}~\bibnamefont
  {Smirnova}}\ and\ \bibinfo {author} {\bibfnamefont {M.}~\bibnamefont
  {Ivanov}},\ }\enquote {\bibinfo {title} {Multielectron high harmonic
  generation: Simple man on a complex plane},}\ in\ \href {\doibase
  10.1002/9783527677689.ch7} {\emph {\bibinfo {booktitle} {Attosecond and XUV
  Physics}}}\ (\bibinfo  {publisher} {Wiley-VCH Verlag GmbH \& Co. KGaA},\
  \bibinfo {year} {2014})\ pp.\ \bibinfo {pages} {201--256}\BibitemShut
  {NoStop}%
\end{thebibliography}%

\end{document}